\documentclass[oneside,12pt]{article}

\usepackage{graphicx,comment}
\usepackage{caption}
\usepackage{subcaption}
\usepackage[top=1in,bottom=1in,left=1in,right=1in]{geometry}
\usepackage[sort&compress]{natbib} 
\usepackage{amsfonts, amsmath, amssymb, amsthm, constants, bbm}
\usepackage{mathrsfs}
\usepackage{authblk}
\usepackage[shortlabels]{enumitem}
\usepackage{booktabs}
\usepackage{float}
\usepackage{placeins}
\usepackage[utf8]{inputenc}

\usepackage{amsthm}
\makeatletter
\def\th@plain{%
  \thm@notefont{}
  \itshape 
}
\def\th@definition{%
  \thm@notefont{}
  \normalfont 
}
\makeatother
\newtheorem{theorem}{Theorem}

\usepackage{xcolor}
\definecolor{darkred}{RGB}{100,0,0}
\definecolor{darkgreen}{RGB}{0,100,0}
\definecolor{darkblue}{RGB}{0,0,150}

\usepackage{hyperref}
\hypersetup{colorlinks=true, linkcolor=darkred, citecolor=darkgreen, urlcolor=darkblue}
\usepackage{url}

\newtheorem{prp}{Proposition}
\newtheorem{lem}{Lemma}
\newtheorem{cor}{Corollary}

\theoremstyle{remark}
\newtheorem{rem}{Remark}

\newcommand{\secref}[1]{Section~\ref{sec:#1}}

\def\beq{\begin{equation}} 
\def\eeq{\end{equation}}
\def\beqn{\begin{eqnarray*}}
\def\eeqn{\end{eqnarray*}}
\def\Bitem{\begin{itemize}\setlength{\itemsep}{.2in}}
\def\bitem{\begin{itemize}\setlength{\itemsep}{.05in}}
\def\eitem{\end{itemize}}
\def\Benum{\begin{enumerate}\setlength{\itemsep}{.2in}}
\def\benum{\begin{enumerate}\setlength{\itemsep}{.05in}}
\def\eenum{\end{enumerate}}
\def\bmult{\begin{multline*}}
\def\emult{\end{multline*}}
\def\bcenter{\begin{center}}
\def\ecenter{\end{center}}
\def\bframe{\begin{frame}}
\def\eframe{\end{frame}}

\DeclareMathOperator*{\argmin}{arg\, min}

\def\cP{\mathcal{P}}

\def\cS{\mathcal{S}}

\def\bX{\mathbf{X}}

\def\bbE{\mathbb{E}}

\def\bbP{\mathbb{P}}

\def\bbR{\mathbb{R}}

\def\symd{\triangle}

\def\Xb{\overline X}

\newcommand{\N}{\mathcal{N}}
\renewcommand{\P}[1]{\mathbb{P}\left(#1\right)}
\newcommand{\Phn}[1]{\mathbb{P}_{\emptyset}\left(#1\right)}
\newcommand{\Pha}[1]{\mathbb{P}_{\cS}\left(#1\right)}
\newcommand{\PP}{\mathbb{P}}
\newcommand{\E}[1]{\mathbb{E}\left( #1 \right)}
\newcommand{\EBig}[1]{\mathbb{E}\Big( #1 \Big)}

\newcommand{\Var}[1]{\text{Var}\left( #1 \right)}
\newcommand{\Cov}[1]{\text{Cov}\left( #1 \right)}

\renewcommand{\exp}[1]{\mathrm{exp}\left(#1\right)}

\usepackage{bbm}
\newcommand{\ind}[1]{\mathbbm{1}\left\{ #1 \right\}}
\newcommand{\Bin}[2]{\text{Bin}(#1,#2)}

\newcommand{\given}{\;\middle|\;}

\usepackage{mathtools}
\DeclarePairedDelimiter\floor{\lfloor}{\rfloor}
\DeclarePairedDelimiter\ceil{\lceil}{\rceil}

\usepackage{mathtools}
\DeclarePairedDelimiter\abs{\lvert}{\rvert}%
\DeclarePairedDelimiter\norm{\lVert}{\rVert}%

\makeatletter
\let\oldabs\abs
\def\abs{\@ifstar{\oldabs}{\oldabs*}}
\let\oldnorm\norm
\def\norm{\@ifstar{\oldnorm}{\oldnorm*}}
\makeatother

\newcommand{\bigO}{\ensuremath{\mathcal{O}}}
\newcommand{\bigOp}[1][\mathbb{P}]{\ensuremath{\bigO_{\scriptscriptstyle{}#1}}}
\newcommand{\smallO}{\ensuremath{o}}
\newcommand{\smallOp}[1][\mathbb{P}]{\ensuremath{\smallO_{\scriptscriptstyle{}#1}}}
\newcommand{\smallOmega}{\ensuremath{\omega}}

\newcommand\numberthis{\addtocounter{equation}{1}\tag{\theequation}}

\urldef{\rivmurl}\url{https://data.rivm.nl/geonetwork/srv/dut/catalog.search#/metadata/1c0fcd57-1102-4620-9cfa-441e93ea5604}

\urldef{\cbsurl}\url{https://opendata.cbs.nl/statline/?dl=2096B#/CBS/nl/dataset/70072NED/table}

\begin{document}

\title{Anomaly Detection for a Large Number of Streams: A Permutation-Based Higher Criticism Approach}
\author[1]{Ivo V. Stoepker}
\author[1]{Rui M. Castro}
\author[2]{Ery Arias-Castro}
\author[1]{Edwin van den Heuvel}
\affil[1]{\small Department of Mathematics and Computer Science, Technische Universiteit Eindhoven, Eindhoven, The Netherlands} 
\affil[2]{Department of Mathematics and Halıcıoğlu Data Science Institute, University of California, San Diego, La Jolla, CA, USA}
\date{}
\maketitle

\thispagestyle{empty}

\begin{abstract}
\noindent Anomaly detection when observing a large number of data streams is essential in a variety of applications, ranging from epidemiological studies to monitoring of complex systems. High-dimensional scenarios are usually tackled with scan-statistics and related methods, requiring stringent modeling assumptions for proper calibration. In this work we take a non-parametric stance, and propose a permutation-based variant of the higher criticism statistic not requiring knowledge of the null distribution. This results in an exact test in finite samples which is asymptotically optimal in the wide class of exponential models. We demonstrate the power loss in finite samples is minimal with respect to the oracle test. Furthermore, since the proposed statistic does not rely on asymptotic approximations it typically performs better than popular variants of higher criticism that rely on such approximations. We include recommendations such that the test can be readily applied in practice, and demonstrate its applicability in monitoring the content uniformity of an active ingredient for a batch-produced drug product.

\vspace{0.3cm}\noindent\textbf{Keywords:} permutation test; minimax hypothesis testing; distribution-free testing
\end{abstract}

\section{Introduction} \label{sec:intro}

We study the problem of anomaly detection when a large number of data streams is observed. The streams themselves are not structured in any special way. In particular, there is no spatial or other proximity measure between streams, unlike in some areas like syndromic surveillance \citep{kulldorff2005stp}. We are interested in sparse alternatives; i.e. anomalies affect only a small fraction of the streams. Our analysis assumes an anomalous stream is affected in its entirety. This is common in literature (see for instance \cite{decentralized, energy-driven, hierarchical-censoring, optimal-distributed, cooperate, MR2450850}) and greatly simplifies the presentation of power guarantees. Note, however, that the proposed methodology is naturally applicable also when the streams are partially affected, and will still have some power.

Formally we consider $n$ streams indexed by $[n] \equiv \{1, \dots, n\}$, that are observed over a time period $[t] \equiv \{1, \dots, t\}$. Specifically we observe $\mathbf{X}\equiv(X_{ij} : i\in[n], j\in[t])$, where $X_{ij}$ is the value from stream $i \in [n]$ collected at time $j \in [t]$. Stream $i\in[n]$ corresponds to a vector denoted by $\mathbf{X}_i\equiv(X_{ij} :  j \in [t])$.

Following standard practice, we place ourselves in a statistical decision theory framework and cast detection as a hypothesis testing problem. For simplicity, we assume the variables to be real valued and independent.

In this paper, we consider a {\em stationary setting} where the distribution of the observations remain unchanged over time. In detail, under the null hypothesis all $X_{ij}$ are independent and identically distributed (i.i.d.) with common distribution function denoted by $F_0$.  The alternative hypothesis is similar, but there is a subset of anomalous streams $\cS \subset [n]$ such that for each $i\in\cS$ the distribution of $X_{ij}$, denoted by $F_{i}$, stochastically dominates $F_0$. In words, larger-than-usual values are observed in the streams indexed by $\cS$.  The subset $\cS$ is unknown and has no particular structure. Note that we are still assuming all the random variables $X_{ij}$ are independent. The sparsity assumption means the number of affected streams is believed to be small in comparison with the total number of streams, i.e. $|\cS| \ll n$. Succinctly, our hypothesis testing problem is therefore:
\begin{align*}
H_0:\qquad &\forall_{i\in[n]} \ X_{ij} \stackrel{\text{i.i.d.}}{\sim} F_0 \numberthis \label{hyp:general}\\
H_1:\qquad & \exists_{\cS \subset [n]}\ : \forall_{i\in\cS} \ X_{ij}\stackrel{\text{i.i.d.}}{\sim} F_{i} \text{ and } \forall_{i\notin\cS} \ X_{ij}\stackrel{\text{i.i.d.}}{\sim} F_0\ ,
\end{align*} 
where for all $i\in\cS$, distribution $F_i$ stochastically dominates $F_0$.

Despite its apparent simplicity, the above formulation and ensuing hypothesis test is quite suitable in a number of scenarios. For instance, in industrial settings the stream observations $\mathbf{X}_i$, $i\in[n]$ might correspond to a failure metric measured from $n$ machines. Under nominal conditions we expect these to be i.i.d.~samples from some distribution, while under the anomalous conditions some of these machines will occasionally display higher-than-normal values for the failure metric. In more complex settings the observations might correspond to residual values, obtained by fitting a model to the streams, as a way to ensure the streams can be meaningfully compared. For instance, in gene expression studies the streams might correspond to the normalized gene expression level of a number of replicates, so that under nominal conditions the expression levels are ``equalized'', and one is trying to determine if genes are differentially expressed under a specific environmental change. Obviously in this and similar settings some pre-processing of the data is needed to ensure all the streams are comparable. In Appendix~\ref{app:covid} we consider yet a different application, where the proposed methodology is applied to the daily counts of diagnosed COVID-19 cases in different municipalities in the Netherlands. This application also showcases how our methodology can be used when observations are dependent, by modeling the dependence and applying our methodology on the corresponding residuals.

\paragraph{Contributions:} Although a thorough characterization of the above testing scenario is far from trivial, it has been addressed quite extensively when the distribution $F_0$ is assumed known, and in particular when this is a normal distribution  \citep{MR1456646,Donoho2004}. Knowledge of that distribution is required to calibrate the proposed tests (e.g., by Monte-Carlo simulation) and ensure the resulting $p$-value computation is done properly. When $F_0$ is unknown there are various ways one can proceed. A natural nonparametric approach is calibration by permutation. This approach is, in fact, quite standard in syndromic surveillance \citep{kulldorff2005stp,19843331,huang2007spatial}.  It is also common in neuroimaging \citep{nichols2002nonparametric} and other contexts \citep{walther2010optimal,MR2792412}.

In this work we propose a variant of the higher criticism statistic that relies on permutations in both its definition and its calibration, leading to an \emph{exact} test. Furthermore, we show that, in the context of a one-parameter exponential family this test is asymptotically optimal, in the sense that it has essentially the same asymptotic performance as the best possible test with full knowledge of the underlying model - referred to as the oracle test. We provide practical recommendations so that the test is readily applicable in practice and demonstrate it has good performance in finite sample scenarios. In terms of testing power our procedure compares quite favorably with other proposed variants of higher criticism, particularly when streams are short (i.e., small $t$), since in its definition it does not rely on asymptotic approximations.

\paragraph{Related work:} Our methodology has two properties rarely encountered jointly in related literature; it can be employed in a distribution-agnostic manner, and it is properly calibrated regardless of sample size. We compare our work with related literature pertaining methodologies with either one of these properties.

A distribution-agnostic approach which is properly calibrated regardless of sample size is considered in \cite{Wu2014}. The setting is different from ours, but with some important similarities: in an association setting in genomics the higher criticism is used to test if some covariates (gene SNP's) have non-null association to the outcome variables (traits). A rare-weak assumption on the number and strength of the association is made. Calibration by permutation of the statistic is not the author's main interest, but is mentioned in the numerical section. The higher criticism statistic itself is computed based on asymptotic approximations. A similar approach has previously been used in applied settings, for example in \cite{Sabatti2009}. The review in \cite{donoho2015} discusses other works in the genomics domain. Calibration by permutation is mentioned here as well. In a different setting, \cite{arias-castro2018a} proposes a permutation-based scan statistic for the detection of structured anomalies (e.g., an interval) in a single stream. This test is shown to match the first order performance of the scan statistic calibrated with knowledge of the null distribution. 

Various authors have proposed distribution-agnostic methodologies which are asymptotically well-calibrated. In \cite{delaigle2009} and \cite{Delaigle2011} methodologies based on the higher criticism are proposed. However, in \cite{delaigle2009}, the setting considered is significantly distinct from ours. The authors consider observations $Z_j$ (with $j=1,\dots,p$) with unknown distribution, and test if each of these have mean equal to the empirical mean of auxiliary observed variables $W_i = (W_{i1},\dots,W_{ip})$ with $i=1,\dots,n_W$. The distribution of the $W_{ij}$ variables is assumed unknown and equal to the distribution of $Z_j$ under the null hypothesis. In that context, we are thus interested in the distribution of $V_j = Z_j - \frac{1}{n_W}\sum_i W_{ij}$. The distribution of $V_{j}$ (under the null) may be approximated based on the repeated observations, and normal approximations naturally arise, and the accuracy of this approximation is characterized through an asymptotic lens. The authors also study the behavior of the higher criticism statistic under dependencies of the component statistics $V_j$, and discuss a classification problem. Conversely, \cite{Delaigle2011} consider a setting closer to ours. The higher criticism statistic is applied to studentized stream means. To compute the higher criticism statistic, the authors estimate the distribution of these studentized stream means through bootstrapping. An asymptotic characterization for the statistic is obtained. For both works there are only asymptotic calibration guarantees.

Moving away from higher criticism, in \cite{arias-castro2017} the problem of detecting sparse heterogeneous mixtures from a nonparametric perspective is considered. The null distribution is assumed to be symmetric and the anomalous observations have a shift in mean, and the proposed tests are calibrated asymptotically. In \cite{Zou2017} a setting similar to ours is considered, where $n$ data streams of length $t$ are observed, and no knowledge on the data distribution is assumed. However, the inference goal is different: instead of hypothesis testing, their goal is to identify the set of anomalous streams. As a performance guarantee, the authors bound the probability of misidentifying the set of anomalous sequences. Their estimator is based on maximum-mean discrepancy and methodologically it is very different from our permutation approach. In \cite{Kurt2020} the focus is on nonparametric change-point detection in a real-time setting and the authors assume that there is a period of time guaranteed without anomalies. The authors show their decision statistics are asymptotically bounded under the null. Power properties are shown empirically.

Finally, \cite{Hall2008, Hall2010} show how the higher criticism statistic behaves under dependencies of the component statistics, assuming distributional knowledge. Our methodology can be applied in ways such that it can deal with some amount of dependencies, though the dependence must then be modeled and our methodology applied on residuals.

\paragraph{Organization:} Section~\ref{sec:intro} introduces and motivates the problem. Section~\ref{sec:normal_model} briefly reviews known results for the normal model and motivates the use of a simple quantized variant of the higher criticism statistic. In Section~\ref{sec:gen-res-exp} we introduce the class of distributional models we use as a benchmark for our theoretical power analysis and provide lower bounds on the anomalous signal strength for any test to be asymptotically powerful in the context of any one-parameter exponential family. Section~\ref{sec:permutation_test} proposes a novel permutation higher criticism test, and we show that it is essentially asymptotically optimal in the context of the one-parameter exponential family. We also propose a permutation max test and establish its power properties. In Section~\ref{sec:experiments} we examine the finite-sample performance of our methodology on simulated data, and also apply our methodology to a real dataset. All the proofs are deferred to Section~\ref{sec:proofs}. Some supporting results are provided in the Appendix to ensure the manuscript is self-contained.

\paragraph{Notation:} Throughout the paper we use standard asymptotic notation. Let $n\to\infty$, then $a_n = \bigO(b_n)$ when $| a_n / b_n |$ is bounded, $a_n = \smallO(b_n)$ when $a_n / b_n \to 0$, and $a_n=\smallOmega( b_n)$ when $b_n = \smallO(a_n)$. We also use probabilistic versions: $a_n = \bigOp(b_n)$ when $| a_n / b_n |$ is stochastically bounded\footnote{That is, for any $\varepsilon>0$ there is a $C_\varepsilon$ and $n_\varepsilon$ such that $\forall n>n_\varepsilon\ \P{| a_n / b_n |>C_\varepsilon}< \varepsilon$.} and $a_n = \smallOp(b_n)$ when $a_n / b_n$ converges to $0$ in probability. Unless otherwise stated, we consider asymptotic behavior with respect to $n \to \infty$. 

\section{An important case: the normal location model}\label{sec:normal_model}

This section considers a specific location model, which serves both as a benchmark and provides important insights needed for generalizations. These are well known results, but serve as a stepping stone to propose our methodology in the coming sections. Consider the setting where $F_0$ is the standard normal distribution with zero mean and unit variance. Under the alternative, there is a set $\cS \subset [n]$ indexing the anomalous streams, and the corresponding observations have an elevated mean $\mu > 0$ corresponding to the signal strength. Specifically, consider the following hypothesis test:
\begin{align*}
H_0:\qquad &\forall_{i\in[n]} \quad X_{ij} \stackrel{\text{i.i.d.}}{\sim} \N(0,1) \ . \numberthis \label{hyp:norm-loc}\\
H_1:\qquad &\exists_{\cS \subset [n]} : \forall_{i\in\cS} \quad X_{ij}\stackrel{\text{i.i.d.}}{\sim}\N(\mu,1), \text{ and } \forall_{i\notin\cS} \quad X_{ij} \stackrel{\text{i.i.d.}}{\sim} \N(0,1)\ .
\end{align*}
A natural question to ask is how large does $\mu$ need to be to ensure one can distinguish the two hypotheses. In this model it is obvious that the collection of stream means
\begin{equation}\label{eq:average}
Y_i(\mathbf{X}) \equiv \frac1{t} \sum_{j \in [t]} X_{ij}
\end{equation}
are jointly sufficient. The setting then reduces to that of the normal means model \citep{MR1456646,Donoho2004}. Note that in those works the alternative hypothesis is a bit different, and often $X_{ij}$ are assumed to be i.i.d.~samples from a sparse normal mixture. Nevertheless the asymptotic characterization of the optimal tests can be easily translated to the setting we are considering.

We take an asymptotic point of view. Let $\psi_n(\bX):\bbR^{nt}\to\{0,1\}$ be a sequence of tests, where the outcome $1$ indicates rejection of the null hypothesis. Consider a scenario where, under the alternative hypothesis, the size of $\cS$ is $s$ (that is, there are exactly $s$ anomalous streams). The \emph{risk} of a test is simply the sum of type I and worst-case type II error, namely
\[
R(\psi_n)\equiv\Phn{\psi_n(\bX)\neq 0} +\max_{\cS: |\cS| = s}\PP_{\cS}\left(\psi_n(\bX)\neq 1\right)\ ,
\]
where we denote $\PP_{\cS}$ the probability under the alternative, with anomalous set $\cS$ having signal strength $\mu$. A sequence of tests is said to be asymptotically powerful if $R(\psi_n)\to 0$ as $n\to\infty$, and it is said to be asymptotically powerless if $R(\psi_n)\to 1$.

To present an asymptotic characterization of the above hypothesis testing problem it is convenient to introduce the following parameterization. Let $\beta\in(0,1)$ be arbitrary (but fixed) and suppose $\cS$ has size
\begin{equation}\label{eqn:parameterization_size}
s = |\cS| = \lceil n^{1-\beta}\rceil\ ,
\end{equation} 
where $\lceil z \rceil$ denotes the smallest integer larger or equal to $z$. With this parameterization, we can identify two main regimes:
\bitem
\item {\em The dense regime: $0< \beta \leq 1/2$.} In this scenario a simple test based on $\sum_{i\in[n]} Y_i(\mathbf{X})$ is optimal. As stated in the beginning we are interested in sparse alternatives, therefore we do not focus much on the dense regime. Whenever appropriate we make comments on how our results and methods extend to this regime.

\item {\em The sparse regime: $1/2<\beta < 1$.} In this case a test based on $\sum_{i\in[n]} Y_i(\mathbf{X})$ will be asymptotically powerless.
Let $r>0$ be fixed and parameterize $\mu$ as
\begin{equation}\label{eqn:parameterization_sparse}
\mu = \sqrt{\frac{2r}{t}\log(n)}\ .
\end{equation}
Define the \emph{detection boundary}
\[
\rho^*(\beta) \equiv \begin{cases} 
\beta - 1/2\ , & 1/2 < \beta \le 3/4\ , \\ 
(1 - \sqrt{1-\beta})^2\ , & 3/4 < \beta < 1\ . 
\end{cases}
\]
It can be shown that, if $r<\rho^*(\beta)$ all tests are asymptotically powerless (see \cite{MR1456646} and Theorem~\ref{th:lower-bound}). Furthermore, if $r>\rho^*(\beta)$ there are tests (introduced below) that are asymptotically powerful. Note that there are actually two sub-regimes within the sparse regime, namely:
\bitem
\item {\em The moderately sparse regime: $1/2<\beta \leq 3/4$.} 
Here \cite{Donoho2004} show that the higher criticism test, based on the maximum of a normalized empirical process of the $(Y_i(\mathbf{X}) ,i\in[n])$ is asymptotically optimal. In fact, this is so in all regimes. Define the stream $p$-values as
\begin{equation}\label{eqn:stream-pval}
\mathfrak{p}_i \equiv  1 - \Phi\left(\sqrt{t}\ Y_i(\mathbf{X})\right)\ ,
\end{equation}
where $\Phi$ denotes the cumulative distribution function of a standard normal random variable. One variant of the higher criticism rejects the null hypothesis for large values of 
\begin{equation}\label{eqn:HC_donoho}
\max_{i\in[\lfloor \alpha_0 n\rfloor]} \sqrt{n}\frac{\frac{i}{n} - \mathfrak{p}_{(i)}}{\sqrt{\mathfrak{p}_{(i)}(1-\mathfrak{p}_{(i)})}}\ ,
\end{equation}
where $\alpha_0\in(0,1)$ and $\mathfrak{p}_{(1)} \le \cdots \le \mathfrak{p}_{(n)}$ are the ordered $p$-values.

\item {\em The very sparse regime: $3/4 < \beta < 1$.} 
Here \cite{Donoho2004} show that the test that rejects for large values of $\max_{i\in[n]} Y_i(\mathbf{X})$ achieves the detection boundary.  This is not the case in the moderately sparse regime, as shown in \cite{Arias-Castro2011a}.
Expressed in terms of $p$-values, the max test is entirely equivalent to multiple testing with Bonferroni correction, rejecting the null hypothesis for small values of $\mathfrak{p}_{(1)} = \min_{i\in[n]} \mathfrak{p}_i$.
\eitem
\eitem

\begin{rem}
Note that the parameterizations in Equations~\eqref{eqn:parameterization_size} and~\eqref{eqn:parameterization_sparse} can be presented in a more general way using asymptotic notation. In particular, we can simply assume $|\cS|=n^{1-\beta+\smallO(1)}$ as $n\to\infty$. For the sparse regime, it suffices to assume $\mu = \sqrt{2(r+\smallO(1))\log(n)/t}$.
\end{rem}

\subsection{A quantized higher criticism statistic}

In this section we show that a quantized higher criticism statistic is also optimal in the normal location model. Quantizing the higher criticism statistic has been done before in different ways - for example, in \cite{Arias-Castro2011a}, \cite{Wu2014} and others, mainly to facilitate the analysis. We discuss a quantization approach because the adaption and analysis of the higher criticism statistic, when using calibration by permutation, is easier for the quantized form introduced here. In contrast, adaptation to the permutation setting of the original higher criticism statistic in~\eqref{eqn:HC_donoho} is not straightforward as the computation of the test statistic itself requires knowledge of the null distribution - through the computation of the $p$-values in Equation~\eqref{eqn:stream-pval}.

We quantize a different variant of higher criticism. This variant was already introduced for analytical purposes in \cite{Donoho2004}. Let $q\geq 0$ and define
\[
N_q(\bX) \equiv \sum_{i\in[n]} \ind{Y_i(\mathbf{X}) \geq \sqrt{\frac{2q}{t}\log(n)}}\ .
\]
This is simply the number of streams whose average is larger or equal to $\sqrt{2(q/t)\log(n)}$. We refer to the latter value as the $q$-threshold. Define also $p_q$, the probability that a stream mean from the null exceeds that same threshold:
\begin{equation}\label{eqn:phc}
p_q \equiv 1- \Phi\left(\sqrt{2q\log(n)}\right)\ .
\end{equation}
Clearly, under the null hypothesis $N_q(\bX)$ is binomial with parameters $n$ and $p_q$, therefore we can standardize $N_q(\bX)$ and define the statistic
\begin{equation}\label{eqn:def-v}
V_q(\bX) \equiv \frac{N_q(\bX) - np_q}{\sqrt{np_q(1-p_q)}}\ .
\end{equation}

The $V_q(\bX)$ statistic, when maximized over $q \in [0,\infty)$, is closely related to the higher criticism statistic (see Remark~\ref{rem:HC_equivalence} below).

The following result summarizes several analytical steps in \cite{Donoho2004}.
\begin{prp}[\cite{Donoho2004}]\label{prop:vq-test}
Let $q\in[0,\infty)$ be fixed but arbitrary, and $h_n > 0$ with $h_n\to\infty$. Then $\Phn{V_q(\mathbf{X})\geq h_n}\leq \frac{1}{h_n^2}\to 0.$
Furthermore, for $\cS$ as in~\eqref{eqn:parameterization_size}, when $1/2<\beta<1$ (sparse regime), $h_n = n^{\smallO(1)}$, and $\mu=\sqrt{2(r/t)\log(n)}$ we have:
\bitem
\item If $r>(1-\sqrt{1-\beta})^2$ then $\Pha{V_1(\mathbf{X})\geq h_n}\to 1$.
\item If $r<1/4$ and $r>\beta-1/2$ then $\Pha{V_{4r}(\mathbf{X})\geq h_n}\to 1$.
\eitem
\end{prp}

The proof of Theorem~\ref{th:perm-hc-discrete-test-nullmean}, which is critical for our main result, hinges crucially in showing that important quantities inside the proof of Proposition~\ref{prop:vq-test} have the same asymptotic characterization when considering calibration by permutation. The heart of the proof of Proposition~\ref{prop:vq-test} is the realization that, under the alternative, $N_q(\bX)$ is the sum of two independent binomial random variables; $N_q(\bX) \sim \Bin{n-s}{p_q} + \Bin{s}{v_q}$, where
\[
p_q = n^{-q + o(1)} \qquad \text{and} \qquad
v_q = \begin{cases}
n^{-(\sqrt{q}-\sqrt{r})^2 + o(1)} &\text{ if } r < q \\
\frac{1}{2} &\text{ if } r = q\\
1 - n^{-(\sqrt{r}-\sqrt{q})^2 + o(1)} &\text{ if } r > q
\end{cases}\ .
\]
We will show a similar result in our setting, implying the result in this proposition can be used to complete the proof of Theorem~\ref{th:perm-hc-discrete-test-nullmean}. For completeness, a proof of Proposition~\ref{prop:vq-test} is also provided in Appendix~\ref{supp:vq}.
\begin{rem}
In the dense regime, it is also possible to show that, when $\mu=(1/\sqrt{t})n^{r-1/2}$ with $r>\beta$, we have $\Pha{V_0(\mathbf{X})\geq h_n}\to 1$. This implies the higher criticism statistic is also optimal in that regime under the normal model.
\end{rem}

Proposition~\ref{prop:vq-test} shows that, depending on the combination of sparsity $|\cS|$ and signal strength $\mu$ we might want to consider a test based on $V_q(\mathbf{X})$ for different values of $q$. Interestingly, the value $q=1$ suffices for the very sparse regime. For the moderately sparse regime the situation is more intricate, and it turns out the choice $q=4r$ is essentially optimal (unless $r \geq 1/4$, as then $q=1$ suffices). Obviously we do not know $r$, so the only way we can capitalize on this knowledge is to scan over a suitable selection of values for $q$, that hopefully include (or approximate) the optimal value. To keep the analysis simple and avoid the use of advanced tools in empirical processes we scan only over a coarse subset of values for $q$. The following result is an almost immediate consequence of Proposition~\ref{prop:vq-test}.
\begin{prp}[Quantized HC]\label{th:hc-discrete-test}
Let $Q\subseteq[0,1]$ be the set
\begin{equation}\label{eq:original-grid}
Q = \left\{0,\frac{1}{k_n}, \frac{2}{k_n}, \dots, 1 \right\}\ ,
\end{equation}
where $k_n \rightarrow \infty$ and $k_n = n^{\smallO(1)}$. Define the test statistic
\[
T(\bX)=\max_{q \in Q} \ V_q(\bX)\ .
\]
Finally, let $h_n = \smallOmega(\sqrt{k_n})$ and $h_n=n^{\smallO(1)}$, and consider the test that rejects the null hypothesis when $T(\bX)\geq h_n$. This test has vanishing type I error. When $\beta>1/2$ and $\mu=\sqrt{2(r/t)\log(n)}$ this test is asymptotically powerful provided $r>\rho^*(\beta)$.
\end{prp}

The proof follows almost immediately by using a union of events bound over the grid elements. Under the alternative, the only intricate case is the moderately sparse regime. However, since the grid $Q$ with $k_n+1$ elements becomes denser as $n$ grows it will contain a value $q$ that differs at most $1/k_n$ from the optimal value $4r$, which suffices for our result.  A proof is provided in Appendix~\ref{supp:hc-disc} for completeness.

\begin{rem}
For the dense regime, one can prove that, if $\mu=(1/\sqrt{t}) n^{r-1/2}$ the test in Proposition~\ref{th:hc-discrete-test} is asymptotically powerful provided $r>\beta$. This follows immediately using the results from Proposition~\ref{prop:vq-test}, since the optimal value 0 is guaranteed to be in the grid.
\end{rem}

\begin{rem}\label{rem:HC_equivalence}
The $V_q(\bX)$ statistic is closely related to the more familiar higher criticism statistic as in~\eqref{eqn:HC_donoho} as
$$\sup_{q\in[0,\infty)} \left\{V_q(\bX)\right\} = \max_{i \in [i_+]} \left\{ \sqrt{n}\frac{\frac{i}{n} - \mathfrak{p}_{(i)}}{\sqrt{\mathfrak{p}_{(i)}(1-\mathfrak{p}_{(i)})}} \right\}\ ,$$
where $i_+$ is the largest value of $i$ for which $\mathfrak{p}_{(i)}<1/2$. See Appendix~\ref{supp:vq-equiv} and Lemma~\ref{lem:hc-vq-equiv}.
\end{rem}

\section{Generalization to exponential families}\label{sec:gen-res-exp}

The results of the previous section were specific for the normal location model. Nevertheless one can envision somewhat natural extensions of the approach and test statistics to the more general context of the exponential family. When studying distribution-free tests, it is customary to compare them with parametric tests \citep{MR758442}. As in \cite{arias2011detection} and \cite{arias-castro2018a}, we consider one-parameter exponential models (in natural form) as parametric models, as these play an important role in the related literature \citep{Donoho2004}.

\paragraph{One-parameter exponential models in natural form:} Let $F_0$ be a probability distribution on the real line, such that all moments are finite. Let $\mu_0$ and $\sigma_0^2$ denote respectively the mean and variance of $F_0$. The distribution might be continuous, discrete or a combination of the two.  In the exponential model there is a parameter $\theta_{i}$ associated with each $i\in[n]$, and the distribution $F_{i}\equiv F_{\theta_{i}}$ is defined through its density $f_{\theta_{i}}$ with respect to $F_0$: for $\theta \in [0, \theta_\star)$, define $f_\theta(x) = \exp{\theta x - \log \varphi_0(\theta)}$, where $\varphi_0(\theta) = \int e^{\theta x} {\rm d}F_0(x)$ and $\theta_\star = \sup\{\theta > 0: \varphi_0(\theta) < \infty\}$, assumed to be strictly positive (and possibly infinite). In other words, $f_{\theta_{i}}$ denotes the Radon-Nykodym derivative of $F_{\theta_{i}}$ with respect to $F_0$. Since a natural exponential family has the monotone likelihood ratio property it follows that $F_\theta$ is stochastically increasing in $\theta$ \cite[Lemma 3.4.2]{MR2135927}. Particular cases of this model are used in a variety of settings. For example, in many signal and image processing applications, a model with $F_\theta$ corresponding to normal distribution with mean $\theta$ and a fixed variance is common. In syndromic surveillance \citep{kulldorff2005stp}, a model with $F_\theta$ corresponding to a Poisson distribution is popular. Bernoulli models \citep{walther2010optimal} are also a particular case of this class, with $F_\theta$ corresponding to a Bernoulli distribution. 

Let $X_{ij}$ be independent observations, where all observations within stream $i$ are i.i.d.~with distribution function $F_{\theta_i}$, as defined above. Under the null hypothesis we have $\theta_i = 0$ for all $i$. Under the alternative, there is a subset of streams for which $\theta_i > 0$. Our hypothesis testing problem is therefore given by:
\begin{align*}
H_0:\qquad &\forall_{i\in[n]} \quad \theta_{i}=0 \ , \numberthis \label{hyp:exp}\\
H_1:\qquad &\exists_{\cS \subset [n]} : \forall_{i\in\cS} \quad \theta_{i} > 0 \text{ and } \forall_{i\notin\cS} \quad \theta_{i}=0\ .
\end{align*}
Keeping the same minimax stance as before, in the remainder, our analysis assumes the anomalous observations have the same distribution so $\theta_i = \theta > 0$ for $i\in\cS$. We keep the parameterization of the anomalous set as in~\eqref{eqn:parameterization_size}, namely $|\cS|=\lceil n^{1-\beta}\rceil$ for $\beta\in(0,1)$, and use the parameterization $\theta=\sqrt{2r/(\sigma_0^2 t)\log(n)}$. Like in the normal model, a detection boundary defines the region in the $(r,\beta)$ plane where any test will be asymptotically powerless:

\begin{theorem}\label{th:lower-bound}
Refer to the hypothesis testing problem in~\eqref{hyp:exp} and the parameterization $|\cS|=n^{1-\beta}$ with $\beta > 1/2$ and $\theta=\sqrt{2r/(\sigma_0^2 t)\log(n)}$ with $r>0$. Provided $t = \smallOmega(\log^3(n))$ any test will be asymptotically powerless when $r<\rho^*(\beta)$.
\end{theorem}

\begin{rem}
Perhaps surprisingly, if the stream lengths are large enough, the detection boundary in this broad family precisely matches the boundary of the normal location model discussed in Section~\ref{sec:normal_model}. The reason is twofold: the collection of stream means is (in our exponential family) a sufficient statistic, and the stream lengths are (with respect to the tails of the exponential family) not too short. This combination results in a setting where an optimal test considers the collection of $n$ stream means which are jointly normal \textit{enough}, such that one can asymptotically do no better than if the stream means were \textit{exactly} normal. If the stream means are not sufficiently normal, the detection boundary may differ from $\rho^*(\beta)$ (e.g, \cite{Donoho2004} present a characterization for the Subbotin family of distributions).
\end{rem}

In the next section, we develop a permutation-based test that is able to attain optimal asymptotic performance, giving a converse to the statement of Theorem~\ref{th:lower-bound}.

\section{Calibration by permutation}\label{sec:permutation_test}
The quantized higher criticism statistic proposed in Section~\ref{sec:normal_model} is asymptotically optimal in the normal location model, but to compute it and calibrate it (i.e. compute a test $p$-value) one needs to assume knowledge of the null distribution. With that knowledge, one could simply calibrate the test by Monte-Carlo simulation.
The goal in this section is to construct a test (and test statistic) that can be computed and properly calibrated without the knowledge of the null distribution. A natural and appealing idea is to calibrate these tests by permutation. The idea of using permutations is not new in the context of comparing multiple groups (e.g, consider the classical Kruskal--Wallis test or any test based on runs). However, such approaches cannot deal with sparse alternatives, as in the setting we are considering (particularly when $\beta>1/2$). Note that the validity of calibration by permutation critically depends on the independence between our observations under the null. Nonetheless, if one is willing to model the dependencies, then our methodology could be applied on the model residuals to draw inference. We exemplify this in Appendix~\ref{app:covid}.

Note that, in the sparse regime we are most interested in, the distribution of the permutation stream means should intuitively follow the true null distribution quite closely. Indeed, in the sparse regime it is well known that the small fraction of non-null signals get ``washed-out'' in the overall mean \citep{Donoho2004}. Heuristically one could similarly argue here that all permutation streams consist of mostly null observations, such that the permutation stream means ``washes-out'' the non-null signals. That being said, it is crucial to thoroughly quantify the extent of the contamination and account for dependencies induced by the permutation process as well. The dense regime is different, and one likely requires much longer streams (e.g., $t$ growing linearly with $n$) to obtain asymptotic power characteristics. Nevertheless, in the dense regime it may be more fruitful to consider instead other statistics to calibrate by permutation (like the Fisher's combined probability test \cite[Section 21.1]{Fisher1934}) since it is known the higher criticism statistic performs poorly in finite samples even with distribution knowledge. These research directions are out of the scope of this paper and remains an interesting avenue for future work.

\subsection{Permutation-based max test} 
As can be expected, using calibration by permutation to calibrate a test based on the value of the maximum stream mean yields a test that cannot be optimal for all values of $\beta$. Nevertheless it is insightful to consider this option for its simplicity.

The max test rejects the null hypothesis for large values of
$$\max_{i\in[n]} Y_i(\mathbf{X})\ , \text{ where }Y_i(\mathbf{X})=\frac{1}{t}\sum_{j\in[t]} X_{ij}\ .$$

Let $\Pi$ denote the set of permutations of $\{(i,j) : i \in [n], j \in [t]\}$. Note that there are $(nt)!$ distinct permutations. Let $\pi\in\Pi$ and $\mathbf{X}^\pi\equiv(X_{\pi(i,j)}, i\in[n], j\in[t])$. In words, $\mathbf{X}^\pi$ corresponds to the permuted data over both streams and time. Calibration by permutation amounts to computing a $p$-value as
\begin{equation}\label{eqn:max_p_value}
\cP_{\text{max-perm}}(\bX) = \frac{1}{(nt)!}\left|\left\{ \pi\in\Pi: \max_i\left\{Y_i(\mathbf{X}^\pi)\right\} \geq \max_i\left\{Y_i(\mathbf{X})\right\} \right\} \right| \ .
\end{equation}
In words, this is the proportion of times the value of the test statistic computed with permuted data is at least as large as the value of the test statistic computed on the original data. Although the computation of the $p$-value might seem computationally prohibitive, we can approximate it accurately by considering instead a uniform sample from the set of permutations. The statistical properties of this test can be characterized as follows:

\begin{theorem}[Permutation max test]\label{th:max-perm-test}
Let $\alpha\in(0,1)$ be arbitrary and refer to the hypothesis testing problem in~\eqref{hyp:exp}. Let $\theta=\sqrt{2r/(\sigma_0^2 t)\log(n)}$ with $r>0$. Consider the test that rejects $H_0$ if $\cP_{\text{max-perm}}(\bX) \leq \alpha$. This test has level at most $\alpha$. Furthermore let $t = \smallOmega(\log^3(n))$ and $\beta>0$. This test has power converging to one provided $r>(1-\sqrt{1-\beta})^2$.
\end{theorem}

This theorem shows the max test calibrated by permutation is only guaranteed to be optimal when $\beta\geq 3/4$, as expected. It is important to note the stream length $t$ starts playing a more prominent role when considering calibration by permutation. Clearly, if $t=1$ it is impossible to calibrate such a test by permutation, as the test statistic is actually invariant under permutations in that case. One expects that, for very small values of $t$ there might therefore be a loss of power in comparison to an oracle test calibrated with knowledge of the null distribution. The requirement for $t$ in the above theorem is needed to ensure the individual stream means have approximately Gaussian tails. For exponential tails this scaling seems necessary. We conjecture that if the tail of $F_0$ is sub-Gaussian the requirement can be relaxed to $t = \smallOmega(\log^2(n))$ and if it has bounded support $t=\smallOmega(\log(n))$ will suffice. 

It might seem the above theorem should follow directly from the results in \cite{arias-castro2018a} as one can interpret the max test as a scan test over intervals of length $t$ within $nt$ observations, but a more careful analysis is needed to get a sharp characterization of the detection boundary.

\subsection{Permutation-based higher criticism test}

Although the simplicity of the max test is appealing, the test is far from optimal in the moderately sparse regime (or the dense regime for that matter). Therefore we turn our attention to the possibility of calibrating the higher criticism statistic by permutation. Note that the statistic in Proposition~\ref{th:hc-discrete-test} critically depends on the location and scale parameters in our normal setting, and since the mean and variance of the null distribution are unspecified in the general setting \eqref{hyp:exp}, this needs to be taken into account. For simplicity, we first introduce a statistic dependent on the null mean and variance, and motivate later how our methodology can be used without such knowledge. 

\subsubsection{Known first two null moments}
Assuming that the first two null moments are known, we define:
\begin{equation}\label{eq:tildeNq_def}
\tilde N_q(\bX) \equiv \sum_{i\in[n]} \ind{Y_i(\mathbf{X}) - \mu_0 \geq \sqrt{\frac{2\sigma_0^2q}{t}\log(n)}}\ ,
\end{equation}
where $Y_i(\mathbf{X})$ is as before. We must also adapt the normalizing term $p_q$ which was previously based on the normal setting. In principle, the computation of this probability requires full knowledge of $F_0$. Instead, we obtain a surrogate value for $p_q$ by permutation as follows:
\begin{equation}\label{eq:tildePq_def}
\tilde P_q(\bX) \equiv \frac{1}{(nt)!}\left|\left\{ \pi\in\Pi:\ \ Y_1(\mathbf{X^\pi}) - \mu_0 \geq \sqrt{\frac{2\sigma_0^2 q}{t}\log(n)} \right\}\right| \ .
\end{equation}
Finally, define the counterpart of $V_q(\bX)$ as:
\begin{equation}\label{eq:tildeVq_def}
\tilde V_q(\bX)\equiv \frac{\tilde N_q(\bX) - n\tilde P_q(\bX)}{\sqrt{n\tilde P_q(\bX)(1-\tilde P_q(\bX))}}\ ,
\end{equation}
where we take the convention that $0/0=0$. For this statistic, we have the following result:

\begin{theorem}[Permutation-based higher criticism test with known null mean and variance]\label{th:perm-hc-discrete-test-nullmean}
Let $Q \subseteq [0,1]$ be the set defined in~\eqref{eq:original-grid} with $k_n \rightarrow \infty$ and $|Q| = k_n = n^{\smallO(1)}$. Define the test statistic:
\begin{equation}\label{eq:permutation_HC_nullstats}
\tilde T(\bX) \equiv \max_{q\in Q} \tilde V_q(\bX)\ .
\end{equation}
Finally, define the higher criticism permutation $p$-value for known null mean and variance as
\[
\tilde\cP_{\text{perm-hc}}(\bX)  \equiv \frac{1}{(nt)!}\left|\left\{ \pi\in\Pi: \tilde T(\bX^\pi) \geq \tilde T(\bX) \right\} \right| \ .
\]

Let $\alpha\in(0,1)$ be arbitrary and refer to the hypothesis testing problem in~\eqref{hyp:exp}. Consider the test that rejects $H_0$ if $\tilde\cP_{\text{perm-hc}}(\bX)\leq \alpha$. This test has level at most $\alpha$. Furthermore, let $\beta > 1/2$ and $\theta=\sqrt{2r/(\sigma_0^2t) \log(n)}$ with $r>0$, $t = \smallOmega(\log^3(n))$ and $t = n^{o(1)}$. This test has power converging to one provided $r>\rho^*(\beta)$.
\end{theorem}

Under the conditions of the theorem we see that the proposed test is optimal for all values $\beta>1/2$, unlike the max test characterized in Theorem~\ref{th:max-perm-test}. It is important to reflect on the conditions on the stream length $t$. The lower bound on the stream length is, as before, required to ensure sufficiently Gaussian tail behavior of the stream means. There is, however, an upper bound on the stream length which was not present in Theorem~\ref{th:max-perm-test}. This restriction is technical, and we conjecture it is an artifact of the proof. Namely, it is used to ensure that the permutation probability $\tilde P_q(\bX)$ is a sufficiently accurate surrogate for the true normalizing probability (corresponding to the probability in Equation~\eqref{eqn:phc} in the normal setting), leading to the result in Lemma~\ref{lem:pq-sharp-bound} (which is stated later and used to prove Theorem~\ref{th:perm-hc-discrete-test-nullmean}). We conjecture the results in Lemma~\ref{lem:pq-sharp-bound} and Theorem~\ref{th:perm-hc-discrete-test-nullmean} also hold for larger values of $t$, but extending their proofs would require very refined integro-local results that, to the best of our knowledge, are currently not available in the statistics and applied probability literature. These results merit investigation on their own, and are outside the scope of this paper. That being said, a simple modification of the methodology allows us to give guarantees for arbitrarily large streams, as stated in Corollary~\ref{cor:longstreams}.

\subsubsection{Unknown null moments}\label{sec:unknown}
The result in Theorem~\ref{th:perm-hc-discrete-test-nullmean} is theoretically appealing, but somewhat impractical as it requires knowledge of the mean and variance of the null distribution. However, this can be viewed as an artifact of scanning the values of $\tilde V_q(\bX)$ over too small of a grid $Q$. To ensure our test has the right power characterization, all that is needed is to ensure the \emph{optimal} value of $q$ is adequately approximated in the grid $Q$.  It turns out it is possible to construct a grid that can account for lack of knowledge of the null mean and variance, while still having adequate approximation properties.

To streamline the presentation we restate our test statistic unencumbered by the $q$-threshold parameterization as:
\begin{equation}\label{eq:stat_reparam}
\check N_\tau \equiv \sum_{i\in[n]} \ind{Y_i(\mathbf{X}) \geq \tau}\ , \check P_\tau(\bX) \equiv \frac{1}{(nt)!}\left|\left\{ \pi\in\Pi:\ \ Y_1(\mathbf{X^\pi}) \geq\tau \right\}\right| \ ,
\end{equation}
and 
\begin{equation}\label{eq:stat_reparam_v}
\check V_\tau(\bX)\equiv \frac{\check N_\tau(\bX) - n\check P_\tau(\bX)}{\sqrt{n\check P_\tau(\bX)(1-\check P_\tau(\bX))}}\ .
\end{equation}
Consider the grid
\begin{equation}\label{eq:complex-grid}
R \equiv \left\{i + \sqrt{\frac{2j^2}{t}\log(n)} : i\in\left\{\frac{k}{\sqrt{t}}\right\}_{k={-\sqrt{t\log(n)}}}^{\sqrt{t\log(n)}} \ , \ j\in\left\{\frac{k}{\sqrt{\log(n)}}\right\}_{k=0}^{\log(n)} \right\} \ .
\end{equation}
Note the expanding nature of this grid; as $n$ increases, the minimum and maximum gridpoint expand. Furthermore, due to the increasing gridsize, the grid elements approximate points within the grid elements with increasing accuracy. This grid is therefore sure to contain or approximate the asymptotically optimal threshold as $n$ increases, while the computation of the statistic $\max_{\tau\in R}\{ \check V_\tau(\bX)\}$ does not require any knowledge of the null distribution. There are other grid constructions that share similar properties, and for which our results hold, but we chose this for concreteness. We can state similar results for this statistic as in Theorem~\ref{th:perm-hc-discrete-test-nullmean}:
\begin{theorem}\label{th:perm-hc-discrete-test}
Let $R$ be the set defined in \eqref{eq:complex-grid}. Define the test statistic:
\begin{equation}\label{eq:permutation_HC}
\check T(\bX) \equiv \max_{\tau\in R} \check V_\tau(\bX)\ .
\end{equation}
Define the higher criticism permutation $p$-value as
\[
\check\cP_{\text{perm-hc}}(\bX) \equiv \frac{1}{(nt)!}\left|\left\{ \pi\in\Pi: \check T(\bX^\pi) \geq \check T(\bX) \right\} \right| \ .
\]
Let $\alpha\in(0,1)$ be arbitrary and refer to the hypothesis testing problem in~\eqref{hyp:exp}. Consider the test that rejects $H_0$ if $\check\cP_{\text{perm-hc}}(\bX)\leq \alpha$. This test has level at most $\alpha$. Furthermore, let $\beta > 1/2$ and $\theta=\sqrt{2r/(\sigma_0^2t) \log(n)}$ with $r>0$, $t = \smallOmega(\log^3(n))$ and $t=n^{o(1)}$. This test has power converging to one provided $r>\rho^*(\beta)$.
\end{theorem}

Note that no further assumptions are needed to attain the same results of Theorem~\ref{th:perm-hc-discrete-test-nullmean}. Importantly, the assumption $t=n^{o(1)}$ now plays two different roles: on the one hand it is needed for technical reasons, as explained earlier. On the other hand, it is required to ensure the grid is small enough (namely of size $n^{o(1)}$). In any case, the ``short'' streams requirement can be avoided by a simple modification of the methodology, leading to the following result:

\begin{cor}\label{cor:longstreams}
Consider the hypothesis testing problem in \eqref{hyp:exp}. Let $\alpha\in(0,1)$. There is a test not requiring knowledge of $F_0$ that has level at most $\alpha$. Furthermore it has power converging to one under the same conditions as in Theorem~\ref{th:perm-hc-discrete-test} except that $t$ may be arbitrarily large.
\end{cor}

The corollary is proved in Section~\ref{sec:proofs-cor:longstreams}. The modified methodology in Corollary~\ref{cor:longstreams} simply maps the original dataset to a new dataset by aggregating small groups of observations within each stream. This effectively creates a scenario with short streams. The methodology described in Section~\ref{sec:unknown} can then be used and its performance characterized by using Theorem~\ref{th:perm-hc-discrete-test}, leading to the stated result.

\subsubsection{A practical alternative}

Instead of considering the grid in \eqref{eq:complex-grid}, a more natural idea would be to approximate the null mean and variance by overall sample estimates. This would correspond to maximizing in \eqref{eq:permutation_HC} over the following data dependent grid:
\begin{equation}\label{eq:data-dep-grid-small}
\hat R(\bX) \equiv \left\{\Xb + \sqrt{\frac{2\sigma_X^2q}{t}\log(n)} : q\in Q \right\} \ ,
\end{equation}
with $Q$ the grid as in \eqref{eq:original-grid},
and
\begin{equation}\label{mean_var}
\Xb\equiv\frac{1}{nt}\sum_{i\in[n],j\in[t]} X_{ij}, \quad \sigma_X^2\equiv\frac{1}{nt} \sum_{i\in[n],j\in[t]} (X_{ij}-\overline X)^2 \ .
\end{equation}
Given the efficiency of the sample estimates above, as well as the relatively coarse approximation required to the optimal gridpoint for our test in Theorem~\ref{th:perm-hc-discrete-test} to work, there is little reason to suspect the data dependent grid approach would incur any loss of power. However, proving this rigorously greatly and nontrivially complicates the analysis; due to the grid being now dependent on the data, there is a need to carefully address the dependencies between the summands of $\check N_\tau(\bX)$, as well as carefully characterize the extra source of randomness introduced in $\check P_\tau(\bX)$. Such an analysis seems to require very refined integro-local results that, to the best of our knowledge, are currently not existing in the literature. Ultimately, the requirement for an approximation for the sample mean and variance stem from the discreteness of the grid $Q$ (so one can use a relatively crude union bound). Dropping the discreteness of the grid would require advanced results from empirical process theory that are beyond the scope of this paper.

\begin{rem}\label{rem:gridsize}
It has been suggested in related works \citep{arias2011detection, Wu2014} that scanning over a discrete grid for this type of statistic is merely a proof artifact, and the results are also valid when $q$ in~\eqref{eq:permutation_HC_nullstats} is optimized over the continuous interval $[0,\infty)$, which would incorporate the true value of the null statistics automatically. We have conducted extensive numerical experiments regarding the choice of grid and observed no deteriorating performance for extremely fine grid choices. However, calibration by permutation without a discrete grid leads to an optimization problem that is very demanding computationally. So, from a practical standpoint the discretization is quite advantageous as well.
\end{rem}

\begin{rem}
Instead of computing $\check P_q(\bX)$ by permutation, one might consider an asymptotic normal approximation as used in \cite{Sabatti2009} and \cite{Wu2014}. However, our numerical experiments (see Section~\ref{sec:comp-approx}) indicate the permutation-based approach in~\eqref{eq:tildePq_def} leads to tests with higher power.
\end{rem}

\begin{rem} 
For calibration by permutation of the $\check T(\bX)$ statistic, we encounter a minor complication; the test statistic \textit{itself} depends on a term $\check P_q(\bX)$ that must be computed by sampling permutations as well. Recognizing that $\check P_q(\bX)$ is invariant under permutations prevents the procedure from being computationally prohibitive; details are deferred to Section~\ref{sec:computation}.
\end{rem}

\begin{rem}
For the dense case (i.e. $\beta\leq1/2$) our preliminary results indicate that when $\theta=(\sigma_0/\sqrt{t})n^{r-1/2}$ with $r>0$ and $t=\omega(n)$ the permutation higher criticism test has power converging to one provided $r>\beta$. However, the analysis of this regime is quite different and out of the scope of this paper.
\end{rem}

\section{Experimental results}\label{sec:experiments}

This section thoroughly explores the potential of the proposed procedure. We begin by making some important computational remarks, followed by results on simulated data, and concluding with the application of the methodology to a real dataset.

\subsection{Computational complexity}\label{sec:computation}
The $p$-value of the permutation test in Theorem~\ref{th:perm-hc-discrete-test} is calculated by considering all possible data permutations. Naturally, this becomes prohibitive even for moderately large datasets, and instead a sample of the permutation distribution can be used to estimate $\check\cP_{\text{perm-hc}}(\bX)$.

At first glance computation and calibration of this test statistic might still seem prohibitive: for each permutation $\pi$ taken from $\Pi$, one computes $\check T(\bX^{\pi})$, which requires us to compute $\check P_q(\bX^{\pi})$. Naturally, we can estimate this quantity in a similar fashion using a random set of permutations. However, this would mean that every permutation sample $\pi$ used to compute $\check T(\bX^\pi)$ requires us to take a new set of permutation samples, which is disastrous computationally. However, note that $\check P_q(\bX)$ is invariant under permutations, namely $\check P_q(\bX) = \check P_q(\bX^\pi)$. Thus we can estimate $\check P_q(\bX)$ using a random sample of permutations, and then use these estimates to compute $\check T(\bX^\pi)$. Moreover, numerical experiments indicate we can use the same set of permutations to both estimate $\check P_q(\bX)$ and compute $\check T(\bX)$, without loss of performance or raising the type I error. This approach has therefore been used in the ensuing numerical experiments.

Finally, note that our test statistic is invariant under time-permutations within individual streams. So, although the size of $\Pi$ is $(nt)!$, there is an equivalence class of permutations that is much smaller, namely of size $\approx (nt)!/(t!)^n$. This further motivates why a Monte-Carlo approach works well even with a relatively small number of permutations.

\subsection{Simulation setting}\label{sec:sim-setup}
For our experimental results, we consider two different models: the normal location model, discussed in Section~\ref{sec:normal_model} which is commonly used as a benchmark, and a model where the underlying distributions are exponential. In the second model, the null distribution $F_0$ is exponential with mean $1/\lambda_0$ (where we chose $\lambda_0 = 1.5$). Following the parameterization set up in Section~\ref{sec:gen-res-exp} the observations in the anomalous streams are exponentially distributed with mean $1/(\lambda_0 - \theta)$, where $\theta\in[0,\theta_*)$ where $\theta_*=\lambda_0$. This second setting allows us to explore the influence of heavier tails. In fact, the exponential tails are the heaviest tails for which we provide asymptotic guarantees.

In all simulations considered we assume the observations in the anomalous streams have distribution $F_\theta$, where
\begin{equation}\label{eq:var-signal}
\theta_\tau = \tau\sqrt{\frac{2\rho^*(\beta)}{\sigma_0^2t}\log(n)}\ ,
\end{equation}
with $\tau\geq 0$. In words, we consider a signal strength that is a multiple of the asymptotically minimal signal strength $\rho^*(\beta)$ needed for detection. Asymptotically we know that when $\tau<1$ all tests are powerless, and when $\tau>1$ the permutation higher criticism test is powerful. In all simulations, test are calibrated at a significance level of $5\%$.

\subsection{Grid choice}\label{sec:grid}
Although the approach characterized in Theorem~\ref{th:perm-hc-discrete-test} is asymptotically sensible, it is somewhat cumbersome in practice and can potentially ignore important information in non-asymptotic settings. The data dependent grid $\hat R(\bX)$ in \eqref{eq:data-dep-grid-small} is easier to employ, and extensive numerical experiments show there is no observable performance loss when it is used. However, even this data dependent grid may leave out important information, if the global maximum of $\check V_\tau(\bX)$ is attained for $\tau > \Xb + \sqrt{2\sigma_X^2\log(n)/t}$. To safeguard the approach against this issue as well, one can extend the data dependent grid $\hat R(\bX)$, such that the grid contains a sufficiently large point $\tau^*$ for which $\check N_{\tau^*}(\bX) = 0$. A specific choice is
\begin{multline}\label{eq:grid-extended}
\hat R'(\bX) \equiv \left\{\Xb + \sqrt{\frac{2\sigma_X^2q}{t}\log(n)} : q\in \hat Q(\bX) \right\} \ , \text{ with } \\ \hat Q(\bX) \equiv \left\{\frac{1}{d_n}, \frac{2}{d_n}, \dots, \left(\frac{\max_{i,j}\{X_{ij}\} - \Xb}{\sigma_X}\right)^2\frac{t}{2\log(n)}\right\} \ ,
\end{multline}
with $d_n \to \infty$. Note that Lemma~\ref{lem:chernoff-term-bounds} ensures that $|\hat Q(\bX)| = \bigO_\PP\left(d_n\log(nt)\right)$, and therefore, provided $d_n = n^{o(1)}$, the grid-size is still small enough to ensure a union-bound argument is valid.

Due to the data-dependency of this grid a statement similar to Theorem~\ref{th:perm-hc-discrete-test} is currently unavailable. Extensive numerical experiments show there is no reason to suspect the data dependent grid performs poorer than its counterpart in Theorem~\ref{th:perm-hc-discrete-test}, and the shown numerical experiments therefore use the extended data dependent grid in \eqref{eq:grid-extended}.

\begin{figure}
\begin{subfigure}[t]{0.49\textwidth}
\includegraphics[width=\textwidth]{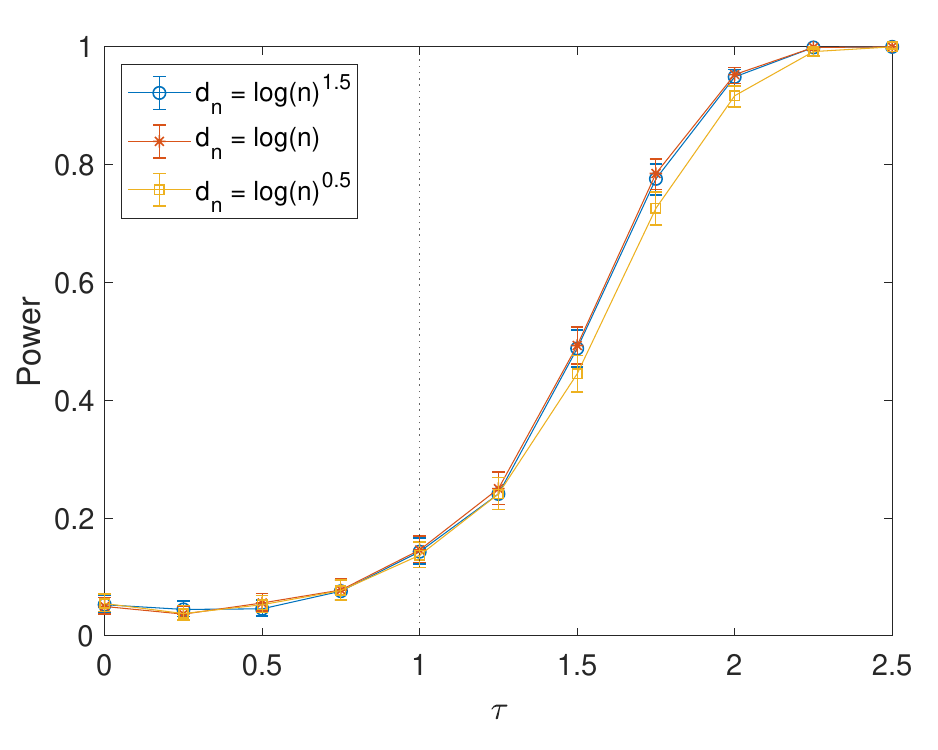}
\caption{Normal setting.}\label{fig:sim-q-norm}
\end{subfigure}
\begin{subfigure}[t]{0.49\textwidth}
\includegraphics[width=\textwidth]{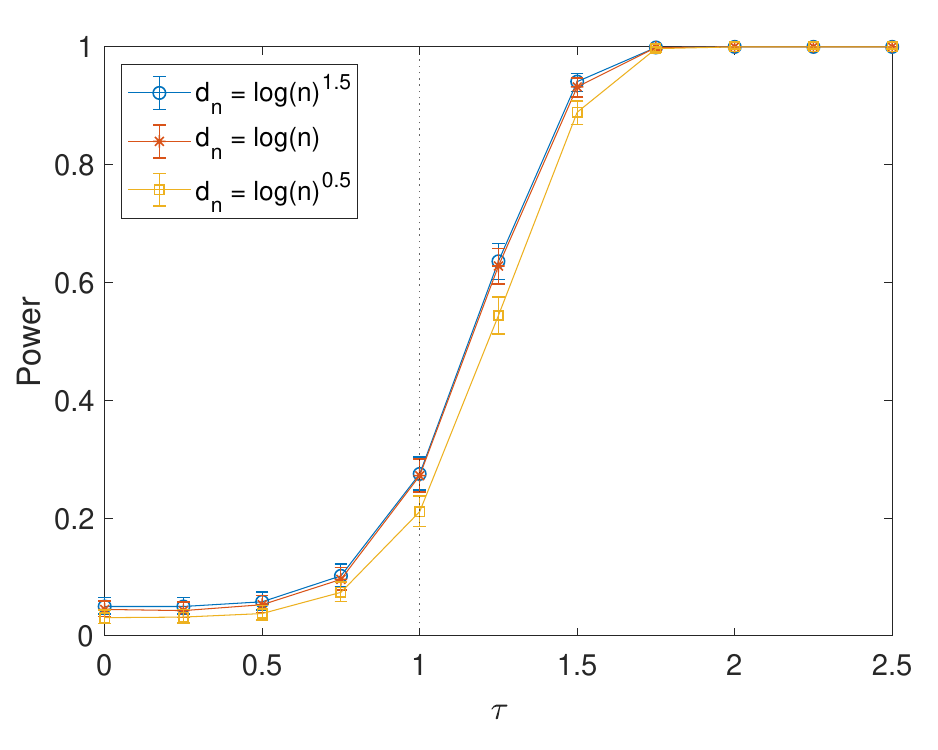}
\caption{Exponential setting.}\label{fig:sim-q-exp}
\end{subfigure}
\caption{Comparison of power of our permutation-based higher criticism test under different choices of $d_n$ as in~\eqref{eq:grid-extended}. The power of the test is depicted as a function of a $\tau$ in the parameterization in~\eqref{eq:var-signal}. We took $n=10^3$ streams with length $t=48$ and set $|\cS|=12$. For each test $10^3$ permutations were used. Each level was repeated $10^3$ times, resulting in the 95\% confidence bars depicted.}\label{fig:sim-q}
\end{figure}

To recommend the reader a choice for $d_n$, we simulate data from both models outlined in Section~\ref{sec:sim-setup} and compare the performance of the tests for different choices of $d_n$ as in~\eqref{eq:grid-extended}. We use $n=10^3$ streams of length $t=\lceil \log^2(n) \rceil = 48$. For the experiments shown we took  $|\cS|=12$, (i.e. $\beta \approx 0.64$, moderately sparse regime), but through extensive experimentation we observed our conclusions are not sensitive to the sparsity. The results are given in Figure~\ref{fig:sim-q}. We observe that a spacing of $1/d_n=1/\log(n)$ is a reasonable choice. In all remaining experiments, we therefore extend the grid as in~\eqref{eq:grid-extended} with $d_n = \log(n)$.

\subsection{Comparison with an oracle test}\label{sec:sim-oracle}
We now compare our methodology with the oracle test making full use of the null distribution. The oracle test statistic is defined in $\eqref{eqn:def-v}$ for the normal model. For the exponential model the stream averages are Gamma-distributed, and so one uses the quantiles of that distribution to compute $p_q$ instead. The oracle-test is then calibrated by Monte-Carlo simulation. For this calibration, $10^4$ simulations were used. In the presentation we use HC-permutation and HC-oracle to refer to the two tests. We compare the oracle test with our methodology with respect to two aspects:
\begin{itemize}
\item {\em Stream lengths.} For $t=1$ calibration by permutation leads to a powerless test. Therefore we expect the performance gap between an oracle and the proposed test to be large for short streams. However, we observe the finite-sample performance is already quite comparable to the oracle test even for very short streams.
\item {\em Performance loss.} Our test makes no use of knowledge of the null distribution. While asymptotically equivalent to first order, we expect some performance loss in finite samples compared to the oracle.
\end{itemize}

\begin{figure}[ht]
\begin{subfigure}[t]{0.49\textwidth}
\includegraphics[width=\textwidth]{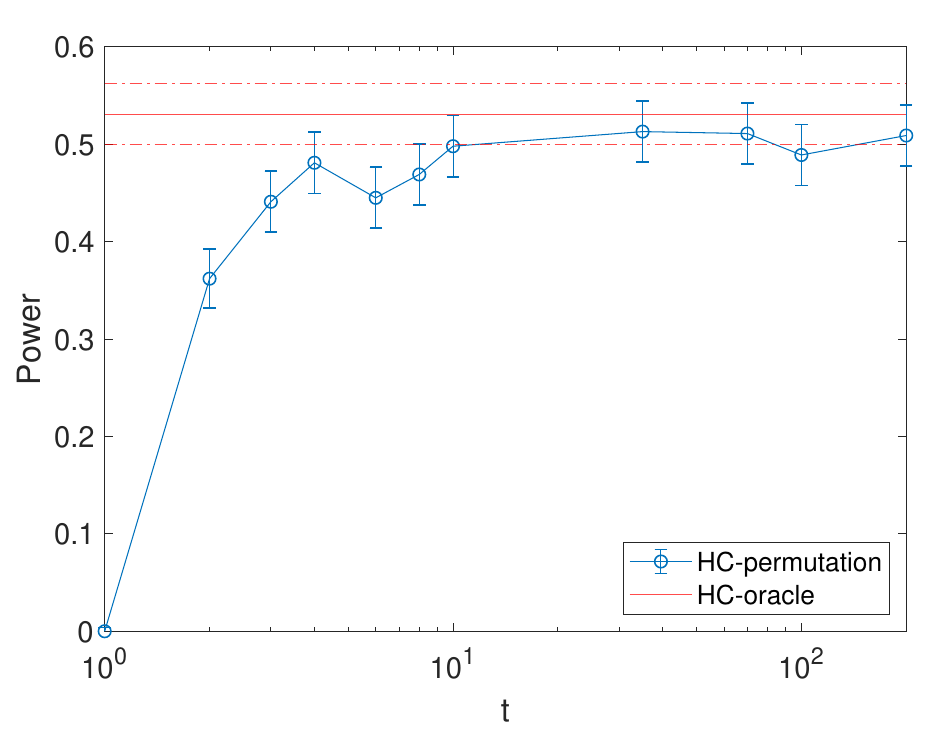}
\caption{Normal setting: $\tau=1.5$. }\label{fig:sim-t-norm}
\end{subfigure}
\begin{subfigure}[t]{0.49\textwidth}
\includegraphics[width=\textwidth]{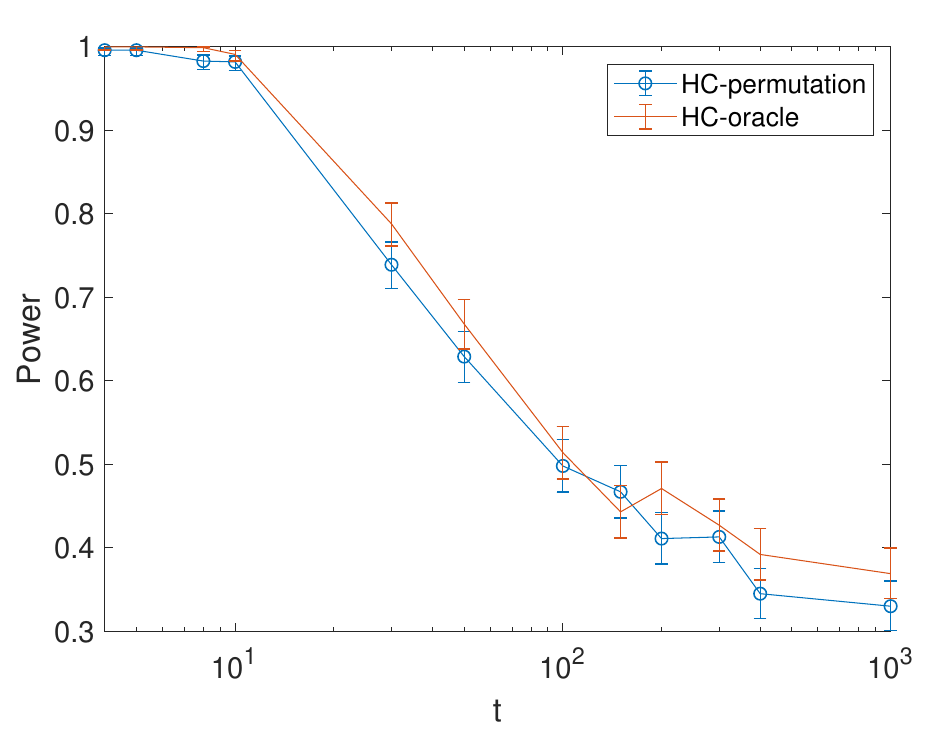}
\caption{Exponential setting: $\tau=1.25$.}\label{fig:sim-t-exp}
\end{subfigure}
\caption{Comparison of power of our permutation-based higher criticism test and the oracle test, as a function of a stream length $t$. We used the parameterization in~\eqref{eq:var-signal} with a fixed level for $\tau$. Note that the strength of the anomalous signal is decreasing in $t$ as discussed in the text. For the exponential model, the parameterization of the alternative is undefined for $t \leq 3$, so the $t$-axis starts at $t=4$. We took $n=10^3$ streams and set $|\cS|=12$. For the permutation tests, $10^3$ permutations were used. For the oracle test, the Monte-Carlo calibration is based on $10^4$ samples. Each level was repeated $10^3$ times, resulting in the 95\% confidence bars depicted.}\label{fig:sim-t}
\end{figure}

We use $n=10^3$ streams in our simulations. To study the influence of stream length we considered $|\cS| = 12$ (i.e. $\beta\approx0.64$, moderately sparse regime) but the results are qualitatively identical for other choices of sparsity. The signal strength was chosen slightly above the asymptotic detectability threshold, but ensuring the difference in performance was highlighted. For the exponential model very short streams would lead to a value $\theta\geq\lambda_0$ in the parameterization in~\eqref{eq:var-signal}, so the shortest stream considered is $t=4$. Figure~\ref{fig:sim-t} shows that the performance is comparable with respect to the oracle tests even at very short lengths, suggesting that finite-sample performance is much more forgiving than what our theoretical characterization might suggest, as there is only a very small loss of power in comparison to an oracle. Naturally, the permutation test has no power for streams of length $t=1$. Note that the strength of the anomalous signal is decreasing in $t$, such that in the normal model the oracle test performs identically for any value of $t$. In contrast, for the exponential model, the difference in null and alternative means is $\theta/(\lambda_0(\lambda_0-\theta))$ and is larger for smaller streams, so all tests perform worse at larger values of $t$.

\begin{figure}
\begin{subfigure}[t]{0.49\textwidth}
\includegraphics[width=\textwidth]{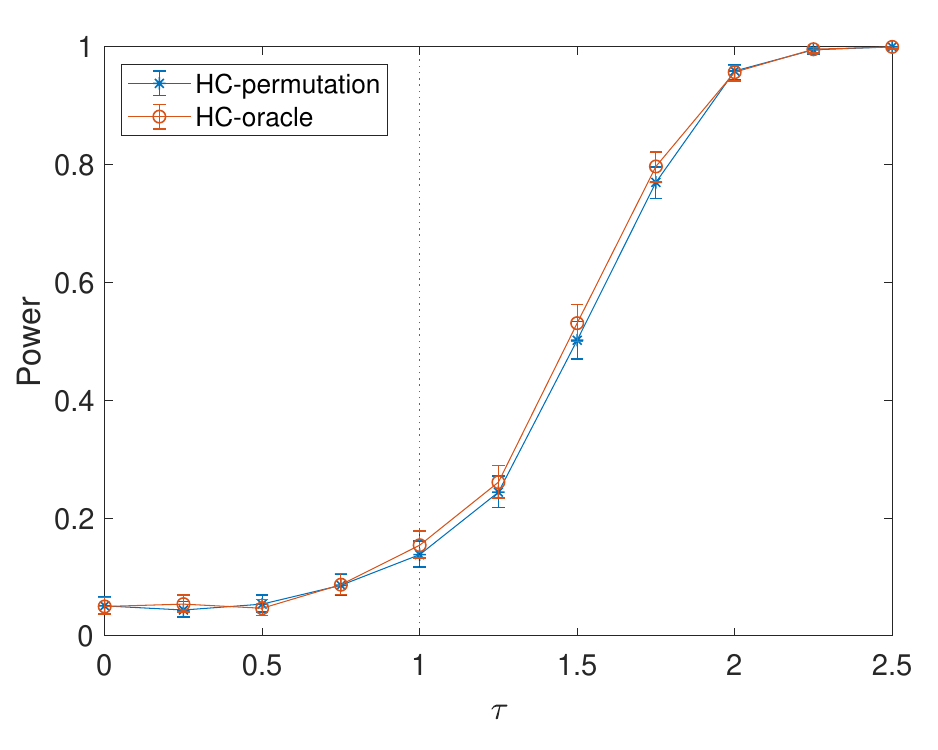}
\caption{Normal setting: $|\cS|=12$. }\label{fig:sim-oracle-exp-dense}
\end{subfigure}
\hfill
\begin{subfigure}[t]{0.49\textwidth}
\includegraphics[width=\textwidth]{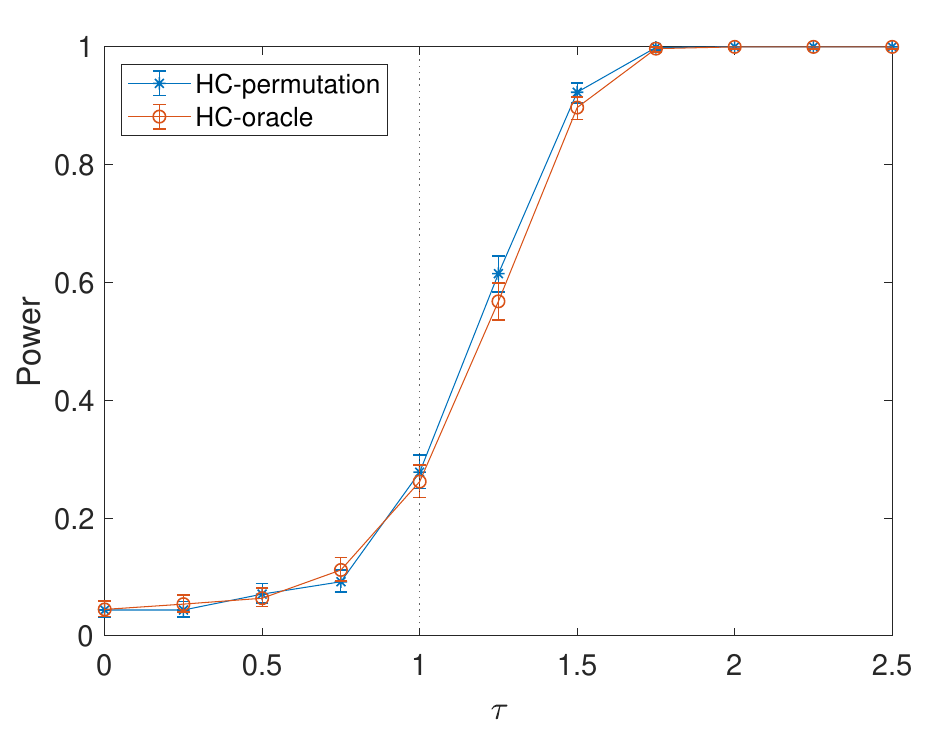}
\caption{Exponential setting: $|\cS|=12$. }\label{fig:sim-oracle-norm-dense}
\end{subfigure}
\begin{subfigure}[t]{0.49\textwidth}
\includegraphics[width=\textwidth]{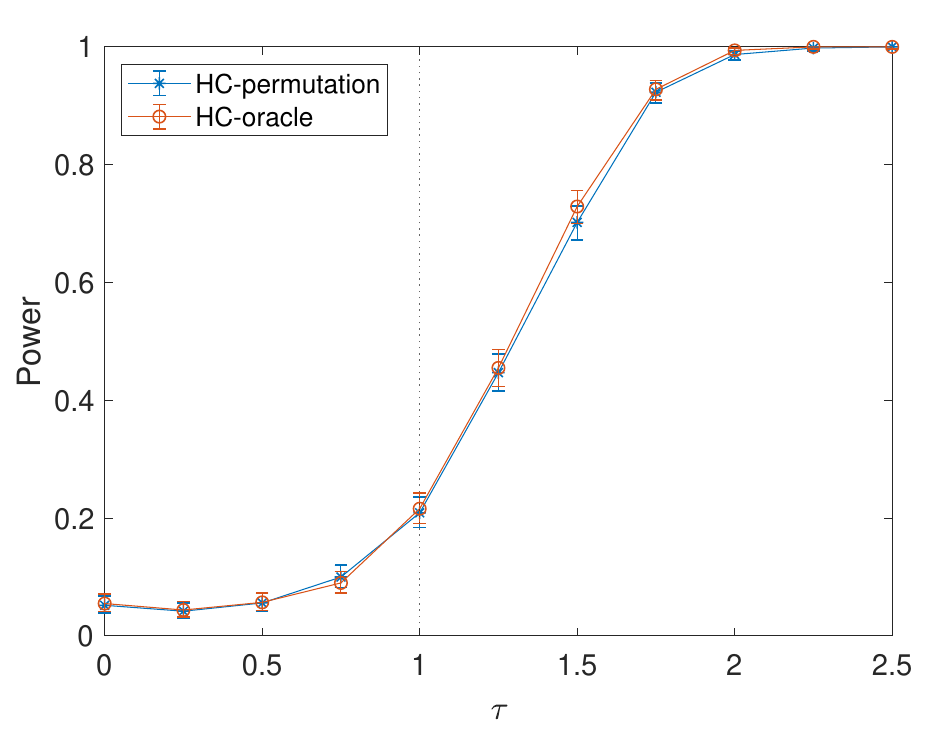}
\caption{Normal setting: $|\cS|=3$. }\label{fig:sim-oracle-norm-sparse}
\end{subfigure}
\hfill
\begin{subfigure}[t]{0.49\textwidth}
\includegraphics[width=\textwidth]{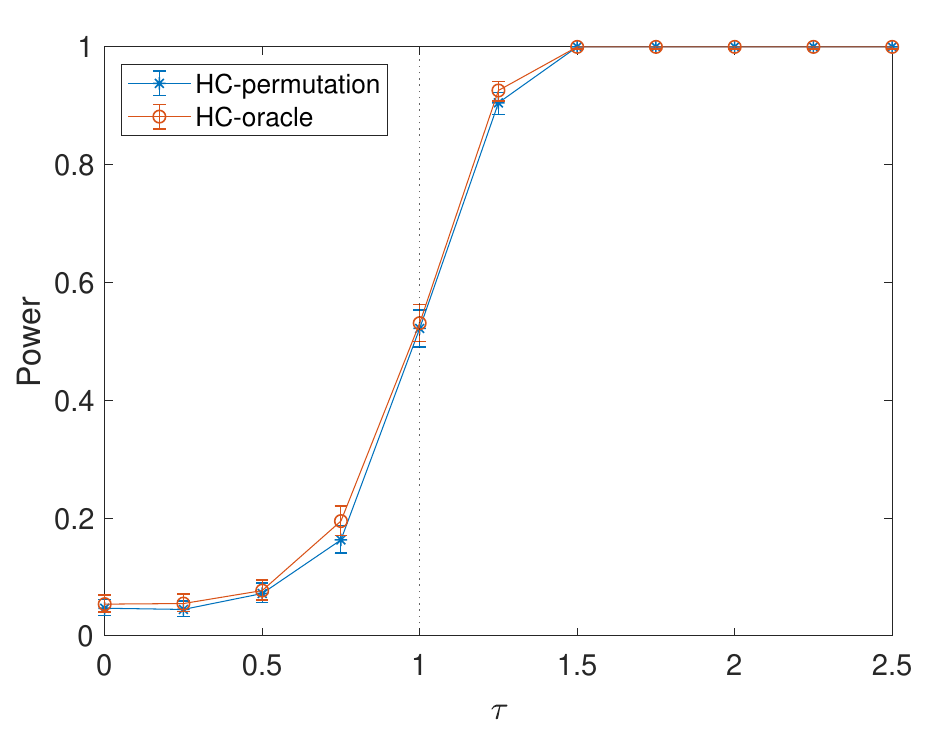}
\caption{Exponential setting: $|\cS|=3$.}\label{fig:sim-oracle-exp-sparse}
\end{subfigure}
\caption{Comparison of power of our permutation-based higher criticism test and the oracle test, as a function of a multiplicative factor $\tau$ in the parameterization in~\eqref{eq:var-signal}. We took $n=10^3$ streams with length $t=48$. For the permutation tests, $10^3$ permutations were used. For the oracle test, the Monte-Carlo calibration is based on $10^4$ samples. Each level was repeated $10^3$ times, resulting in the $95\%$ confidence bars depicted.}\label{fig:sim-oracle}
\end{figure}

We now turn our attention to the power characteristics of the proposed tests. We fix $n=10^3$ and $t=\lceil \log^2(n) \rceil = 48$ and consider $|\cS| = 12$ (i.e. $\beta\approx0.64$, moderately sparse regime), and $|\cS| = 3$ (i.e. $\beta\approx0.84$, very sparse regime). The results are plotted in Figure~\ref{fig:sim-oracle}, and we see that the proposed permutation higher criticism test has only minimal loss of power when compared to the oracle test.

\subsection{Comparison with approximation methods}\label{sec:comp-approx}

As mentioned earlier, several authors have considered practical variants of higher criticism where the permutation stream means are assumed to be approximately normal, and therefore the test statistic is computed based on a normal approximation (e.g., see \cite{Sabatti2009,Wu2014} and references therein). In essence, instead of computing (or, more appropriately, estimating) $\check P_q(\bX)$ as we do in Equation~\eqref{eq:stat_reparam}, one could instead compute:
\[
p_q^\Phi = 1 - \Phi(\sqrt{2q\log(n)}) \ ,
\]
which arises from the assumption that the standardized permutation stream means $Y_i(\bX^\pi)$ are approximately standard normal. The statistic can then be computed like before, and calibrated by permutation. We refer to this methodology as HC-approximation in what follows. We also present results for the max test.

\begin{figure}
\begin{subfigure}[t]{0.49\textwidth}
\includegraphics[width=\textwidth]{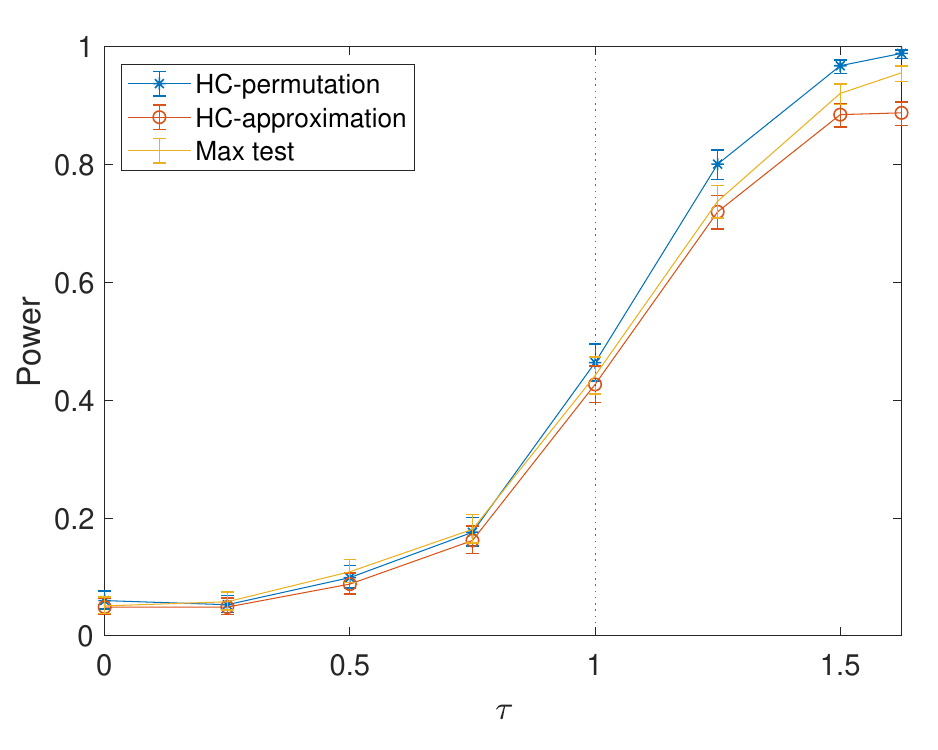}
\caption{Exponential setting: $t=4$.}\label{fig:sim-approx-1}
\end{subfigure}
\hfill
\begin{subfigure}[t]{0.49\textwidth}
\includegraphics[width=\textwidth]{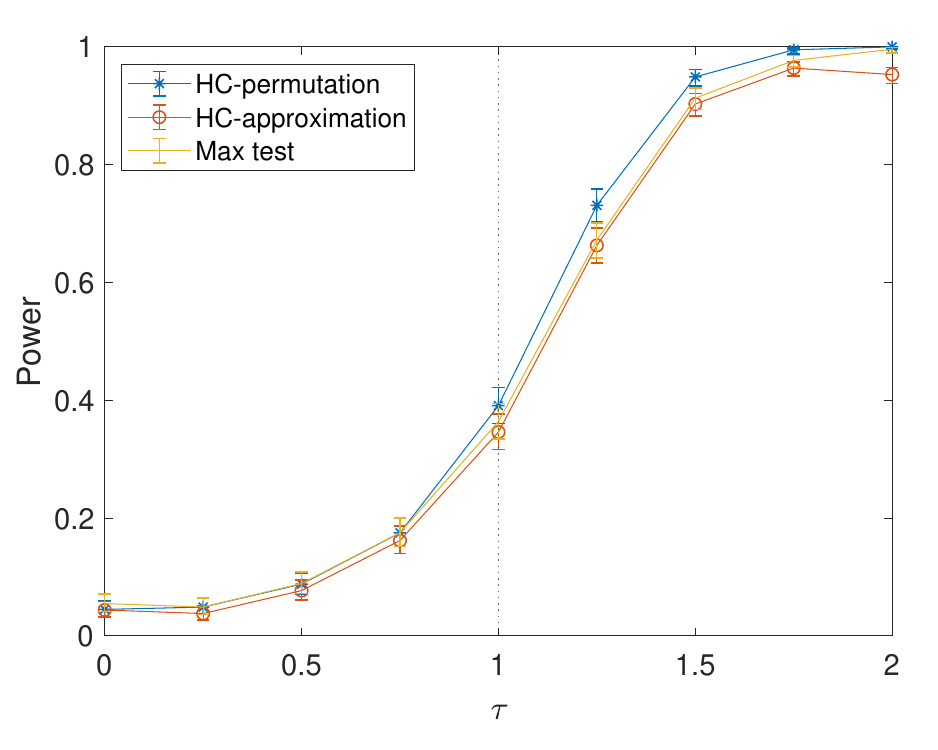}
\caption{Exponential setting: $t=6$.}\label{fig:sim-approx-2}
\end{subfigure}
\caption{Comparison of power of our permutation-based higher criticism test and a variant using asymptotic approximations. The power of the test is depicted as a function of a $\tau$ in the parameterization in~\eqref{eq:var-signal}. We took $n=10^2$ streams and set $|\cS|=12$. For each test $10^3$ permutations were used. Each level was repeated $10^3$ times, giving rise to the $95\%$ confidence bars depicted.}\label{fig:sim-approx}
\end{figure}

To compare the two approaches, we consider the exponential setting outlined in Section~\ref{sec:sim-setup} with $n=100$ streams and $t=4$ and $t=6$ as stream length. The reason to take a smaller number of streams and shorter stream length than before is to clearly emphasize the differences between the normal approximation and permutation approaches, as discussed below. We set $|\cS|=12$ (i.e. $\beta\approx0.64$), but similar results can be obtained in the other regimes. Figure~\ref{fig:sim-approx} depicts the results.

We see that there is a clear improvement in finite-sample performance when computing $\hat P_q(\bX)$ by permutation, as opposed to the normal approximation. As expected, the difference in power decreases as the streams get longer, since the normal approximation becomes more accurate. In any case, the approximation $p_q^\Phi$ underestimates the true value of $p_q$ when $q$ is large, as the tail of a Gamma distribution is much heavier than that of a normal distribution. Since $\hat P_q(\bX)$ does not rely on asymptotic approximations it more accurately approximates $p_q$, resulting on a more powerful test. Extensive experimentation showed that the permutation-based approach appears to be always superior when the underlying model is not normal.

\subsection{An analysis of content uniformity of a batch-produced drug product}\label{sec:cdc}
To showcase a possible application of our methodology, we consider a retrospective monitoring of the content uniformity of an active ingredient for a batch-produced drug product. Pharmaceutical companies are required to report on the quality of the production batches in an annual product quality review in line with good manufacturing practice. One aspect of this product quality review is the identification of potential production issues, which may be identified by a proper anomaly detection approach.

Batch production processes are well-controlled due to stringent quality requirements, but slight variability in the active ingredient within batches of the product may be expected. In the absence of production anomalies, this variability should be homogeneous across batches. Therefore, under nominal settings, batch-to-batch variation should come from the sampling variation within batches. Our methodology is therefore suitable to detect potential anomalies during the production process. 

\begin{figure}
\centering
\begin{subfigure}[t]{0.48\textwidth}
\includegraphics[width=\textwidth]{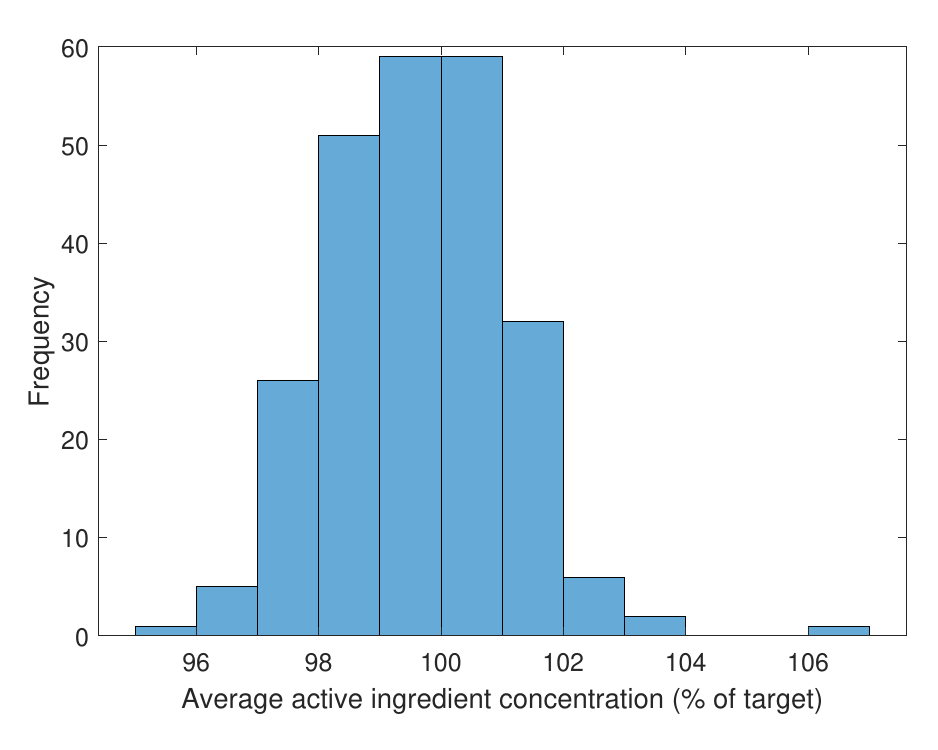}
\caption{Histogram of the average active ingredient concentration of each of the 242 batches.}\label{fig:cu-histogram}
\end{subfigure}
\hfill
\begin{subfigure}[t]{0.48\textwidth}
\includegraphics[width=\textwidth]{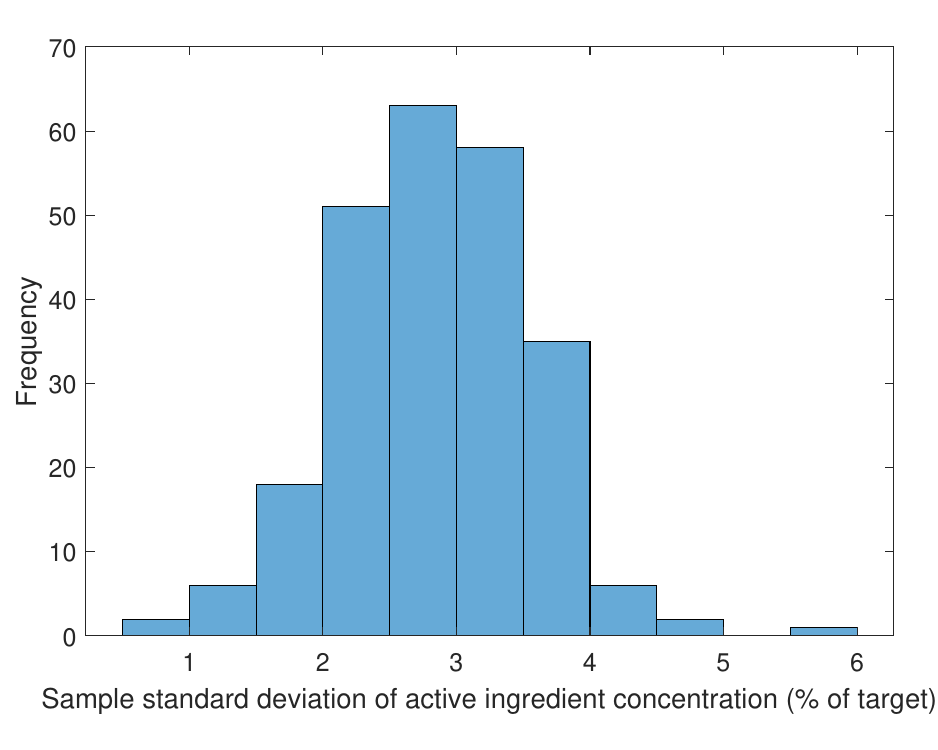}
\caption{Histogram of the sample standard deviation of the active ingredient concentration of each of the 242 batches.}\label{fig:cu-histogram-sd}
\end{subfigure}
\caption{Content uniformity data summary.}
\end{figure}

We consider data containing 242 production batches, each containing 10 measurements of the active ingredient, given as a percentage of the target concentration. Details pertaining to the drug or the active ingredient are not reported here due to the confidential nature of the data. The anonymized data is published as supplementary material. A histogram of the average active ingredient concentration per batch is given in Figure~\ref{fig:cu-histogram}, and a histogram of the sample standard deviation of the active ingredient concentration per batch is given in Figure~\ref{fig:cu-histogram-sd}.

Clearly, the batch with average concentration exceeding 106\%, corresponding to batch number 64, is remarkably large in light of the other observations, and a powerful test is unnecessary to warrant further investigation of this batch and the production process. Unexpectedly, application of our methodology on this data results in a $p$-value of 0. Therefore, it is more interesting to see if this clear anomaly is solely responsible for the detection, or if other signs of anomalies remain in the data after clear outliers are removed.

We may formalize the notion of ``clear'' outliers by using the max test statistic as described in Theorem~\ref{th:max-perm-test}. We use the 95\% quantile of the permutation maximum batch average active ingredient distribution, and identify all batches with average active ingredient exceeding this quantile as ``clear'' outliers. For our data, this 95\% quantile corresponds to an active ingredient concentration of 103.25\%, which leads to identifying batch 64 as a ``clear'' outlier, but also two other batches with large averages; batch 68 and 242.

Now, the application of our methodology on the remaining 239 batches leads to a $p$-value of $0.027$, indicating that, after removal of ``clear'' outliers, there are still signs of anomalies present in the data. Note that the data is reasonably normal, and the normal approximation alternative as proposed in Section~\ref{sec:comp-approx} leads to a $p$-value of $0.029$; indicating that our test may not lead to much power loss compared to normality-based methods when such assumptions are reasonable. Application of the max test as described in Theorem~\ref{th:max-perm-test} leads to a $p$-value of $0.095$, showcasing the benefit of using the higher criticism statistic in this context.

When the $p$-values of our test are small as above, one would ideally like to subsequently identify the anomalous batches. This requires significantly stronger signals than needed for detection alone. In \cite{Zou2017} a method for identification is given, with a bound on the probability of misidentifying the set of anomalous streams. Alternatively, in our context, a natural approach would be to compute $p$-values of each batch by permutation as
\begin{equation}\label{eq:per-pval}
p_i^\pi \equiv \frac{1}{(nt)!}\left|\left\{ \pi\in\Pi:\ \ Y_i(\mathbf{X}^\pi) \leq Y_i(\mathbf{X}) \right\}\right| \ ,
\end{equation}
and apply suitable multiple-testing methods to control the family-wise error rate (FWER) or the false discovery rate (FDR). A sensible way to control the FWER at level $\alpha$ is to use a Bonferroni correction and identify anomalous streams as those with $p$-values lower than $\alpha/n$. This method is intimately related to the method using the $(1-\alpha)$-quantile of the max test statistic presented above. For the FDR control one needs to take into consideration that the permutation $p$-values in~\eqref{eq:per-pval} are dependent. Standard methods as \cite{benjamini1995} have been adapted to work under arbitrarily dependent $p$-values \citep{benjamini2001}, but are known to be conservative.

Note, however, that there are situations where detection of anomalies is possible, but identification is not (see \cite{Donoho2004}). In the application we consider, knowledge of the existence of anomalous batches can be of use in the quality control process to gather additional data (such as retesting of batches) to fix production anomalies.

\begin{rem}
Note that the data in the analysis above seems to follow a normal distribution. In Appendix~\ref{app:covid}, we include an analysis of COVID-19 cases in the Netherlands where such normality is questionable, and in this context our methodology exhibits more power than the methodology based on normal approximations. However, to apply our methodology in this context one needs to address the serial dependencies present in the data. We discuss this in the appendix, noting that the validity of the conclusions depends naturally on the appropriateness of the model capturing such dependencies.
\end{rem}

\section{Discussion}\label{sec:discussion}
This paper considered the problem of detecting anomalous streams of data among many observed streams. This problem is closely related to the detection of sparse mixtures, where it is known that a test based on a higher criticism statistic is asymptotically optimal. Computation of this test statistic and calibration of the ensuing test requires the knowledge of the null (nominal) distribution. In this work we propose a distribution-free version of the statistic that is based on permutation, and we show that it is also asymptotically optimal in the context of one-parameter exponential families. The proposed test is exact, and appears to be superior in the sparse settings considered to alternative proposals that make use of asymptotic approximations. The numerical results in Section~\ref{sec:sim-oracle} show that the calibration by permutation results in nearly no loss of power in finite samples compared to an oracle test constructed and calibrated with full distribution knowledge. Its usefulness on real statistical problems is showcased in Section~\ref{sec:cdc} and Appendix~\ref{app:covid}.

Our asymptotic results are proven for the class of exponential models. In \cite{Delaigle2011} a less restrictive class of models is considered, for which the proposed testing methodology attains a similar detection boundary when streams are of length $t = n$. We conjecture that, when we consider a similar class of models (i.e., a location family where the base distribution satisfies some tail-decaying condition) the detection boundary $\rho^*(\beta)$ is attained as well, provided the streams are long enough. Specifically, when the tails of the base distribution are decaying exponentially, a stream length $t = \omega(\log^3(n))$ suffices, as shown in our work. For Pareto tails, we conjecture that at least $t = \omega(n)$ is needed to attain the $\rho^*(\beta)$ detection boundary, using large and moderate deviations theory results from \cite{mikosch:98}. One should keep in mind, however, that in an arbitrary location family the stream means are in general not a sufficient statistic for the location parameter, therefore the optimal detection boundary might not be characterized in the same way as for the normal model. For example, if $F_\theta$ is the uniform distribution over $[\theta,\theta+1]$ a test based on the stream means will surely be suboptimal.

Further research could discuss the possibility of adapting the Berk-Jones statistic \citep{Berk1979} in a similar fashion. Alternatively, one could investigate the possibility of replacing the observations by their ranks. This has numerical advantages and in some cases leads only to a minor loss of power \citep{arias-castro2018a}. A thorough analysis of this proposal is challenging as one must carefully account for the dependencies induced by the use of ranks, and remains an interesting avenue for future work. Another area of future research could discuss the identification of anomalous streams. In case permutation $p$-values are used, their dependence must be carefully characterized and accounted for. Such research could characterize the signal regimes in which anomalies are identifiable, merely detectable, or neither. In high-dimensional problems, the FWER may be a too stringent measure to control and other measures, such as FDR control, may be more appropriate. Work related to this problem can be found, for example, in \cite{Romano2007, romano2008}.

\subsection*{Acknowledgments}
I.V.S. would like to thank Richard A.J. Post for interesting discussions on applications of our methods. R.M.C. and E.A-C. would like to thank Ervin T\'{a}nczos and Meng Wang for helpful discussions in the early stages of this project.

\section{Proof of the main results}\label{sec:proofs}

In this section we provide the proofs of our main results. We begin with a simple technical lemma that greatly facilitates the presentation. The proof is a trivial consequence of standard tail-bounds for the normal approximation \citep[Section~7.1, Lemma~2]{feller_1968}.
\begin{lem}\label{lem:normalcdf-order}
Let $\Phi$ be the cumulative distribution function of the standard normal distribution and $x\geq 0$. Then
\[
1 - \Phi\left(\sqrt{2(x+\smallO(1))\log(n)}\right) = n^{-x + o(1)} \text{ as } n\to\infty\ .
\]
\end{lem}
Next, we argue that for the proofs of our main results, we can assume $F_0$ has zero mean and unit variance.

\subsection{Simplifying assumption}\label{app:gen-assumption}

To prove the results in this paper it suffices to consider the case where the nominal distribution $F_0$ has zero mean and unit variance. To see this suppose $F_0$ has arbitrary mean and variance $\mu_0$ and $\sigma_0^2$. Define $\tilde F_0(x)=F_0(\mu_0+\sigma_0 x)$. It is easy to see the distribution $\tilde F_0$ has zero mean and unit variance. Using this we can easily re-parameterize the hypothesis test in~\eqref{hyp:exp}.

Let $X$ be a random variable with distribution $F_\theta$ for some $\theta\in[0,\theta_*)$ and define $\tilde X=\frac{X-\mu_0}{\sigma_0}$. Define also $\tilde \varphi_0(\tilde\theta) = \int e^{\tilde\theta x} {\rm d}\tilde F_0(x)$, the moment generating function of $\tilde F_0$. It is easy to check that $\tilde X$ has density with respect to $\tilde F_0$ given by $\exp{\tilde \theta x-\log(\tilde\varphi(\tilde \theta))}$ where $\tilde \theta=\sigma_0 \theta$ (equivalently $\theta=\frac{1}{\sigma_0}\tilde \theta$). Therefore, statements in $\tilde\theta$ pertaining a zero mean and unit variance distribution can be translated to a general distribution by simple multiplication by a factor $1/\sigma_0$.

\subsection{Proof of Theorem~\ref{th:lower-bound}}
\begin{proof}
Without loss of generality and as explained in Section~\ref{app:gen-assumption} we assume that $F_0$ has mean zero and variance one, as this makes the arguments easier and less cluttered.

Let $\psi(\bX):\bbR^{nt}\to\{0,1\}$ denote an arbitrary test function. We begin by bounding the worst case risk of this test by the average risk, namely
\begin{align*}
R(\psi)&=\Phn{\psi(\bX)\neq 0} +\max_{\cS:|\cS|=s}\PP_{\cS}\left(\psi(\bX)\neq 1\right)\\
&\geq \Phn{\psi(\bX)\neq 0} +\frac{1}{\binom{n}{s}}\sum_{\cS:|\cS|=s}\PP_{\cS}\left(\psi(\bX)\neq 1\right)\ .
\end{align*}
The average risk can naturally be interpreted as the risk of testing the simple null hypothesis against a simple alternative, where $\cS$ is chosen uniformly at random over the class of all subsets of $[n]$ with cardinality $s$. Since we are doing a test between two simple hypotheses the optimal test (i.e., the test minimizing the average risk) is given by the Neyman-Pearson lemma, namely $\psi(\bX)=\ind{L\geq 1}$ where $L$ is the likelihood ratio given by
$$L\equiv \frac{1}{\binom{n}{s}}\sum_{\cS:|\cS|=s} \exp{\theta X_\cS-ts\log(\varphi_0(\theta))} \text{ with } X_\cS\equiv \sum_{i\in\cS,j\in[t]} X_{ij}\ .$$
The risk of this test can be easily expressed as $1-\frac{1}{2}\E{|L-1|}$, where the expectation is with respect to the null hypothesis (so all $X_{ij}$ are i.i.d.~with distribution $F_0$). To proceed we need to get an upper bound on $\E{|L-1|}$. A simple, but often useful way to proceed is to use Jensen's inequality to get
$$\E{|L-1|}\leq \sqrt{\E{(L-1)^2}}=\sqrt{\E{L^2}-1}\ ,$$
where the equality above follows since $\E{L}=1$. This approach is generally referred to as the \emph{second moment method}. To show any test is asymptotically powerless it suffices therefore to show that $\E{L^2}$ converges to one as $n\to\infty$.

To simplify the presentation let $\cS$ and $\cS'$ denote two independent random variables, and both independent from $\bX$. Both $\cS$ and $\cS'$ are sampled uniformly from the set $\{\cS\subset [n]:|\cS|=s\}$. Then clearly $L=\E{\exp{\theta X_\cS-ts\log(\varphi_0(\theta))}|\bX}$ and therefore
\begin{align*}
\E{L^2} &= \E{\exp{\theta X_\cS-ts\log(\varphi_0(\theta))}\exp{\theta X_{\cS'}-ts\log(\varphi_0(\theta))}}\\
&= \E{\exp{\theta (X_\cS+X_{\cS'})-2ts\log(\varphi_0(\theta))}}\\
&= \E{\exp{t |\cS \cap \cS'| (\log \varphi_0(2\theta) - 2\log \varphi_0(\theta))}}\ .
\end{align*}
For the last equality we used the fact that for all $\cS$ and $\cS'$ we have $X_\cS + X_{\cS'} = 2 X_{\cS \cap \cS'} + X_{\cS \symd \cS'}$ (in the previous expression $\symd$ denotes the symmetric set difference).

The beauty of the above result is that is reduces quantification of the risk to a statement about the moment generating function of the random variable $K\equiv |\cS \cap \cS'|$. Given the distribution $\cS$ and $\cS'$ we conclude that $K$ has an hypergeometric distribution with parameters $(n, s, s)$, and therefore $K$ is stochastically bounded from above by the binomial distribution with parameters $(s, \tfrac{s}{n-s})$. Using the well-know expression for the moment generating function of a binomial distribution we conclude that
$$\E{L^2} \leq \left(1 - \tfrac{s}{n-s} + \tfrac{s}{n-s} \kappa(\theta)^t\right)^s\ ,$$
where $\kappa(\theta) \equiv \varphi_0(2\theta)/\varphi_0(\theta)^2$. Therefore $\E{L^2}\to 1$ provided
$$\frac{s^2}{n-s} (\kappa(\theta)^t - 1) \to 0\ .$$

Consider now the specific parameterizations of $s$ and $\theta$ in the theorem statement. Note that when $t=\omega(\log^3 n)$ then necessarily $\theta\to 0$, so we can conveniently use a Taylor expansion of the moment generating function $\varphi_0(\theta)$ around $\theta=0$:
\begin{equation}\label{eq:taylor-phi0}
\varphi_0(\theta) = \varphi_0(0) + \theta\varphi_0'(0) + \frac{\theta^2}{2}\varphi_0''(0) + \bigO(\theta^3) \ ,
\end{equation}
as $\theta \rightarrow 0$. Using the fact that $F_0$ has mean zero and unit variance we get $\varphi_0(\theta) = 1 + \tfrac12 \theta^2 + \bigO(\theta^3)$ as $\theta \to 0$.  Simple asymptotic algebra yields that $\kappa(\theta) = 1 + \theta^2 + \bigO(\theta^3)$. Since $1+x\leq e^x$ we conclude that $\kappa(\theta)\leq \exp{\theta^2 + \bigO(\theta^3)}$.

When $\beta>1/2$ we conclude that
\begin{align*}
\frac{s^2}{n-s} (\kappa(\theta)^t - 1) &= (1+\smallO(1))n^{1-2\beta} (\kappa(\theta)^t - 1)\\
&\leq (1+\smallO(1))n^{1-2\beta} \left(\exp{t\theta^2 + \bigO(t\theta^3)}-1\right)\\
&= (1+\smallO(1))\exp{(1-2\beta)\log n} \left(\exp{2r\log n + \smallO(1)}-1\right)\ .
\end{align*}
The last expression converges to $0$ provided $1-2\beta+2r<0$ meaning that when $r<\beta-1/2$ any test is asymptotically powerless. This lower bound is tight when $\beta\in(1/2,3/4]$ (the moderately sparse regime) but it is a bit loose for the very sparse regime. However, a modification of the above argument allows us to get a tight lower bound when $\beta>3/4$.

\paragraph{The very sparse regime:} the main limitation of the second moment method as presented above has to do with the fact that the likelihood ratio statistic $L$ might take rather large values. Although this might be a rare occurrence, it can be enough to ensure the second moment is much larger than the first moment. A way to mitigate this issue is to consider a so-called truncated second moment method. Let $\Omega$ denote an arbitrary event and define the truncated likelihood ratio $\tilde L\equiv L\ind{\Omega}$. Clearly $\tilde L\leq L$ and therefore
\begin{align*}
\E{|L-1|} &= \E{|L-\tilde L+\tilde L-1|}\\
&\leq \E{|\tilde L-1|}+1-\E{\tilde L}\\
&\leq \sqrt{\E{\tilde L^2}-2\E{\tilde L}+1}+1-\E{\tilde L}\ ,
\end{align*}
where we used the triangle inequality and the fact that $\E{L}=1$, followed by Jensen's inequality. Therefore, to show a test is powerless it suffices to show that both $\E{\tilde L}$ and $\E{\tilde L^2}$ converge to one as $n\to\infty$. The choice of event $\Omega$ is therefore quite crucial. In the present context we are going to consider the event
\beq
\Omega = \left\{\max_{i \in [n]} Y_i < \underbrace{\sqrt{\frac{2(1+\eta) \log n}{t}}}_{\equiv \tau(\eta)}\right\} ,
\eeq 
where $\eta>0$ must be carefully chosen.

\paragraph{Truncated first moment:} Note first that $\E{\tilde L}=\E{L \ind{\Omega}}$ is the probability of $\Omega$ under the alternative hypothesis (where there is a set $\cS$ of anomalous streams and $\cS$ is chosen uniformly at random over the subsets of $[n]$ with cardinality $s$). Given the symmetry of the definition of $\Omega$ we see that $\E{\tilde L}=\bbP_\cS\left(\Omega\right)$ where $\cS$ is an arbitrary set with cardinality $s$. Without loss of generality let $\cS=[s]$. Then
\begin{align*}
\E{\tilde L} &= \bbP_\cS\left(\Omega\right)\\
&= 1-\bbP_\cS\left(\max_{i \in [n]} Y_i \geq \tau(\eta)\right)\\
&= 1-s\bbP_\cS\left(Y_1 \geq \tau(\eta)\right)-(n-s)\bbP_\emptyset\left(Y_1 \geq \tau(\eta)\right)\ .
\end{align*}
using the union bound in the last line. Using Lemma~\ref{lem:sum-normality} we conclude that
$$\bbP_\emptyset\left(Y_1 \geq \tau(\eta)\right)=n^{-1+\eta+\smallO(1)}\ ,$$
and provided $r\leq 1+\eta$
$$\bbP_\cS\left(Y_1 \geq \tau(\eta)\right)=n^{-(\sqrt{1+\eta}-\sqrt{r})^2+\smallO(1)}\ .$$
Therefore, when $r\leq 1+\eta$
\begin{align*}
\E{\tilde L} &= 1-n^{1-\beta} n^{-(\sqrt{1+\eta}-\sqrt{r})^2+\smallO(1)}-(n-s)n^{-1+\eta+\smallO(1)}\\
&= 1-n^{1-\beta-(\sqrt{1+\eta}-\sqrt{r})^2+\smallO(1)}-\bigO(1)\to 1\ ,
\end{align*}
provided $r<(\sqrt{1+\eta}-\sqrt{1-\beta})^2$. This means the first truncated moment converges to one for any $\eta>0$, provided $r<(1-\sqrt{1-\beta})^2$.

\paragraph{Truncated second moment:} bounding this term requires significantly more work. Begin by noting that
\begin{align*}
\E{\tilde L^2} &= \E{\exp{2\theta X_{\cS\cap \cS'}+\theta X_{\cS\symd\cS'}-ts\log(\varphi_0(\theta))}\ind{\Omega}}\\
&= \varphi_0(\theta)^{-st}\E{\exp{2\theta X_{\cS\cap \cS'}}\exp{\theta X_{\cS\symd\cS'}}\prod_{i\in[n]}\ind{Y_i<\tau(\eta)}}\\
&\leq \varphi_0(\theta)^{-st}\E{\exp{2\theta X_{\cS\cap \cS'}}\exp{\theta X_{\cS\symd\cS'}}\prod_{i\in \cS\cup\cS'}\ind{Y_i<\tau(\eta)}}\\
&\leq \varphi_0(\theta)^{-st}\E{\exp{t |\cS \cap \cS'| (\log \tilde\varphi_0(2\theta))-t |\cS \symd \cS'| (\log \tilde\varphi_0(\theta)}}\ ,
\end{align*}
where $\tilde\varphi_0(\theta)^t \equiv \E{\exp{\theta t Y_1} \ind{Y_1 < \tau(\eta)}}$. The steps above mimic the derivation for the regular second moment, and the main difference is that we now need to consider the moment generating function of a truncated distribution, instead of the original distribution. Clearly $\tilde\varphi_0(\theta)\leq\varphi_0(\theta)$ and so we conclude that
$$\E{\tilde L^2} \leq \E{\exp{t |\cS \cap \cS'| (\log \tilde\varphi_0(2\theta)-2\log \varphi_0(\theta))}}\ .$$
Define $\tilde\kappa(\theta) \equiv \tilde\varphi_0(2\theta)/\varphi_0(\theta)^2$. As before, to show the truncated moment converges to zero it suffices to show that
$$\frac{s^2}{n-s} (\tilde\kappa(\theta)^t - 1) \to 0\ .$$

Note that, the argument based on the untruncated second moment method indicates all tests are powerless if $r<\beta-1/4$. Since we are considering the case $\beta\geq 3/4$ this means that it suffices to treat only the case where $r\geq 3/4-1/2=1/4$. To get an upper bound on $\tilde\varphi_0(2\theta)^t$ we make use of the following technical result.
\begin{lem}\label{lem:integration}
Let $X$ be a real-valued random variable and let $f : \bbR \to [0,\infty)$ be one-to-one increasing and differentiable. 
Then, for any $\tau \in \bbR$,
\beq
\E{f(X) \ind{X \le \tau}} 
= \int_{-\infty}^\tau \bbP(X > x) f'(x) {\rm d}x\ .
\eeq
\end{lem}

To use this lemma we must get a good upper bound on $\Phn{Y_1>x}$ for $x\leq \tau(\eta)$. When $x\leq 0$ we trivially bound this probability by one, and for $0\leq x <\theta_*$ we make use of a simple Chernoff bound. In a similar fashion to the proof of Lemma~\ref{lem:sum-normality} we have
\begin{align*}
\Phn{Y_1> x} &\leq \Phn{\sum_{j\in[t]} X_{1j}\geq xt}\\
&\leq \exp{-t\left[\sup_{\lambda\in[0,\theta_*)} \{\lambda x-\log(\varphi_0(\lambda))\}\right]}\\
&= \exp{-t\left[\sup_{\lambda\in[0,\theta_*)} \{\lambda x-\log(1+\lambda^2/2+\bigO(\lambda^3))\}\right]}\\
&\leq \exp{-t\left[x^2-\log(1+x^2/2+\bigO(x^3))\}\right]}\\
&\leq \exp{-t(x^2/2+\bigO(x^3))}\ .
\end{align*}
Let $n$ be large enough so that $\tau(\eta)<\theta_*$. Applying the above result and Lemma~\ref{lem:integration} we get
\begin{align*}
\tilde\varphi_0(2\theta)^t &= \int_{-\infty}^{\tau(\eta)} \P{Y_1>x} 2\theta t \exp{2\theta t x} dx\\
&\leq \int_{-\infty}^0  2\theta t \exp{2\theta t x} dx \ + \ \int_0^{\tau(\eta)} \exp{-t(x^2/2+\bigO(x^3))}2\theta t \exp{2\theta t x} dx\\
&\leq 1 \ + \ 2\theta t \int_0^{\tau(\eta)} \exp{-t(x^2/2+\bigO(x^3))} \exp{2\theta t x} dx\\
&=1 \ + \ 2\theta t \exp{2\theta^2 t}\exp{\bigO(t \tau^3(\eta))} \int_0^{\tau(\eta)} \exp{-t(x-2\theta)^2/2)}  dx\\
&=1 \ + \ \sqrt{8\pi}(1+\smallO(1))(\theta\sqrt{t}) \exp{2\theta^2 t} \int^{(\tau(\eta)-2\theta)\sqrt{t}}_{-2\theta\sqrt{t}} \frac{1}{\sqrt{2\pi}}\exp{-y^2/2}  dy\\
&\leq 1 \ + \ \sqrt{8\pi}(1+\smallO(1))(\theta\sqrt{t}) \exp{2\theta^2 t} \Phi((\tau(\eta)-2\theta)\sqrt{t})\ ,
\end{align*}
 where in the second to last step we used the fact that $t=\omega(\log^3 n)$. At this point note that
 $$(\tau(\eta)-2\theta)\sqrt{t}=-\sqrt{2\left(\sqrt{4r}-\sqrt{1+\eta}\right)^2\log n}<0$$
when $r\geq 1/4$, provided we choose $\eta>0$ sufficiently small. Therefore using Lemma~\ref{lem:normalcdf-order} we conclude that
\begin{align*}
\tilde\varphi_0(2\theta)^t &\leq 1 \ + \ \sqrt{8\pi}(1+\smallO(1))\sqrt{2r\log n} n^{4r} n^{-(\sqrt{4r}-\sqrt{1+\eta})^2+\smallO(1)}\\
&= 1 \ + \ \sqrt{8\pi}(1+\smallO(1))\sqrt{2r\log n}\ n^{4r-(\sqrt{4r}-\sqrt{1+\eta})^2+\smallO(1)}\\
&= n^{4r-(\sqrt{4r}-\sqrt{1+\eta})^2+\smallO(1)}\ .
\end{align*}
With an analogous argument to the one used for $\tilde \varphi(2\theta)^{t}$ one can show that $(\varphi_0(\theta))^{2t}=n^{2r+\smallO(1)}$. Therefore
\begin{align*}
\frac{s^2}{n-s} (\tilde\kappa(\theta)^t - 1) &= \smallO(1)+n^{1-2\beta+2r-(\sqrt{4r}-\sqrt{1+\eta})^2+\smallO(1)}\ .
\end{align*}
The above converges to zero when $1-2\beta+2r-(\sqrt{4r}-\sqrt{1+\eta})^2<0$. This is the case $\eta$ is small enough and $r < (1 - \sqrt{1-\beta})^2$, since in that case $1-2\beta+2r - (\sqrt{4r}-1)^2 < 0$, completing the proof.
\end{proof}

\subsection{Proof of Theorem~\ref{th:max-perm-test}}

\begin{proof}
By the arguments in Section~\ref{app:gen-assumption}, the proof continues under the assumption that $F_0$ has zero mean and unit variance, without loss of generality.

Under the null and for a given (but arbitrary) permutation $\pi\in\Pi$ it is clear that $\bX^\pi$ and $\bX$ have exactly the same distribution. Therefore $\max_i\left\{Y_i(\bX)\right\}$ is uniformly distributed on the set $\{ \max_i\{Y_i(\bX^\pi\}, \pi \in \Pi \}$ (with multiplicities) conditionally on the order statistics of $\bX$. So, for a given $\alpha>0$
\begin{align*}
\Phn{\cP_{\text{max-perm}}(\bX)  \leq \alpha} &= \Phn{ \left| \{\pi \in \Pi : \max_i\{Y_i(\bX^\pi)\} \ge \max_i\{Y_i(\bX)\}\} \right| \leq \alpha(nt)!} \\ 
&\leq \frac{\floor{\alpha (nt)!}}{(nt)!} \leq \alpha\ ,
\end{align*}
If there are no ties, the first inequality above is an equality, but with ties present the test becomes slightly more conservative. This argument is completely standard and for more details on permutation tests the reader is referred to \citep{MR2135927}.

What remains to be proven is the behavior of the test under the alternative. Namely we must show that, provided $r$ is large enough (as stated in the theorem) then for any $\alpha>0$ 
\[
\Phn{\cP_{\text{max-perm}}(\bX)  >\alpha} \longrightarrow 0\ .
\]
For convenience, let $\pi$ be a uniformly distributed permutation of $\Pi$ and let this be independent from $\bX$. We can rewrite our permutation $p$-value as a conditional probability:
\[
\cP_{\text{max-perm}}(\bX)  = \P{\max_i Y_i^\pi\geq \max_i Y_i \given \bX}\ .
\]
To get a good upper-bound on the permutation $p$-value we use the following concentration inequality (see \cite{ShoWel} and \cite{arias-castro2018a}, for instance).
\begin{lem}[Bernstein bound for sampling without replacement]\label{lem:bernstein}
Let $(Z_1,\dots,Z_m)$ be sampled without replacement from the set $\{z_1,\dots,z_n\}$. Define $z_{\text{max}} = \max_j\{z_j\}$, $\overline z = \frac{1}{n}\sum_{j=1}^nz_j$, $\overline Z = \frac{1}{m}\sum_{j=1}^n Z_j$ and $\sigma_z^2 = \frac{1}{n}\sum_{j=1}^n (z_j - \overline z)^2$. Then, for all $\tau \geq 0$:
\[
\P{\overline Z \geq \overline z + \tau} \leq \exp{-\frac{m\tau^2}{2\sigma_z^2 + \frac{2}{3}(z_{\text{max}} - \overline z)\tau}}.
\]
\end{lem}

Using this lemma, we find that
\begin{align*}
\cP_{\text{max-perm}}(\bX)  &= \P{\max_i Y_i(\bX^\pi)\geq \max_i Y_i(\bX) \given \bX}\\
&\leq \sum_{k\in[n]}  \P{Y_k(\bX^\pi)\geq \max_i Y_i(\bX) \given \bX}\\
&= \sum_{k\in[n]} \P{\frac{1}{t}\sum_{j\in[t]} X_{k,j}^\pi\geq \Xb + \left(\max_i Y_i(\bX)-\Xb\right) \given \bX}\\
&\leq \sum_{k\in[n]} \exp{- \frac{t \left(\max_i Y_i(\bX)-\Xb\right)^2}{2 \sigma_X^2 + \frac23 (\max_{i,j} X_{ij} - \Xb) \left(\max_i Y_i(\bX)-\Xb\right)}}\\
&= n \cdot \exp{- \frac{t \left(\max_i Y_i(\bX)-\Xb\right)^2}{2 \sigma_X^2 + \frac23 (\max_{i,j} X_{ij} - \Xb) \left(\max_i Y_i(\bX)-\Xb\right)}}\ .
\end{align*}
The first inequality is a consequence of a simple union bound, and we used Lemma~\ref{lem:bernstein} in the second inequality.

At this point it is clear that, to control the $p$-value of our test we need to characterize the behavior of $\Xb$, $\sigma^2_X$, $\max_{i,j} X_{ij}$ and $\max_i Y_i(\bX)$ under the alternative hypothesis. Since $|\cS|=\smallO(n)$ most of the elements of $\bX$ are samples from the null distribution. Therefore we intuitively expect that $\bX$ and $\sigma^2_X$ should be good estimators for the mean and variance of $F_0$. The behavior of the term $\max_{i,j} X_{ij}$ is a bit more delicate, but one can see that for the given parameterization of the null the dominant contribution is still given by the null distribution. In contrast, the term $\max_i Y_i(\bX) -\Xb$ really depends on the alternative - the largest stream mean is surely driven by the anomalous observations. Formally, we can show the following result.
\begin{lem}\label{lem:chernoff-term-bounds}
Let $\beta\in(0,1)$, $\theta=\sqrt{2r(\log n)/t}$ with $r>0$ and consider the alternative hypothesis in~\eqref{hyp:exp}. Assume $F_0$ has zero mean and variance one and $t=\omega(\log(n))$. Then
\begin{itemize}
\item[(i)] If $\beta > 1/2$, then $\Xb=
\bigOp\left(\frac{1}{\sqrt{nt}}\right) \text{ and } \sigma_X^2=
1+\bigOp\left(\frac{1}{\sqrt{nt}}\right)$
\item[(ii)] Let $c \in (0,\theta_*-\theta)$. Then,
\[
\Pha{\max_{i,j} X_{ij}-\Xb \leq \frac{3}{c}\log (nt)} \to 1 \text{ as } n\to\infty\ .
\]
\item[(iii)] Assume further that $t=\omega(\log^3(n))$ and let $\varepsilon>0$. Provided $r>(\sqrt{1+\varepsilon} - \sqrt{1-\beta})^2$ we have
\[
\Pha{\max_i Y_i(\bX) - \Xb \geq \sqrt{\frac{2(1+\varepsilon)}{t}\log(n)}} \to 1
\]
as $n\to\infty$.
\end{itemize}
\end{lem}

The first result of the lemma provides a rate at which the bound on our sample variance can decrease. For the analysis of the max test a much simpler result (already proved in \cite{arias-castro2018a} which holds for $\beta > 0$) suffices: for any $\varepsilon>0$ (i) implies that
\[
\Pha{\sigma_X^2\leq (1+\varepsilon/2)}\to 1\ .
\]

Note also that the bound in Lemma~\ref{lem:bernstein} is monotonically decreasing in $\tau$. This ensures that for $\varepsilon>0$ and with probability tending to one under the alternative, provided $r>(\sqrt{1+\varepsilon} - \sqrt{1-\beta})^2$, the overall $p$-value of test satisfies
\begin{align*}
\cP_{\text{max-perm}}(\bX)  &\leq n \cdot \exp{- \frac{2(1+\varepsilon)\log(n)}{2(1+\varepsilon/2)+2\varepsilon+ \frac{2}{c} \log(nt)\sqrt{2(1+\varepsilon)\frac{1}{t}\log(n)} }}\ .
\end{align*}
Written differently, with probability tending to 1 under the alternative:
\[
\log \cP_{\text{max-perm}}(\bX)  \leq \log(n)\left(1- \frac{1+\varepsilon}{1+\varepsilon/2+ \frac{1}{c}(\log(n)+\log(t))\sqrt{2(1+\varepsilon)\frac{\log(n)}{t}} }\right)\ .
\]
To ensure $\cP_{\text{max-perm}}(\bX) \to 0$, or equivalently, $\log \cP_{\text{max-perm}}(\bX)  \to -\infty$ it suffices to ensure
\[
\frac{1}{c}(\log(n)+\log(t))\sqrt{2(1+\varepsilon)\frac{\log(n)}{t}}< \varepsilon/2\ .
\]
However, since we assume $t=\omega(\log^3(n))$ this is immediately satisfied, since the l.h.s. converges to zero.

We have just proved that, for $\varepsilon>0$ and $r>(\sqrt{1+\varepsilon} - \sqrt{1-\beta})^2$, $\cP_{\text{max-perm}}(\bX) \to 0$ as $n\to\infty$. Since $\varepsilon>0$ is arbitrary this implies the result in the theorem, concluding the proof.
\end{proof}

\subsection{Proof of Theorem~\ref{th:perm-hc-discrete-test-nullmean}}
\begin{proof} By the arguments in Section~\ref{app:gen-assumption}, the proof continues under the assumption that $F_0$ has zero mean and unit variance. Furthermore the conservativeness of this test follows by the standard argument already presented in the proof of Theorem~\ref{th:max-perm-test}.

For the rest of the proof consider alternative hypothesis. We must show that
$$\Pha{\tilde\cP_{\text{perm-hc}}(\bX) \leq \alpha } \to 1$$
as $n \to \infty$. Like before, we can write our permutation $p$-value as a conditional probability as follows:
\[
\tilde\cP_{\text{perm-hc}}(\bX) = \Pha{ \tilde T(\bX^\pi)\geq \tilde T(\bX) \given \bX} \ ,
\]
where $\pi$ is independent from $\bX$ and uniformly distributed over $\Pi$. The first step is to understand and simplify the role of $\pi$ in the above expression. This mirrors the analysis under the null hypothesis in the proof of Theorem~\ref{th:hc-discrete-test}, as we need to show that the values of the test statistic computed with permuted data are a good surrogate for the values of the test statistic under the null. However, the argument becomes more complex due to the dependencies introduced by the permutation. Like before, we will use the union bound for the max-operator in the permuted statistic, inducing a multiplicity by the grid-size:
\begin{equation}\label{eq:perm-hc-union}
\tilde\cP_{\text{perm-hc}}(\bX) = \Pha{ \max_{q \in Q} \tilde V_q(\bX^\pi) \geq \tilde T(\bX) \given \bX } \leq \sum_{q\in Q} \Pha{ \tilde V_q(\bX^\pi) \geq \tilde T(\bX) \given \bX } \ .
\end{equation}
To proceed recall that quantifying $\tilde V_q(\bX^\pi)$ requires the quantification of two terms: $\tilde N_q(\bX^\pi)$ and $\tilde P_q(\bX^\pi)$. 
Note, however, that $\tilde P_q(\bX)$ is invariant under permutations of $\bX$, and therefore $\tilde P_q(\bX^\pi)=\tilde P_q(\bX)$, as explained before. As such the only random quantity inside the probability operator above (conditionally on $\bX$) is $\tilde N_q(\bX^\pi)$. Noting that $\E{\tilde N_q(\bX^\pi) \given \bX} = n\tilde P_q(\bX)$ we have
\begin{equation}\label{eq:hc-perm-p-int1}
\tilde\cP_{\text{perm-hc}}(\bX)\leq \sum_{q\in Q} \Pha{\tilde  N_q(\bX^\pi) - \E{\tilde N_q(\bX^\pi)} \geq \tilde T(\bX)\sqrt{n \tilde P_q(\bX)(1- \tilde P_q(\bX))} \given \bX} \ .
\end{equation}
To apply Chebyshev's inequality we need the right-hand-side of the inequality inside the probability to be positive, and $\tilde T(\bX)$ might be negative. In the latter case we simply bound the probability by one. Therefore, using Chebyshev's inequality we get
\begin{equation}\label{eq:hc-perm-p-int2}
\tilde\cP_{\text{perm-hc}}(\bX) \leq \ind{\tilde T(\bX) \leq 0}+\frac{\ind{\tilde T(\bX) > 0}}{\tilde T(\bX)^2}\sum_{q\in Q} \frac{\Var{\tilde N_q(\bX^\pi) \given \bX}}{n\tilde P_q(\bX)(1-\tilde P_q(\bX))}  \ ,
\end{equation}
where we convention that $0/0=0$. To continue, we must quantify the conditional variance of $\tilde N_q(\bX^\pi)$. The permutation on $\bX$ causes dependencies, but these are benign when realizing the conditional permuted stream means are negatively associated conditional on the data. Using Theorem 2.11 and the properties $\text{P}_6$ and $\text{P}_4$ in \cite{Joag-Dev1983}, we find that $Y_i(\bX^\pi)|\bX$ and $Y_j(\bX^\pi)|\bX$ are negatively associated if $i\neq j$. For ease of notation, define
\begin{equation}\label{eq:zq}
z_q \equiv \sqrt{\frac{2q}{t}\log(n)}\ .
\end{equation}
Then,
\begin{align*}
\Var{\tilde N_q(\bX^\pi) \given \bX} &= \sum_{i\in[n]} \Var{\ind{Y_i(\bX^\pi) \geq z_q}\given \bX}\\
&\quad\qquad + \sum_{i\in[n]}\sum_{j\neq i} \Cov{\ind{Y_i(\bX^\pi) \geq z_q}, \ind{Y_j(\bX^\pi) \geq z_q} \given \bX} \\
& \leq n\tilde P_q(\bX)(1-\tilde P_q(\bX)) \ ,
\end{align*}
where we used the definition of negative association (Definition 2.1 from \cite{Joag-Dev1983}). In conclusion we get the following simple bound for the $p$-value:
\begin{align}
\tilde\cP_{\text{perm-hc}}(\bX) &\leq \frac{1}{\tilde T(\bX)^2}\left(\sum_{q\in Q} 1\right)\ind{\tilde T(\bX) > 0} + \ind{\tilde T(\bX) \leq 0}\nonumber\\
&=\frac{k_n+1}{\tilde T(\bX)^2}\ind{\tilde T(\bX) > 0} + \ind{\tilde T(\bX) \leq 0}\ .\label{eq:hc-perm-p-int3}
\end{align}

To continue the proof we must show that $\tilde T(\bX)$ is of order larger than $k_n=n^{\smallO(1)}$. This mimics the approach in Proposition~\ref{prop:vq-test} and Theorem~\ref{th:hc-discrete-test} under the alternative. Recall that $\tilde T(\bX)=\max_{q\in Q} \tilde V_q(\bX)$, so it suffices to show that $\tilde V_q(\bX)$ is larger than $k_n$ with high probability for particular values of $q\in Q$. At the final stretch of the proof, it will become clear that one only needs to consider sequences of values $q_n\in Q$ which converge to a fixed value $q$, so let $q_n \in Q$ with $q_n \to q$ and $q > 0$. To start, note that:
\begin{equation}\label{eq:vq-prob-aq}
\Pha{\tilde V_{q_n}(\bX) \geq k_n} = \Pha{\tilde N_{q_n}(\bX) - \E{\tilde N_{q_n}(\bX)} \geq A_{q_n}(\bX) } \ ,
\end{equation}
where we have defined for convenience
\[
A_{q_n}(\bX) \equiv k_n\sqrt{n\tilde P_{q_n}(\bX)(1-\tilde P_{q_n}(\bX)} + \left(n\tilde P_{q_n}(\bX) - \E{\tilde N_{q_n}(\bX)}\right) \ .
\]
To bound the above probability using Chebyshev's inequality, we first need to find a high-probability upper bound for the random quantity $A_{q_n}(\bX)$. Note that the second term in $A_{q_n}(\bX)$ will typically be negative when anomalies are present. To characterize this quantity, define:
\begin{align*}
w_{i,q}\equiv\Pha{ Y_i(\bX) \geq  \sqrt{\frac{2q}{t}\log(n)} } \ ,
\end{align*}
and let $\tilde p_q \equiv w_{i,q}$ if $i\notin \cS$ and $\tilde v_q \equiv w_{i,q}$ if $i\in\cS$.  Now, note that under the alternative $\E{\tilde N_q(\bX)} = (n-s)\tilde p_q + s\tilde v_q$, such that
\begin{equation}
A_{q_n}(\bX) = k_n\sqrt{n\tilde P_{q_n}(\bX)(1-\tilde P_{q_n}(\bX))} + s(\tilde p_{q_n}-\tilde v_{q_n}) + n\left(\tilde P_{q_n}(\bX) - \tilde p_{q_n}\right)\ .
\end{equation}
Note that using Lemma~\ref{lem:sum-normality}, we can easily characterize $\tilde p_{q_n}$ and $\tilde v_{q_n}$, and conclude that $\tilde p_{q_n}=n^{-q + o(1)}$ and
\[
\tilde v_{q_n}=\left\{\begin{array}{ll}
n^{-(\sqrt{q} - \sqrt{r})^2 + o(1)} & \text{ if } r<q\\
n^{o(1)} & \text{ if } r\geq q
\end{array}\right. \ .
\]

At this point, the expression above looks remarkably similar to Equation~\eqref{eqn:surrogate_expression} in the proof of Proposition~\ref{prop:vq-test}. However, we have an extra term $n\left(\tilde P_{q_n}(\bX) - \tilde p_{q_n}\right)$ that also needs to be controlled. If we ignore that term then it would suffice to show that $\tilde P_{q_n}(\bX)\approx n^{-q+o(1)}$ to complete the proof. However, such guarantee is not enough to control the last term, and a much more refined result is required to ensure $\tilde P_{q_n}(\bX)$ is a sufficiently accurate surrogate for $\tilde p_{q_n}$. In detail, we require the first term in $A_{q_n}(\bX)$ to be at most $n^{(1-q)/2 + o(1)}$ with high probability, and the third term cannot outweigh the preceding two. The accuracy of the approximation $\tilde P_{q_n}(\bX)$ to $\tilde p_{q_n}$ is captured in the following lemma:

\begin{lem}\label{lem:pq-sharp-bound}
Consider the setting of Lemma~\ref{lem:chernoff-term-bounds} with $\beta > 1/2$ and let $q_n \to q$ with $q\in(0,1]$, $t = \omega(\log^3(n))$ and $t = n^{o(1)}$. Then, for any $\varepsilon > 0$, there exists sequence $g_n\to0$ such that under both the null and alternative hypothesis
\[
\P{ \tilde P_{q_n}(\bX) - \tilde p_{q_n} \leq n^{\max\{-\frac{1+q}{2},-\beta-q\}+\varepsilon +g_n} } \to 1 \ .
\]
\end{lem}
Note that this lemma, together with the characterization of $\tilde p_q$, implies that there exists a sequence $g_n \to 0$ such that 
\[
\Pha{ \tilde P_{q_n}(\bX) \leq n^{-q + g_n}} \to 1 \ .
\]
Putting all the facts together, we conclude that for any $\varepsilon > 0$, there exists a deterministic sequence $a_n$ with characterization
\[
a_n \equiv \begin{cases}
n^{\max\{\frac{1-q}{2}, 1-\beta-q\} + \varepsilon + o(1)} - n^{1-\beta-(\sqrt{q}-\sqrt{r})^2 + o(1)} \text{ if } r < q \ , \\
n^{\max\{\frac{1-q}{2}, 1-\beta-q\} + \varepsilon + o(1)} - n^{1-\beta + o(1)} \text{ if } r \geq q \ ,
\end{cases}
\]
such that for the event $\Omega \equiv \{A_{q_n}(\bX) \leq a_n\}$ we have $\P{\Omega}\to 1$. Note that $a_n$ is nearly the same term as encountered in the proof of Proposition~\ref{prop:vq-test} - although there we were able to characterize the counterpart of $\tilde v_q$ in a sharper way, but this does not affect the final result.
We can now proceed as follows:
\begin{align*}
\Pha{ \tilde V_{q_n}(\bX) \leq k_n} &= \Pha{ \tilde N_{q_n}(\bX) - \E{\tilde N_{q_n}(\bX)} \leq  A_{q_n}(\bX)} \\
&\leq \Pha{ \tilde N_{q_n}(\bX) - \E{\tilde N_{q_n}(\bX)} \leq  a_n \given \Omega}  + \Pha{\Omega^{c}}\\
&\leq \Pha{\Omega}^{-1}\Pha{ \tilde N_{q_n}(\bX) - \E{\tilde N_{q_n}(\bX)} \leq  a_n}  + \Pha{\Omega^{c}}\\
&= \Pha{\Omega}^{-1}\Pha{-\left(\tilde N_{q_n}(\bX) - \E{\tilde N_{q_n}(\bX)}\right)  \geq -a_n} + \Pha{\Omega^{c}}\\
&\leq  \Pha{\Omega}^{-1}\Pha{ \abs{\tilde N_{q_n}(\bX) - \E{\tilde N_{q_n}(\bX)}} \geq -a_n} + \Pha{\Omega^{c}} \\
&\leq \Pha{\Omega}^{-1}a_n^{-2}\Var{ \tilde N_{q_n}(\bX)} + \Pha{\Omega^{c}} \ , \numberthis\label{eq:chebyshev-mid}
\end{align*}

where the last inequality follows from Chebyshev's inequality provided $a_n < 0$. Note that $a_n < 0$ for $n$ sufficiently large provided:
\begin{equation}\label{eq:conditions-1}
\begin{cases}
1-\beta - (\sqrt{q}-\sqrt{r})^2 - \max \left\{\tfrac{1-q}{2}, 1-\beta - q \right\} - \varepsilon > 0 &\text{ if } r < q \ , \\
1-\beta - \max \left\{\tfrac{1-q}{2}, 1-\beta - q \right\} - \varepsilon > 0 &\text{ if } r \geq q \ .
\end{cases}
\end{equation}
Recall that $\Pha{\Omega} \to 1$, and therefore for $n$ sufficiently large we have $\Pha{\Omega} \geq 1/2$. Furthermore, the summands in $\tilde N_q(\bX)$ are independent, such that
\[
\Var{\tilde N_{q_n}(\bX)} = (n-s)\tilde p_{q_n}(1-\tilde p_{q_n}) + s\tilde v_{q_n}(1-\tilde v_{q_n}) \ .
\]
Assuming \eqref{eq:conditions-1} and using \eqref{eq:chebyshev-mid} we conclude that for large enough $n$
\[
\Pha{ \tilde V_{q_n}(\bX) \leq k_n} \leq 2 a_n^{-2}\Big((n-s)\tilde{p}_{q_n}(1-\tilde{p}_{q_n}) + s\tilde{v}_{q_n}(1-\tilde{v}_{q_n})\Big) + \P{\Omega^{c}} \ .
\]
Note that $\Pha{\Omega^c} \to 0$. Using the asymptotic characterization of $\tilde p_{q_n}$ and $\tilde v_{q_n}$, the first term converges to 0 provided:
\begin{equation}\label{eq:conditions-2}
\begin{cases}
\max \{1-q, 1-\beta - (\sqrt{q}-\sqrt{r})^2 \} - 2(1-\beta - (\sqrt{q}-\sqrt{r})^2) < 0 &\text{ if } r < q \ , \\
\max \{1-q, 1-\beta \} - 2(1-\beta) < 0 &\text{ if } r \geq q \ .
\end{cases}
\end{equation}
Note that the conditions in \eqref{eq:conditions-1} and \eqref{eq:conditions-2} are nearly identical to those obtained when proving Proposition~\ref{prop:vq-test}, with the former holding for any $\varepsilon > 0$. Now, similar algebra as used in the proof of that proposition boils down to the same resulting requirements in the statement of that proposition, i.e. if
\begin{equation}\label{eq:cases-r}
\begin{cases}
r > (1-\sqrt{1-\beta})^2 &\text{ if } q=1 \ ,\\
r < 1/4 \text{ and } r > \beta - 1/2 &\text{ if } q = 4r \ ,
\end{cases}
\end{equation}
then there exists an $\varepsilon > 0$ such that \eqref{eq:conditions-1} and \eqref{eq:conditions-2} hold, and thus $\Pha{\tilde V_{q_n}(\bX) \leq k_n} \to 0$.

At this point we can simply follow the arguments of the proof of Proposition~\ref{th:hc-discrete-test} almost verbatim. Suppose that $r \leq 1/4$ and $r \geq \beta - 1/2$. Consider the gridpoint $q^*_n \equiv \min_{q\in Q}\abs{q-4r}$. Since the size of the grid is increasing with $n$, we have $q^*_n = 4r + o(1)$. Therefore:
\[
\Pha{\max_{q\in Q}\left\{\tilde V_q(\bX)\right\} \leq k_n} \leq \Pha{\tilde V_{q^*_n}(\bX) \leq k_n} \to 0 \ .
\]
The other case, when $r > (1-\sqrt{1-\beta})^2$, follows analogously with $q_n=1$, since this value is included in the grid $Q$. Then, this result trivially implies that $\tilde T(\bX)\to 1$ and $(k_n+1)/\tilde T^2(\bX)\to 0$ with probability tending to one, and therefore
\[
\Pha{ \tilde\cP_{\text{perm-hc}}(\bX) \leq \alpha} \to 1\ ,
\]
completing the proof.
\end{proof}

\subsection{Proof of Theorem~\ref{th:perm-hc-discrete-test}}
\begin{proof}
The conservativeness of this test follows by the standard argument already presented in the proof of Theorem~\ref{th:max-perm-test}. For the alternative, the proof relies on the results shown in Theorem~\ref{th:perm-hc-discrete-test-nullmean}. The proof of that theorem requires a grid no larger than $n^{o(1)}$, which contains two sequences $q_{1,n}, q_{2,n} \in Q$, such that $q_{1,n} = 4r + o(1)$ if $r < 1/4$ and $q_{2,n} = 1 + o(1)$ otherwise. In the context of our restated statistic, this means our result follows if there exists two sequences $\tau_{1,n}, \tau_{2,n} \in R$ such that:
\begin{align*}
\tau_{1,n} &= \mu_0 + \sqrt{\frac{2\sigma_0^2}{t}(1+o(1))\log(n)} \ , \\
\tau_{2,n} &= \mu_0 + \sqrt{\frac{2\sigma_0^2}{t}(4r+o(1))\log(n)} \ .
\end{align*}
We show that the first sequence exists in $R$; the second sequence then follows analogously by replacing $\sigma_0^2$ by $4r\sigma_0^2$. Note that the requirement for $\tau_{1,n}$ can be rewritten as:
\[
\tau_{1,n} - \left(\mu_0 + \sqrt{\frac{2\sigma_0^2}{t}\log(n)}\right) = o\left(\sqrt{\frac{\log(n)}{t}}\right) \ .
\]
Since $\sqrt{\log(n)} \to \infty$, there exists an $n_0$ such that for $n \geq n_0$ there exists sequences $i_n^*\in \left\{\frac{k}{\sqrt{t}}\right\}_{k=-\sqrt{t\log(n)}}^{\sqrt{t\log(n)}}$ and $j\in\left\{\frac{k}{\sqrt{\log(n)}}\right\}_{k=0}^{\log(n)}$ such that
\[
\abs{i_n^* - \mu_0} \leq \frac{1}{\sqrt{t}}\ , \hspace{1cm}
\abs{j_n^* - \sigma_0} \leq \frac{1}{\sqrt{\log(n)}} \ .
\]
Now, defining $\tau_{1,n} = i_n^* + \sqrt{\frac{2(j_n^*)^2}{t}\log(n)}$, we have that:
\begin{align*}
\abs{\tau_{1,n} - (\mu_0 + \sqrt{\frac{2}{t}\sigma_0^2\log(n)})} \leq \abs{i_n^* - \mu_0} + \abs{j_n^* - \sigma_0}\sqrt{\frac{2}{t}\log(n)} \leq \frac{1+\sqrt{2}}{\sqrt{t}} = o\left(\sqrt{\frac{\log(n)}{t}}\right)\ ,
\end{align*}
so $\tau_{1,n}$ is sufficiently close to the optimal value.
Now, the size of the grid is of order $\bigO\left(\sqrt{t\log^3(n)}\right)$, and since $t$ is of order $n^{o(1)}$, the grid is not too large.
\end{proof}

\subsection{Proof of Lemma~\ref{lem:integration}}
\begin{proof}
We have
\begin{align*}
\E{f(X) \ind{X \le \tau}}
&= \int_0^\infty \bbP(f(X) \ind{X \le \tau} > t) {\rm d}t\\
&= \int_0^{f(\tau)} \bbP(f(X) > t) {\rm d}t \\
&= \int_0^{f(\tau)} \bbP(X > f^{-1}(t)) {\rm d}t \\
&= \int_{-\infty}^\tau \bbP(X > x) f'(x) {\rm d}x,
\end{align*}
where in the last line we changed the integration variable to $x := f^{-1}(t)$.
\end{proof}

\subsection{Proof of Lemma~\ref{lem:chernoff-term-bounds}}

\begin{proof}
Let us start by characterizing the overall sample mean $\Xb$. By Chebyshev's inequality we have
\[
\Xb=\E{\Xb}+\bigOp\left(\frac{1}{\sqrt{nt}}\right)\ ,
\]
as $n\to\infty$. Now note that, under the alternative hypothesis
\[
\E{\Xb} = \frac{|\cS|}{n}\E{X} = \frac{|\cS|}{n}\int \frac{x\exp{\theta x}}{\varphi(\theta)} dF_0(x)\ ,
\]
where $X\sim F_\theta$. A Taylor expansion of the function inside the integral around $\theta=0$ yields
\begin{align*}
\int \frac{x\exp{\theta x}}{\varphi(\theta)} dF_0(x) &= \int x + x^2\theta + \bigO(\theta^2) dF_0(x) \\
&= \theta + \bigO(\theta^2)
\end{align*} 
Finally $\theta= \bigO\left(\sqrt{\log(n)/t}\right) \to 0$ since $t = \omega(\log(n))$. Putting all this together yield the first result stated in (i).

For the second result in (i) note that
\begin{align*}
\sigma_X^2 &= \frac{1}{nt}\sum_{i,j} (X_{ij}-\Xb)^2= \frac{1}{nt}\sum_{i,j} X_{ij}^2 -  \Xb^2  \\
&= \left(\frac{1}{nt}\sum_{i,j} \E{X_{ij}^2}\right) + \left(\frac{1}{nt}\sum_{i,j} X_{ij}^2 - \E{X_{ij}^2}\right) -\Xb^2
\end{align*}
For the first term we see that
\begin{align*}
\frac{1}{nt}\sum_{i,j} \E{X_{ij}^2} &= \frac{1}{nt}\sum_{i\notin\cS,j\in[t]} \Var{X_{ij}}+\frac{1}{nt}\sum_{i\in\cS,j\in[t]} \Var{X_{ij}}+\E{X_{ij}}^2\\
&=\frac{n-|\cS|}{n}+\frac{|\cS|}{n}\left(1 + \bigO(\theta) \right) \\
&= 1 + \bigO\left(n^{-\beta}\sqrt{\frac{\log n}{t}}\right)\ ,
\end{align*}
as $n\to\infty$. In the above the variance of the anomalous streams was characterized with a Taylor expansion of $\theta$ around 0, similarly to what was done for the average term.

For the second term note first that all the moments of $F_0$ are finite, in particular the fourth moment.  Therefore by Chebyshev's inequality we have
\[
\frac{1}{nt}\sum_{i,j}X_{ij}^2 - \E{X_{ij}}= \bigOp\left(\frac{1}{\sqrt{nt}}\right)\ .
\]

Finally, we know that $\Xb^2=\bigOp(1/nt)$ when $\beta>1/2$. Putting everything together yields the second result stated in (i).

The argument needed to prove (ii) is the same already used in \cite{arias-castro2018a}, and presented here for completeness. 

Letting $x>0$, a union bound gives
\begin{align}
\Pha{\max_{i,j} X_{ij}> x} &\leq \Pha{\max_{i\in\cS,j\in[t]} X_{ij}>x}+\Pha{\max_{i\notin\cS,j\in[t]} X_{ij}>x}\nonumber\\
&\leq |\cS|t(1-F_{\theta}(x))+(n-|\cS|)t(1-F_0(x))\ .\label{eq:max_bound}
\end{align}

Now, let $c \in (0, \theta_*-\theta)$. We have that:
\begin{align*}
1-F_{\theta}(x) & = \frac{1}{\varphi_0(\theta)} \int_x^\infty \exp{\theta u} dF_0(u) \\
&= \frac{1}{\varphi_0(\theta)} \int_x^\infty \exp{(\theta+c) u} \exp{-cu} dF_0(u) \\
&\leq \frac{1}{\varphi_0(\theta)} \exp{-cx}  \int_x^\infty \exp{(\theta+c) u} dF_0(u) \\
&\leq  \frac{\varphi_0(\theta + c)}{\varphi_0(\theta)} \exp{-cx} \ .
\end{align*}
Therefore, with considerable slack, we can take $x=\frac{2}{c}\log (nt)$ guarantee that both terms in~\eqref{eq:max_bound} converge to zero. Together with the characterization of $\Xb$ in (i) we conclude that
\[
\Pha{\max_{i,j} X_{ij}-\Xb \leq \frac{3}{c}\log (nt)} \to 1\ .
\]

To show part (iii) very different argument is needed as this is a lower-bound on the tail probability, rather than an upper bound. The following lemma gives a precise characterization of the tail probability.

\begin{lem}\label{lem:sum-normality} Consider the setting of Lemma~\ref{lem:chernoff-term-bounds}. Let $i \in \cS$ and let $q_n\to q>r$ as $n\to\infty$ and $t=\omega(\log^3 n)$. Then
\[
\Pha{Y_i(\bX) \geq \sqrt{\frac{2q_n\log n}{t}}} = n^{-(\sqrt{q}-\sqrt{r})^2 + o(1)}\ .
\]
If $q \leq r$ we have $\Pha{Y_i(\bX) \geq \sqrt{\frac{2q_n\log n}{t}}} = n^{o(1)}$.
\end{lem}

To show (iii) begin by noting that
\begin{align*}
\lefteqn{\Pha{\max_{i\in[n]} Y_i(\bX) \geq \sqrt{\frac{2(1+\varepsilon)}{t}\log(n)}}}\\
&\geq \Pha{\max_{i\in\cS} Y_i(\bX) \geq \sqrt{\frac{2(1+\varepsilon)}{t}\log(n)}}\\
&=1 - \left(1-\Pha{Y_i(\bX) \geq \sqrt{\frac{2(1+\varepsilon)}{t}\log(n)}}\right)^{|\cS|} \ . \numberthis{\label{eq:max-prod-reduction}}
\end{align*}

Consider the case where $(\sqrt{1+\varepsilon}-\sqrt{1-\beta})^2<r<1+\varepsilon$. Note that, in that case, we can use Lemma~\ref{lem:sum-normality} with~\eqref{eq:max-prod-reduction} which gives:
\begin{align*}
\Pha{\max_{i\in[n]} Y_i(\bX) \geq \sqrt{\frac{2(1+\varepsilon)}{t}\log(n)}} &= 1 - \left(1-n^{-(\sqrt{1+\varepsilon}-\sqrt{r})^2+\smallO(1)}\right)^{|\cS|}\\
&= 1 - \exp{n^{1-\beta} \log\left(1-n^{-(\sqrt{1+\varepsilon}-\sqrt{r})^2+\smallO(1)}\right)}\\
&\geq 1 - \exp{-n^{1-\beta} n^{-(\sqrt{1+\varepsilon}-\sqrt{r})^2+\smallO(1)}}\ ,
\end{align*}
where in last inequality we simply used the fact that $\log(1+x)\geq x$. Finally, provided $(\sqrt{1+\varepsilon}-\sqrt{1-\beta})^2<r<1+\varepsilon$ we guarantee that $1-\beta-(\sqrt{1+\varepsilon}-\sqrt{r})^2+\smallO(1)>0$. The statement for $r>1+\varepsilon$ follows immediately since this probability is monotonically increasing in $r$. To get the statement in (iii) we just need to use this result together with the characterization of $\Xb$, concluding the proof.
\end{proof}

\subsection{Proof of Lemma~\ref{lem:pq-sharp-bound}}
\begin{proof}
Note that $\tilde p_q$ and $\tilde P_q(\bX)$ differ in two major aspects; first, $\tilde p_q$ depends on the null distribution, while $\tilde P_q(\bX)$ depends on the distribution of permutation stream means - which may be ``contaminated'' by anomalous observations. Secondly, $\tilde P_q(\bX)$ is random, while $\tilde p_q$ is deterministic.

To characterize the ``contamination'' effect, we define the following quantity corresponding to the probability of a permutation stream exceeding $z_q\equiv\sqrt{2q(\log n)/t}$ when precisely $k$ anomalous observations are sampled in the permutation stream:
\[
\tilde P_{k,q}'(\bX) \equiv \P{Y_1(\bX^{\pi_k}) \geq z_q \given \bX} \ ,
\]
where $\pi_k$ is uniformly distributed over the set $\Pi^{(k)}$ independent from $\bX$, and the set $\Pi^{(k)} \subseteq \Pi$ is defined as the set of permutations with exactly $k$ observations sampled from anomalous streams in permutation stream with index 1. Specifically:
\[
\Pi^{(k)} \equiv \bigg\{\pi \in \Pi : \sum_{i\in\cS}\sum_{j\in[t]}\sum_{h\in[t]}\ind{\pi(i,j) = (1,h)} = k \bigg\} \ .
\]
Note that $\Pi^{(k)}$ does not depend on the data, but merely on the index set $\cS$. By definition $\E{\tilde P_{0,q}'(\bX)} = \tilde p_q$. Now consider the following decomposition of our quantity of interest:
\begin{equation}\label{eq:decomp-pqx}
\tilde P_{q_n}(\bX) - \tilde p_{q_n} = \left(\tilde P_{q_n}(\bX) - \tilde P_{0,q_n}'(\bX)\right) + \left(\tilde P_{0,q_n}'(\bX) - \E{\tilde P_{0,q_n}'(\bX)}\right) \ .
\end{equation}
Intuitively, bounding the first term amounts to characterizing the effect of ``contamination'' by anomalous observations when computing $\tilde P_{q_n}(\bX)$. The second term focusses mainly on the random fluctuations (when contamination is not present).

We start with the first term in \eqref{eq:decomp-pqx}. Intuitively, with high probability, the permutation stream $Y_1(\bX^\pi)$ consists solely of nominal observations, especially for small $t$. With modest probability, a few anomalous observations are sampled, but their influence on the ensuing distribution of the stream mean $Y_1(\bX^\pi)$ should be minimal, especially in light of the exponential tails of the distributions in question. Finally, sampling a large number of anomalies might unduly influence the distribution of $Y_1(\bX^\pi)$, but the probability of this happening is very small. For our purposes, the effect of the anomalous observations is modest on $Y_1(\bX^\pi)$ provided there are no more than $\log(n)$ contaminating samples.

To proceed, we first condition on the number of anomalous observations sampled in the first permutation stream, which allows us to bound the first component in \eqref{eq:decomp-pqx} as:
\begin{equation}\label{eq:conditioning-comp1}
\tilde P_{q_n}(\bX) - \tilde P_{0,q_n}'(\bX) \leq \sum_{k=1}^{\floor{\log(n)}} \tilde P_{k,q_n}'(\bX)\P{\pi\in\Pi^{(k)}} + \sum_{k=\ceil{\log(n)}}^t \P{\pi\in\Pi^{(k)}} \ . 
\end{equation}
Note that the bound above is only sensible when $\P{\pi \in \Pi^{(0)}}$ is close to 1. By assuming $t=n^{o(1)}$ this is indeed the case.

To characterize the bound in \eqref{eq:conditioning-comp1}, we start by characterizing the probability $\tilde P_{k,q_n}'(\bX)$. For small enough $k$, we can bound this term in a nontrivial way through the use of Lemma~\ref{lem:bernstein} and Lemma~\ref{lem:chernoff-term-bounds}. Note that $Y_1(\bX^{\pi_k})$ arises from two sampling processes; $t-k$ samples from the null streams, and $k$ samples from the anomalous streams. We will bound the contribution of the anomalous streams crudely by their maximum. Define $U_m(\bX)$ as the sum of a sample of size $m$ without replacement from the nominal observations of $\bX$, i.e. from the set $\bX_0 \equiv \{X_{ij} : i\not\in\cS, j\in[t]\}$. Now, we can bound:
\begin{align*}
\tilde P'_{k,q_n}(\bX) &\equiv \P{Y_1(\bX^{\pi_k}) \geq z_{q_n} \given \bX} \\
&\leq \P{\frac{1}{t}\left(U_{t-k}(\bX) + k\max_{i,j}\{X_{ij}\} \right) \geq z_{q_n} \given \bX} \ . \numberthis\label{eq:comp1-bound1}
\end{align*}
To continue, we will use the Bernstein bound of Lemma~\ref{lem:bernstein}. Define $\Xb_0$ and $\sigma_{\bX_0}^2$ the sample mean and variance of $\bX_0$. Direct application of the lemma results in complicated expressions, so we first define for convenience:
\[
d_k(\bX) = \frac{\sigma_{\bX_0}}{\sigma_{\bX}}\left(\sqrt{\frac{t-k}{t}} - \Xb_0\sqrt{\frac{t-k}{2\sigma_{\bX_0}q\log(n)}} + \sqrt{\frac{t-k}{2\sigma_{\bX_0}q\log(n)}}\frac{k}{t-k}\max_{i,j}\{X_{ij}\}+ \bigO\left(\frac{1}{t}\right)\right) \ .
\]
While this term is complex, it is ultimately a nuisance term as one should note that, since $t = \omega(\log(n)^3)$ and $k \leq \log(n)$, Lemma~\ref{lem:chernoff-term-bounds} applied to the set $\bX_0$ implies there exists a sequence $g_n \to 0$ such that
\begin{equation}\label{eq:convergence-dk}
\P{\abs{d_k(\bX)} \leq 1 + g_n} \to 1 \ .
\end{equation}
Now, we can continue from \eqref{eq:comp1-bound1} as:
\begin{align*}
&\P{\frac{1}{t}\left(U_{t-k}(\bX) + k\max_{i,j}\{X_{ij}\} \right) \geq z_{q_n} \given \bX} \\ 
&= \P{\frac{U_{t-k}(\bX)}{t-k} \geq \frac{t}{t-k}z_{q_n} -  \frac{k}{t-k}\max_{i,j}\{X_{ij}\} \given \bX} \\
&= \P{\frac{U_{t-k}(\bX)}{t-k}  \geq \overline X_0 + d_k(\bX)\sqrt{\frac{2\sigma_{\bX_0}^2q_n}{t-k}\log(n)} \given \bX }\\
&\leq \exp{ -\frac{ 2d^2_k(\bX)\sigma_{\bX_0}^2q_n\log(n)}{2\sigma_{\bX_0}^2 + \frac{2}{3}(\max_{i\not\in\cS,j\in[t]}\{X_{ij}\}-\overline X_0)\sqrt{\frac{2d^2_k(\bX)\sigma_{\bX_0}^2q_n}{t-k}\log(n)}}} \\
&= \exp{ -\frac{ d_k^2(\bX)q_n\log(n)}{1 + \frac{1}{3}(\max_{i\not\in\cS,j\in[t]}\{X_{ij}\}-\overline X_0)\sqrt{\frac{2d^2_k(\bX)q_n}{(t-k)\sigma_{\bX_0}^2}\log(n)}}} \equiv B(\bX) \ , \numberthis\label{eq:comp1-bound2}
\end{align*}
where the inequality is due to Lemma~\ref{lem:bernstein}. Now, due to the result in \eqref{eq:convergence-dk}, as well as application of Lemma~\ref{lem:chernoff-term-bounds} to the set $\bX_0$, we have that:
\begin{align*}
\Pha{B(\bX) \leq \exp{ -\frac{ q_n(1+o(1))\log(n)}{1 + o(1)}}} \to 1 \ . 
\end{align*}
Since this high probability upper bound can be restated as $n^{-q+o(1)}$, we ultimately obtain that
\begin{equation}\label{eq:comp1-bound3}
\Pha{\tilde P_{k,q_n}(\bX) \leq n^{-q + o(1)}} \to 1 \ .
\end{equation}
Next, we characterize the probability $\P{\pi \in \Pi^{(k)}}$. Note that $\P{\pi \in \Pi^{(k)}} = \P{H = k}$, with $H$ a hypergeometric distribution with population size $nt$, number of success states $st$ and sample size $t$. A coupling argument can be used to show $H$ is stochastically dominated by a binomial random variable with $t$ trials with success probability $st/nt$, and $H$ stochastically dominates a binomial random variable with $t$ trials with success probability $(st-t)/nt$. For convenience, denote $p_L = (st-t)/nt$, and $p_U = st/nt$. We have that:
\begin{align*}
\sum_{k=1}^{\floor{\log(n)}}\P{\pi \in \Pi^{(k)}} &= \P{H \leq \floor{\log(n)}} - \P{H = 0} \\
&\leq \P{\text{Binomial}(t,p_L) \leq \floor{\log(n)}} - \P{\text{Binomial}(t,p_U) = 0} \\
&\leq (1-p_L)^{t} - (1-p_U)^{t} + \sum_{k=1}^{\floor{\log(n)}}\binom{t}{k} p_L^k \ .
\end{align*}
Now, the first part of the expression above can be upper bounded as:
\begin{align*}
(1-p_L)^{t} - (1-p_U)^{t} &= \left(1-\frac{st-t}{nt}\right)^t - \left(1-\frac{s}{n}\right)^t \\
&= \left(1-\frac{s}{n}\right)^t\left(\left(1 + \frac{1}{n-s} \right)^t - 1\right) \\
&\leq \left(1 + \frac{1}{n-s}\right)^t - 1 \\
&\leq \frac{1}{1-\frac{t}{n-s}}-1 \\ 
&= \frac{t}{n-s-t} = n^{-1+o(1)} \ ,
\end{align*}
where the second inequality is due to Bernoulli's inequality. The second part of the expression can be upper bounded as:
\begin{align*}
\sum_{k=1}^{\floor{\log(n)}}\binom{t}{k} p_L^k &\leq \sum_{k=1}^{\floor{\log(n)}}\left(\frac{et}{k}\right)^k p_L^k \\
&\leq \sum_{k=1}^{\floor{\log(n)}} n^{-k\beta  + k\frac{\log(t)}{\log(n)} + o(1)} \\
&\leq n^{-\beta + \frac{\log(t)}{\log(n)} + o(1)} + \sum_{k=2}^{\floor{\log(n)}} n^{-k\beta + k\frac{\log(t)}{\log(n)} + o(1)} \\
&\leq n^{-\beta + o(1)} + \sum_{k=2}^{\floor{\log(n)}} n^{-k\frac{\beta}{2} + o(1)} \\
&\leq n^{-\beta + o(1)} \ ,
\end{align*}
where the first inequality is due to Stirling's inequality, and the fourth inequality is due to $t=n^{o(1)}$ such that $\log(t)/\log(n) = o(1)$ and for sufficiently large $n$, we thus have $\log(t)/\log(n) \leq \beta/2$. We can conclude that for any $\varepsilon > 0$,
\begin{equation}\label{eq:comp1-bound4}
\sum_{k=1}^{\floor{\log(n)}}\P{\pi \in \Pi^{(k)}} \leq n^{-\beta + o(1)} \ .
\end{equation}
For the other probability term in \eqref{eq:conditioning-comp1}, we have that:
\begin{align*}
\sum_{k=\ceil{\log(n)}}^t\P{\pi \in \Pi^{(k)}} &= \P{H \geq \ceil{\log(n)}} \leq \P{\text{Binomial}(t,p_U) \geq \ceil{\log(n)}} \\
&\leq \sum_{k=\ceil{\log(n)}}^{t}\left(\frac{et}{k}\right)^k \left(\frac{s}{n}\right)^k \\
&=\sum_{k=\ceil{\log(n)}}^t n^{-k\beta + k\frac{\log(t)}{\log(n)} + o(1)} \\
&\leq \sum_{k=\ceil{\log(n)}}^t n^{-k\beta + k\frac{\beta}{2}+ o(1)} = \sum_{k=\ceil{\log(n)}}^t n^{-k\frac{\beta}{2} + o(1)} \\
&\leq tn^{-\log(n)\frac{\beta}{2}} \leq n^{-\beta-q} \ , \numberthis\label{eq:comp1-bound5}
\end{align*}
where we have assumed that $t = n^{o(1)}$, and subsequently that the fraction $\frac{\log(t)}{\log(n)} \leq \frac{\beta}{2}$ for $n$ sufficiently large. We have also used that $\beta > 1/2$ and $q \leq 1$. 

Putting the results from \eqref{eq:comp1-bound3},  \eqref{eq:comp1-bound4}, and \eqref{eq:comp1-bound5} in \eqref{eq:conditioning-comp1}, we have that there exists a deterministic sequence $g_{n} \to 0$ such that:
\begin{equation}\label{eq:comp1}
\Pha{\tilde P_{q_n}(\bX) - \tilde P_{0,q_n}'(\bX) \leq n^{-\beta-q+g_{n}}} \to 1 \ .
\end{equation}
For the second term in \eqref{eq:decomp-pqx}, we must show that $\tilde P_{0,q}'(\bX)$ concentrates well around its mean. We use Chebyshev's inequality, requiring a characterization of the variance of $\tilde P'_{0,q}(\bX)$, obtained by carefully characterizing the dependency between two permutation streams. In many cases, these streams do not share any observations, and thus are independent. To characterize this rigorously, let $\pi_*$ be some arbitrary fixed permutation from $\Pi^{(0)}$. We then partition $\Pi^{(0)}$ in sets $\{\Pi^{(0)}_k(\pi_*)\}_{k=0}^t$ as follows: we have $\pi \in \Pi^{(0)}_k(\pi_*)$ if $\pi$ permutes precisely $k$ of the same coordinates as $\pi_*$ to the first stream. Specifically:
\[
\Pi^{(0)}_k(\pi_*) \equiv \left\{ \pi \in \Pi : \sum_{j=1}^t \ind{\pi ^{-1}(1,j) = \pi^{-1}_*(1,j)} = k\right\} \ .
\]
In particular, note that if $\pi\in\Pi^{(0)}_0(\pi_*)$ with $\pi_*$ some fixed permutation, this means that the random variables $Y_1(\bX^\pi)$ and $Y_1(\bX^{\pi_*})$ are independent (given the two permutations). Note that, since $n \gg t$, intuitively $\Pi \approx \Pi_0(\pi_*)$ for any permutation $\pi_*$, such that one should expect a very weak dependency between $Y_1(\bX^\pi)$ and $Y_1(\bX^{\pi_*})$ when $\pi \in \Pi$ uniformly at random.

Now, we first find the second moment of $\tilde P'_{0,q}(\bX)$. Let $\pi_1$ and $\pi_2$ be permutations uniformly and independently at random from $\Pi^{(0)}$, such that:
\begin{align*}
\E{\tilde P'_{0,q}(\bX)^2} &= \bbE\bigg[\EBig{\ind{Y_1(\bX^{\pi_1}) \geq z_q} \mathrel{\Big|} \bX}\EBig{\ind{Y_1(\bX^{\pi_2}) \geq z_q} \mathrel{\Big|} \bX}\bigg] \\
&= \bbE\bigg[ \EBig{\ind{Y_1(\bX^{\pi_1}) \geq z_q}\ind{Y_1(\bX^{\pi_2}) \geq z_q} \mathrel{\Big|} \bX} \bigg] \\
&= \bbE\bigg[ \ind{Y_1(\bX^{\pi_1}) \geq z_q}\ind{Y_1(\bX^{\pi_2}) \geq z_q} \bigg] \\
&= \sum_{\pi \in \Pi^{(0)}}\sum_{k=0}^t\sum_{\xi\in\Pi^{(0)}_k(\pi)}\bbE\bigg[ \ind{Y_1(\bX^{\pi_1}) \geq z_q}\ind{Y_1(\bX^{\pi_2}) \geq z_q} \mathrel{\Big|} \pi_1 = \xi, \pi_2 = \pi \bigg]\\&\hspace{8.5cm}\cdot\P{\pi_1 = \xi}\P{\pi_2 = \pi} \ .\numberthis\label{eq:sec-mom-pq-interm}
\end{align*}
At this point, it is convenient to look at the summands of $k$ in \eqref{eq:sec-mom-pq-interm} individually. First, note that for $k=0$ we have that, due to independence:
\begin{align*}
& \sum_{\pi \in \Pi^{(0)}}\sum_{\xi\in\Pi_0^{(0)}(\pi)} \bbE\bigg[ \ind{Y_1(\bX^{\pi_1}) \geq z_q}\ind{Y_1(\bX^{\pi_2}) \geq z_q} \mathrel{\Big|} \pi_1 = \xi, \pi_2 = \pi \bigg]\P{\pi_1=\xi}\P{\pi_2 = \pi} \\
&= \sum_{\pi \in \Pi^{(0)}}\sum_{\xi\in\Pi_0^{(0)}(\pi)}  \bbE\bigg[ \ind{Y_1(\bX^{\xi}) \geq z_q} \bigg] \bbE\bigg[\ind{Y_1(\bX^{\pi}) \geq z_q} \bigg]\P{\pi_1=\xi}\P{\pi_2 = \pi} \\
&\leq \sum_{\pi \in \Pi^{(0)}} \Bigg\{ \bbE\bigg[\ind{Y_1(\bX^{\pi}) \geq z_q} \bigg]\P{\pi_2 = \pi} \sum_{\xi\in\Pi^{(0)}}  \bbE\bigg[ \ind{Y_1(\bX^{\xi}) \geq z_q} \bigg] \P{\pi_1=\xi} \Bigg\} \\
&= \sum_{\pi \in \Pi^{(0)}} \left\{ \bbE\bigg[\ind{Y_1(\bX^{\pi}) \geq z_q} \bigg]\P{\pi_2 = \pi}  \E{\tilde P'_{0,q}(\bX)} \right\} \\
&= \E{\tilde P'_{0,q}(\bX)}^2 \ .
\end{align*}

Now, before proceeding to bound the term in \eqref{eq:sec-mom-pq-interm} for $k\geq1$, we first define
\[
\rho_k \equiv \P{\xi \in \Pi^{(0)}_k(\pi)} \ ,
\]
when $\xi$ is sampled uniformly at random from $\Pi^{(0)}$, and $\pi \in \Pi^{(0)}$ arbitrarily and fixed. Note that this is a hypergeometric probability; the probability corresponds to choosing $t$ indexes, of which $k$ indexes should match the $t$ indexes placed in the first stream by the permutation $\pi$, out of the $(n-s)t$ indexes in total, without replacement. To characterize this probability, we use a stochastic domination argument like before; note that a hypergeometric random variable $H$ with $(n-s)t$ total states, with $t$ success states and a sample size of $t$, is stochastically dominated by a binomial random variable with $t$ draws with success probability~$t/((n-s)t)$. Therefore:
\begin{align*}
\sum_{k=1}^t \rho_k &= \P{H \geq 1} \leq \P{\text{Bin}\left(t,\frac{1}{n-s}\right) \geq 1} \leq \sum_{k=1}^t\binom{t}{k}\left(\frac{1}{n-s}\right)^k \\
&\leq \sum_{k=1}^t\left(\frac{et}{nk}\right)^k\left(1 + \bigO\left(\frac{s}{n}\right)\right)^{k} \leq \sum_{k=1}^t n^{-k + k\frac{\log(t)}{\log(n)} + o(1)} \\
&\leq n^{-1+o(1)} + \sum_{k=2}^tn^{-k(1-\varepsilon) + o(1)} \leq n^{-1+o(1)}\ , \numberthis\label{eq:order-rhok}
\end{align*}
where the fifth inequality holds for any $\varepsilon > 0$ since $t = n^{o(1)}$, and thus for sufficiently large $n$ we have $\log(t)/\log(n) \leq \varepsilon$. Then, for the summands $k \geq 1$ in \eqref{eq:sec-mom-pq-interm}, we now have that:
\begin{align*}
& \sum_{\pi \in \Pi^{(0)}}\sum_{\xi\in\Pi^{(0)}_k(\pi)} \bbE\bigg[ \ind{Y_1(\bX^{\pi_1}) \geq z_q}\ind{Y_1(\bX^{\pi_2}) \geq z_q} \mathrel{\Big|} \pi_1 = \xi, \pi_2 = \pi \bigg]\P{\pi_1=\xi}\P{\pi_2 = \pi} \\
&\leq \sum_{\pi \in \Pi^{(0)}} \bbE\bigg[ \ind{Y_1(\bX^{\pi_2}) \geq z_q} \mathrel{\Big|} \pi_2 = \pi \bigg]\P{\pi_2 = \pi} \sum_{\xi\in\Pi^{(0)}_k(\pi)}  \P{\pi_1=\xi} \\
&= \rho_k\E{\tilde P'_{0,q}(\bX)} \ .
\end{align*}
For the second moment, we therefore have that:
\[
\E{\tilde P'_{0,q}(\bX)^2} \leq \E{\tilde P'_{0,q}(\bX)}^2 + \E{\tilde P'_{0,q}(\bX)}\sum_{k=1}^t\rho_k = \E{\tilde P'_{0,q}(\bX)}^2 + \tilde p_q\sum_{k=1}^t\rho_k \ ,
\]
where the equality holds by definition of $\tilde P'_{0,q}(\bX)$. Now, letting $q_n \to q$, the variance of $\tilde P'_{0,q_n}(\bX)$ can be bounded by:
\begin{align*}
\Var{\tilde P'_{0,q_n}(\bX)} &\leq \tilde p_{q_n}\sum_{k=1}^t\rho_k \leq n^{-1-q+o(1)} \ ,
\end{align*}
where the second inequality is due to Lemma~\ref{lem:sum-normality} and the result in \eqref{eq:order-rhok}. Now, Chebyshev's inequality implies that, for any $\varepsilon > 0$, we have:
\begin{equation}\label{eq:comp2}
\P{ \tilde P'_{0,q_n}(\bX) - \E{\tilde P'_{0,q_n}(\bX)} \geq n^{-\frac{1+q}{2}+\varepsilon}} \leq \frac{\Var{\tilde P'_{0,q_n}(\bX)}}{n^{-1-q+2\varepsilon}} \leq n^{-\varepsilon} \to 0  \ .
\end{equation}
We have now bounded both components of  Equation \eqref{eq:decomp-pqx}. Combining our results of \eqref{eq:comp1} and \eqref{eq:comp2} implies that, for any $\varepsilon > 0$, there exists a sequence $g_n \to 0$ such that:
\[
\P{\tilde P_q(\bX) - \tilde p_q \leq n^{\max\{-\beta-q, n^{-\frac{1+q}{2}}\}+\varepsilon +g_n} } \to 1 \ ,
\]
concluding the proof.
\end{proof}

\subsection{Proof of Lemma~\ref{lem:sum-normality}}
To streamline the presentation let $W_1,\ldots,W_t$ be i.i.d.~with distribution $F_\theta$ and denote by $\varphi_\theta(x)$ the moment generating function of $F_\theta$. Define also $\tau = \sqrt{(2q_n/t)\log n}$. We start by getting an upper bound for the said probability when $r<q$. A simple Chernoff bounding argument yields
\begin{equation}\label{eq:tail_probability}
\Pha{\frac{1}{t}\sum_{j\in[t]} W_j\geq \tau} \leq \exp{-t\left[\sup_{\lambda\in[0,\theta_*-\theta)} \{\lambda\tau-\log(\varphi_\theta(\lambda))\}\right]}\ .
\end{equation}
We must now characterize $\varphi_\theta(\lambda)$. First note that $\varphi_\theta(\lambda)=\varphi_0(\lambda+\theta)/\varphi_0(\theta)$. Now, we develop a Taylor expansion of $\varphi_0(\lambda)$ around $\lambda=0$, as we did in Equation~\eqref{eq:taylor-phi0}. Note that $F_0$ has zero mean and unit variance. We obtain:
\begin{equation}\label{eq:varphi_taylor}
\varphi_0(\lambda)=1+\frac{\lambda^2}{2}+\bigO(\lambda^3)\ ,
\end{equation}
as $\lambda\to 0$. A similar expansion can be developed for $\varphi_0( \lambda + \theta)$ around $\lambda + \theta = 0$. Combining all this yields
\[
\varphi_\theta(\lambda) = \frac{1 + 1/2(\lambda+\theta)^2 + \bigO((\lambda+\theta)^3)}{1 + \theta^2/2 + \bigO(\theta^3)} = 1+\theta\lambda+\frac{\lambda^2}{2}+\bigO((\lambda+\theta)^3)\ ,
\]
as both $\lambda,\theta\to 0$, where we used a Taylor expansion for the fraction around $\theta^2/2 + \bigO(\theta^3) = 0$. This suggests the choice $\lambda^*=\tau-\theta$, which is positive provided $n$ is large enough since $r<q$. This choice yields the bound
\begin{align*}
\sup_{\lambda\in[0,\theta_*-\theta)} \{\lambda\tau-\log(\varphi_\theta(\lambda))\} &\geq \tau^2-\log(\varphi_\theta(\lambda))\\
&\geq \frac{1}{2}(\tau-\theta)^2+\bigO\left(\tau^3\right)\ ,
\end{align*}
since $\tau>\theta$ and we used the basic inequality $\log(1+x)\leq x$. When $t=\omega(\log^3 n)$ the first term dominates, and therefore we conclude that
\begin{align*}
\Pha{\frac{1}{t}\sum_{j\in[t]} W_j\geq \tau} &\leq \exp{-(\sqrt{q_n}-\sqrt{r})^2 (\log n)+\smallO(1)}\\
&= n^{-(\sqrt{q}-\sqrt{r})^2+\smallO(1)}\ .
\end{align*}

To lower-bound the probability in~\eqref{eq:tail_probability} we use a tilting argument. Let $\theta_\tau$ be such that $F_{\theta_\tau}$ has mean $\tau$. Such a choice exists for $n$ large enough and necessarily $\theta_\tau>\theta$ for large $n$, since $r<q$. Define $\tilde W_1,\ldots, \tilde W_t$ to be i.i.d.~with distribution $F_{\theta_\tau}$. Then

\begin{align*}
\lefteqn{\Pha{\frac{1}{t}\sum_{j\in[t]} W_j\geq \tau} = \int \ind{\frac{1}{t}\sum_{j\in[t]} w_j\geq \tau} dF_\theta(w_1)\cdots dF_\theta(w_t)} \\
&= \int \ind{\frac{1}{t}\sum_{j\in[t]} w_j\geq \tau} \prod_{j=1}^t \exp{\theta w_j - \log\varphi_0(\theta)} dF_0(w_1)\cdots dF_0(w_t)\\
&= \int \ind{\frac{1}{t}\sum_{j\in[t]} w_j\geq \tau} \prod_{j=1}^t \exp{(\theta- \theta_\tau) w_j - \log\left(\frac{\varphi_0(\theta)}  {\log\varphi_0(\theta_\tau)}\right)} dF_{\theta_\tau}(w_1)\cdots dF_{\theta_\tau}(w_t)\\
&= \left( \frac{\varphi_0(\theta_\tau)}{\varphi_0(\theta)} \right)^t \E{\ind{\frac{1}{t}\sum_{j\in[t]} \tilde W_j \geq \tau} \exp{-(\theta_\tau-\theta)\sum_{j\in[t]} \tilde W_j}}\ .
\end{align*}
With this change of measure we can conveniently use the central limit theorem to get a meaningful bound. Begin by noting that
\begin{align*}
\lefteqn{\Pha{\frac{1}{t}\sum_{j\in[t]} W_j\geq \tau}}\\
&\geq \left( \frac{\varphi_0(\theta_\tau)}{\varphi_0(\theta)} \right)^t \E{\ind{0\leq \frac{1}{\sqrt{t}}\sum_{j\in[t]} \frac{\tilde W_j-\tau}{\sigma_{\theta_\tau}} \leq 1} \exp{-(\theta_\tau-\theta)\sum_{j\in[t]} \tilde W_j}}\\
&\geq \left( \frac{\varphi_0(\theta_\tau)}{\varphi_0(\theta)} \right)^t \exp{-(\theta_\tau-\theta)(t\tau+\sqrt{t}\sigma_{\theta_\tau})}\P{0\leq \frac{1}{\sqrt{t}}\sum_{j\in[t]} \frac{\tilde W_j-\tau}{\sigma_{\theta_\tau}} \leq 1}\ ,
\end{align*}
where $\sigma^2_{\theta_\tau}$ denotes the variance of $F_{\theta_\tau}$. By the central limit theorem we know that 
\[
\frac{1}{\sqrt{t}}\sum_{j\in[t]} \frac{\tilde W_j-\tau}{\sigma_{\theta_\tau}}
\] 
converges in distribution to a standard normal distribution and therefore the probability in the expression above converges to $\Phi(1)-\Phi(0)\approx 0.34>1/4$. We conclude that, for $n$ large enough
$$\Pha{\frac{1}{t}\sum_{j\in[t]} W_j\geq \tau} \geq \frac{1}{4}\left( \frac{\varphi_0(\theta_\tau)}{\varphi_0(\theta)} \right)^t \exp{-(\theta_\tau-\theta)(t \tau+\sqrt{t}\sigma_{\theta_\tau})}\ .$$
To control the remaining terms recall that $\varphi_0(\lambda)=1+\lambda^2/2+\bigO(\lambda^3)$ as $\lambda\to 0$ (see Equation~\eqref{eq:varphi_taylor}). Note also that
$\tau=\theta_\tau+\bigO(\theta^2_\tau)$, which implies (after some manipulation) that $\theta_\tau=\tau+\bigO(\tau^2)$. Finally, note that both $\tau$ and $\theta$ have the same order of magnitude. Putting all this together we conclude that
\[
\log\left(\left( \frac{\varphi_0(\theta_\tau)}{\varphi_0(\theta)} \right)^t\right) = \frac{t}{2}\left(\tau^2-\theta^2 + \bigO(\theta^3)\right)\ .
\]
For the other term note that $\sigma_{\theta_\tau}=1+\smallO(1)$, and therefore $\sigma_{\theta_\tau}/\sqrt{t}=\smallO(\theta)$. This implies that
$$-(\theta_\tau-\theta)(t \tau+\sqrt{t}\sigma_{\theta_\tau})=-t\left(\tau(\tau-\theta)+\bigO(\theta^3)\right)\ .$$
In conclusion
$$\Pha{\frac{1}{t}\sum_{j\in[t]} W_j\geq \tau} \geq \frac{1}{4}\exp{-\frac{t}{2}\left((\tau-\theta)^2+\bigO(\theta^3)\right)}\ .$$
When $t=\omega(\log^3 n)$ the term $(\tau-\theta)^2$ term dominates, and we see we get the asymptotic behavior as in the upper bound, concluding the proof.

For the case $r\geq q$ we see that necessarily $\P{\frac{1}{t}\sum_{j\in[t]} W_j\geq \tau}\geq n^{\smallO(1)}$, so the tail probability cannot be extremely small. In fact, when $r>q$ this probability will be lower bounded by a constant.

\subsection{Proof of Corollary~\ref{cor:longstreams}}\label{sec:proofs-cor:longstreams}
\begin{proof}
To prove this corollary we simply map the original dataset to a new dataset with short streams, for which we can apply Theorem~\ref{th:perm-hc-discrete-test}. Partition the set $[t]$ into $\tilde t$ sets of size $k$ such that $\tilde t \equiv \frac{t}{k} = n^{o(1)}$ and $\tilde t = \omega(\log^3(n))$. For simplicity, we assume $t$ is divisible by $k$ and define
\[
\tilde X_{ij} \equiv \frac{1}{\sqrt{k}}\sum_{\ell = (j-1)k + 1}^{jk} X_{i\ell} \ .
\]
Recall that $X_{ij}$ belongs to a natural exponential family with natural parameter $\theta_i$. Note that the distribution of $\tilde X_{ij}$ also belongs to a natural exponential family. Particularly, let $\tilde F_0$ be the distribution of $\tilde X_{ij}$ when $\theta_i=0$. Then the density of $\tilde X_{ij}$ with respect to $\tilde F_0$ is given by $\exp{\tilde\theta_i x - \log(\tilde\varphi_0(\tilde\theta_i))}$ where $\tilde \varphi_0\left(\theta\right) \equiv \varphi_0(\theta/\sqrt{k})^k$ is the moment generating function of $\tilde F_0$ and $\tilde\theta_i \equiv \theta_i\sqrt{k}$. Note that $\tilde F_0$ has variance $\sigma_0^2$ and so, using the parameterization in Theorem~\ref{th:perm-hc-discrete-test} we have for $i\in\cS$
\[
\tilde \theta_i = \theta\sqrt{k} =  \sqrt{2r / (\sigma^2_0 t) \log(n)}\sqrt{k} = \sqrt{2r / (\sigma^2_0\tilde t) \log(n)} \ .
\] 
We can now apply the test as in Theorem~\ref{th:perm-hc-discrete-test} on the set of observations $\tilde \bX \equiv \{ \tilde X_{ij} : i\in[n], j\in[\tilde t] \}$. Since $\tilde t = n^{o(1)}$ and $\tilde t = \omega(\log^3(n))$, Theorem~\ref{th:perm-hc-discrete-test} implies the test has power converging to one provided $r > \rho^*(\beta)$. If $t$ is not divisible by $k$ one can simply ignore the last observations in each stream of the original data and proceed as above.
\end{proof}

\newpage
\appendix
\section{Appendix}

\subsection{Proof of Proposition~\ref{prop:vq-test}}\label{supp:vq}
\begin{proof}
For convenience, define  $|\cS| = s$. The first statement in the proposition is a simple consequence of Chebyshev's inequality. Under the null hypothesis $N_q(\bX) \sim \text{Binomial}(n,p_q)$, therefore
\begin{align*}
\Phn{ V_q(\bX) \geq h_n} &= \Phn{ \frac{N_q(\bX) - np_q}{\sqrt{np_q(1-p_q)}} \geq h_n }\\
& = \Phn{N_q(\bX) - np_q \geq h_n\sqrt{np_q(1-p_q)}} \\
& \leq \Phn{\abs{N_q(\bX) - np_q} \geq h_n\sqrt{np_q(1-p_q)}} \\
& \leq \frac{1}{h_n^2}\ .
\end{align*}
Since $h_n\to\infty$ the statement of the proposition follows.

Consider first the sparse regime $1/2<\beta<1$. Under the alternative hypothesis $N_q(\bX)$ is a sum of two independent binomial random variables, that is, $N_q(\bX) \sim \Bin{n-s}{p_q} + \Bin{s}{v_q}$. Each of the variables corresponds respectively to the counts of the null and anomalous stream means exceeding the threshold $\sqrt{2q\log(n)}$, where
\[
v_q = 1-\Phi\left(\sqrt{2\log(n)}(\sqrt{q}-\sqrt{r})\right)\ .
\]
In words, $v_q$ is the probability an anomalous stream mean exceeds the threshold $\sqrt{2q\log(n)}$. For convenience of presentation define also
\begin{equation}\label{eqn:surrogate_expression}
a_n \equiv s(v_q -p_q) - h_n\sqrt{np_q(1-p_q)}\ .
\end{equation}
Using this definition we see that
\begin{align*}
\Pha{ V_q(\bX) < h_n} &= \Pha{ N_q(\bX) \leq h_n\sqrt{np_q(1-p_q)} + np_q} \\
&= \Pha{N_q(\bX) - \E{N_q(\bX)}  \leq h_n\sqrt{np_q(1-p_q)} + s(p_q - v_q) } \\
&= \Pha{N_q(\bX) - \E{N_q(\bX)}  \leq -a_n} \\
&= \Pha{ -\left(N_q(\bX) - \E{N_q(\bX)} \right) \geq a_n} \\
&\leq  \Pha{ \abs{N_q(\bX) - \E{N_q(\bX)}} \geq a_n}\ . \numberthis \label{eq:vq-model-intermediate}
\end{align*}
When $a_n>0$ we can easily bound the above expression using Chebyshev's inequality. To make the presentation and derivations clear and simple we make use of asymptotic notation. All the statements below are taken when $n\to\infty$. Using Lemma~\ref{lem:normalcdf-order} we see that
\begin{equation}\label{eq:order-probs}
p_q = n^{-q + o(1)} \qquad \text{and} \qquad
v_q = \begin{cases}
n^{-(\sqrt{q}-\sqrt{r})^2 + o(1)} &\text{ if } r < q \\
\frac{1}{2} &\text{ if } r = q\\
1 - n^{-(\sqrt{r}-\sqrt{q})^2 + o(1)} &\text{ if } r > q
\end{cases}\ .
\end{equation}
Now, since $h_n = n^{o(1)}$, we have that $a_n$ is given by:
\begin{equation}\label{eq:a_n}
a_n = \begin{cases}
n^{1-\beta-(\sqrt{q}-\sqrt{r})^2 + o(1)} - n^{\max\{\frac{1-q}{2},1-\beta-q\} + o(1)} &\text{ if } r < q \\
n^{1-\beta + o(1)} - n^{\max\{\frac{1-q}{2},1-\beta-q, 1-\beta-(\sqrt{q}-\sqrt{r})^2\} + o(1)} &\text{ if } r \geq q \\
\end{cases}.
\end{equation}
We can now easily retrieve the conditions under which $a_n>0$ for large enough $n$, namely the exponent of the positive terms must be larger than the exponent of the negative terms in the right-hand-side of Equation~\eqref{eq:a_n}. As $a_n$ differs depending on whether $r < q$ or $r \geq q$ we obtain a different set of conditions of each of these cases. However, many of such conditions are satisfied trivially, and for each case we obtain only one single nontrivial condition:
\begin{align}
\text{ If $r < q$ then $a_n\to\infty$ provided}\quad & 1-\beta -(\sqrt{q}-\sqrt{r})^2 > \frac{1-q}{2}\ ; \label{eq:condition1}\\
\text{ If $r \geq  q$ then $a_n\to\infty$ provided}\quad & 1-\beta > \frac{1-q}{2}. \label{eq:condition2}
\end{align}
Under these conditions we can continue from~\eqref{eq:vq-model-intermediate} via Chebyshev's inequality and obtain
\[
\Pha{ V_q(\bX) < h_n} \leq \frac{\Var{N_q(\bX)}}{a_n^2}\ .
\]
To analyze the bound, first observe that
\[
\Var{N_q(\bX)} = (n-s)p_q(1-p_q) + sv_q(1-v_q) \leq np_q + sv_q(1-v_q).
\]
Note that $v_q$ and $1-v_q$ are equal for $r < q$ and $r > q$ respectively, and thus $v_q(1-v_q)$ is equal for both cases asymptotically. Moreover, the case $r=q$ can be merged with $r\neq q$ by observing that $\frac{1}{4}n^{1-\beta} = \frac{1}{4}n^{1-\beta - (\sqrt{q}-\sqrt{r})^2} = n^{1-\beta - (\sqrt{q}-\sqrt{r})^2 + o(1)}$. We therefore obtain for all $r > 0$:
\[
\Var{N_q(\bX)} = n^{ \max \{1-q, 1-\beta - (\sqrt{q}-\sqrt{r})^2 \} + o(1)}.
\]
Consequently
\[
\Pha{ V_q(\bX) < h_n} \leq 
\begin{cases}
n^{ \max \{1-q, 1-\beta - (\sqrt{q}-\sqrt{r})^2 \} - 2(1-\beta - (\sqrt{q}-\sqrt{r})^2) + o(1)} &\text{ if } r < q, \\
n^{ \max \{1-q, 1-\beta - (\sqrt{q}-\sqrt{r})^2 \} - 2(1-\beta) + o(1)} &\text{ if } r \geq q.
\end{cases}
\]
All that needs to be done at this point is to identify under which conditions the above bound converges to zero. Clearly, when $r \geq q$ this is the case. If $r \leq q$ we require that:
\begin{align}
&1-q - 2(1-\beta - (\sqrt{q}-\sqrt{r})^2) < 0 \text{ and } \label{eq:condition3} \\
&1-\beta - (\sqrt{q}-\sqrt{r})^2 > 0\ . \label{eq:condition4}
\end{align}
Therefore, for the results in the proposition to hold we require that conditions~\eqref{eq:condition1},~\eqref{eq:condition2},~\eqref{eq:condition3} and~\eqref{eq:condition4} are simultaneously satisfied. We can easily see that~\eqref{eq:condition1} and~\eqref{eq:condition3} are equivalent. After rewriting condition~\eqref{eq:condition3} slightly, we are then left with the following three conditions:
\begin{align}
\text{ if } r \geq q, \text{ then }& 1-\beta > \frac{1-q}{2}. \label{eq:condition1a} \\
\text{ if } r < q, \text{ then }& 1-\beta - (\sqrt{q}-\sqrt{r})^2 > \frac{1-q}{2}, \label{eq:condition2a} \\
\text{ if } r < q, \text{ then }& 1-\beta - (\sqrt{q}-\sqrt{r})^2 > 0. \label{eq:condition3a}
\end{align}
At this point it is a matter of algebra to check these conditions are satisfied for the statements in the proposition. Note that we do not provide results for $\beta=1$, as this would require taking $q>1$.
\end{proof}

\subsection{Proof of Proposition~\ref{th:hc-discrete-test}}\label{supp:hc-disc}
\begin{proof}
Consider first the analysis under the null hypothesis. A simple application of the result in Proposition~\ref{prop:vq-test} and a union of events bound yields
\begin{align*}
\Phn{T(\bX)\geq h_n} &= \sum_{q\in Q} \Phn{V_q(\bX)\geq h_n}\\
&\leq \frac{k_n+1}{h_n^2}\to 0\ ,
\end{align*}
where the last statement follows from the assumptions on $h_n$ and $k_n$.

Now we turn our attention to the alternative hypothesis, and must show that
$$\Pha{T(\bX)\geq h_n}\to 1\ ,$$
as $n\to\infty$, under the conditions in the theorem for $r$ and $\beta$. For the regimes where the choices $q=0$ or $q=1$ are optimal it is straightforward to ensure that $\Pha{V_q\geq h_n}\to 1$, so the theorem result is obviously true. For instance,
$$\Pha{\max_{q\in Q} V_q(\bX) \geq h_n}\geq \Pha{V_1 \geq h_n}\to 1$$
provided $\mu=\sqrt{2r\log(n)}$ and $r>(1-\sqrt{1-\beta})^2$. Likewise a similar argument holds in the dense case, where the choice $q=0$ is the best one. 

The only situation that is slightly more intricate is that when we must consider the choice $q=4r$, as in general that value is not in the grid $Q$. However, since $k_n\to\infty$ we will be able to find a value in $Q$ that is close enough to $4r$. Consider the case $\mu=\sqrt{2r\log(n)}$, $r<1/4$ and $r>\beta-1/2$. Let $q^*=\min_{q\in Q} |4r-q|$. Clearly $|q^*-4r|\leq\frac{1}{k_n}$ so $q^*=4r+\smallO(1)$. Therefore $p_{q^*}=n^{-4r+\smallO(1)}$ and $v_{q^*}=n^{-r+\smallO(1)}$ and we can follow exactly the same steps as in the proof of Proposition~\ref{prop:vq-test}. Therefore
\[
\Pha{\max_{q\in Q} V_q(\bX) \geq h_n}\geq \Pha{V_{q^*} \geq h_n}\to 1\ ,
\]
which is what we wanted to show.
\end{proof}

\subsection{Statement and proof regarding Remark~\ref{rem:HC_equivalence}}\label{supp:vq-equiv}
The following lemma formalized the equivalence between the traditional higher-criticism statistic as in~\eqref{eqn:HC_donoho} and the $V_q(\bX)$ statistic introduced in~\eqref{eqn:def-v}.

\begin{lem}\label{lem:hc-vq-equiv}
\begin{equation}\label{eqn:hc-vq-equiv}
\sup_{q\in[0,\infty)} \left\{V_q(\bX)\right\} = \max_{i \in [i_+]} \left\{ \sqrt{n}\frac{\frac{i}{n} - \mathfrak{p}_{(i)}}{\sqrt{\mathfrak{p}_{(i)}(1-\mathfrak{p}_{(i)})}} \right\},
\end{equation}
where $i_+$ is the largest value of $i$ for which $\mathfrak{p}_{(i)}<1/2$.
\end{lem}

The proof follows by noting that the supremum of $V_q(\bX)$ is attained at a finite number of points. This result is essentially stated in \cite{Donoho2004} but without a complete proof. Note that the right-hand-side is not entirely equal to the higher criticism statistic, but it is very similar (particularly when taking $\alpha_0=1/2$, which is a common recommendation). Taking $\alpha_0$ means that essentially almost all streams that have $p$-values larger than $1/2$ will be ignored.

\begin{proof}
We show our assertion~\eqref{eqn:hc-vq-equiv} in two steps. First, we will show that the supremum of $V_q(\bX)$ is attained at a finite number of points. We will then show that the maximum of these points equals the HC statistic, if $\alpha_0$ is chosen appropriately. For ease of notation, define:
\[
h_{i}(p) = \sqrt{n}\frac{\frac{i}{n} - p}{\sqrt{p(1-p)}},
\]
such that we can succinctly write our classical higher-criticism statistic as the maximum over $h_i(\mathfrak{p}_{(i)})$:
\[
\max_{i \in [n]} \left\{ \sqrt{n}\frac{\frac{i}{n} - \mathfrak{p}_{(i)}}{\sqrt{\mathfrak{p}_{(i)}(1-\mathfrak{p}_{(i)})}} \right\} = \max_{i \in [n]} \left\{ h_i(\mathfrak{p}_{(i)}) \right\}
\]
To start, define:
\[
q_{i} = \frac{t\Xb^2_{i}}{2\log(n)}.
\]
for all $Y_i > 0$. Denote $\Xb_{(i)}$ the ordered stream means, with $\Xb_{(1)}$ being the largest. Note that $q_i$ is monotone in $Y_i > 0$, and thus the ordering is preserved.

Note that $N_q(\bX)$ is a stepwise, nonincreasing function in $q$, constant on intervals $(q_{(i)}, q_{(i+1)}]$. Next, note that $V_q(\bX)$ is decreasing in $p_q$, and $p_q$ is decreasing in $q$. Therefore, as $N_q(\bX)$ is constant on $(q_{(i)}, q_{(i+1)}]$, $V_q(\bX)$ is increasing on $(q_{(i)}, q_{(i+1)}]$.

Finally, as $N_q(\bX)$ is nonincreasing, $V_{q_{(i)}}(\bX) \geq V_{q_{(i)}+\delta}(\bX)$. Therefore, the supremum of $V_q(\bX)$ must be attained at the points $q_{i}$.

Now, note that  $N_{q_{(i)}}(\bX) = i$ and $p_{q_{(i)}} = \mathfrak{p}_{(i)}$, and thus $V_{q_{(i)}} = h_{i}(\mathfrak{p}_{(i)})$. However, we only have these $q_{(i)}$ defined for $Y_i > 0$, so:
\[
\sup_{q\in[0,\infty)} \left\{V_q(\bX)\right\} = \max_{i \in [i_+]} \{ V_{q_{(i)}}\}= \max_{i \in [i_+]} \left\{ h_i(\mathfrak{p}_{(i)}) \right\},
\]
where $i_+ = \argmin_{i\in[n]} \{ Y_i : Y_i > 0 \}$.
\end{proof}

\subsection{An analysis of daily COVID-19 diagnoses across municipalities in the Netherlands}\label{app:covid}
To showcase another possible application of our methodology we consider a way to monitor the number of new daily diagnoses of COVID-19, aiming to quickly identify localized outbreaks. During the current COVID-19 pandemic countries experienced successive waves with large numbers of cases, interspersed with periods with more stable disease dynamics, typically due to measures put in place to limit its spread. In the latter regime it is of high importance for policy makers to quickly detect signs of new impending outbreaks.

One possible way to proceed is to monitor the daily number of diagnoses per capita in each separate municipality of the country over a short time frame, and test if a (small) subset of municipalities has a higher-than-usual number of cases. As such, the framework we have introduced in this paper can be useful here to detect signs of local outbreaks, as these would lead to rejection of our null hypothesis~\eqref{hyp:general}. A municipality may turn anomalous during the observed stream and not right from the start. However, recall that our methodology still has some power even when anomalous streams are only partially affected.

Within small time windows, it may be sensible (to some extent) to apply the proposed methodology directly to the raw data. Nevertheless, we also consider a more sophisticated approach by first fitting a model to small time windows of the raw data and then analyzing the residuals instead, as explained in Section~\ref{sec:intro}. Note, however, that our methodology is in principle not suitable for serially dependent data, and validity of the conclusions based on this residual analysis hinge crucially on the validity of the fitted model. In addition, estimated residuals are dependent (but usually only weakly provided an adequate model is chosen). In rigor, one should carefully address the influence of this dependency on the validity of the test conclusions, but this is outside of the scope of this manuscript.

\begin{figure}[ht]
\centering
\includegraphics[width=0.8\textwidth]{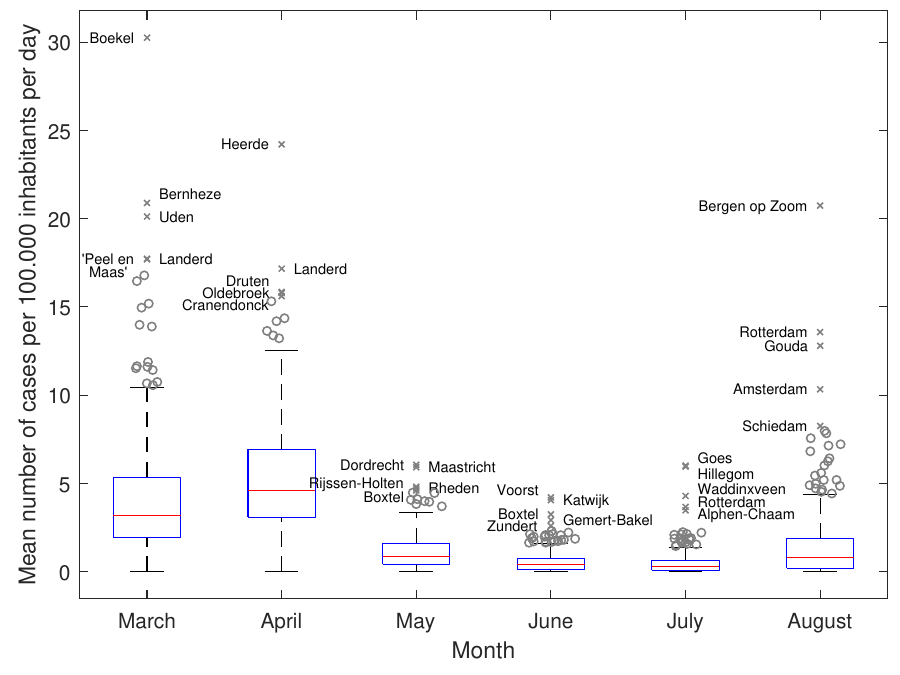}
\caption{Boxplots depicting the normalized number of monthly cases per municipality. For each municipality, the total number of monthly cases per 100.000 inhabitants is normalized by the number of days in the month. The five municipalities with the highest mean number of cases are depicted with a cross and labeled. Data in March pertains only the 13th up until the 31st of March, and August only pertains the 1st up until the 10th of August.}\label{fig:covid-descriptive}
\end{figure}

We consider data from The Netherlands. In this country, it is not unreasonable to assume municipalities are sufficiently comparable: the country is very small and population density is not too different among municipalities. We use data on newly diagnosed COVID-19 cases per 100.000 inhabitants from 13th of March up until 10th of August 2020, for each of the $n=355$ municipalities of the Netherlands. This data was processed from two sources; the data on the number of diagnoses (uncorrected for municipality population) was retrieved from the Dutch national institute for public health and the environment (RIVM) at \url{https://data.rivm.nl}\footnote{The specific hyperlink to this data is \rivmurl}, and the data on municipality population was retrieved from the Dutch central agency for statistics (CBS) at \url{https://opendata.cbs.nl}\footnote{The specific hyperlink to this data is \cbsurl}. These were combined to obtain the number of newly diagnosed COVID-19 cases per 100.000 inhabitants. Let $i\in\{1,\ldots,355\}$ denote a municipality and $j\in\{1,\ldots,151\}$ denote a day in the period above. Then $y_{i,j}$ denotes the \emph{daily rate of new cases per municipality} (daily-rate, for short). Specifically, the ratio between the number of new positive diagnoses on that day divided by the number of inhabitants of the municipality (as a multiplicative factor of 100.000). To give some insight into this data, we depict the number of new cases per municipality in each month in Figure~\ref{fig:covid-descriptive}. As can be seen, some municipalities clearly have a large number of cases when compared to the others, but they do not stay so large consistently throughout the months considered.

As we are interested in quickly detecting outbreaks our analysis focuses on very short time frames, namely windows of five consecutive days. This is motivated by the knowledge of the incubation time of the disease, believed to be on average around 5 days, and typically between two and fourteen days \citep{Lauer2020}. Within such a short time frame both the independence and stationarity assumptions might not be terribly unreasonable under the null; in a stable regime without local outbreaks, relatively few cases are distributed somewhat evenly over the population, and those infectious individuals tend not to infect many others (within that limited time frame). For larger windows of time one naturally expects the validity of such assumptions to be more questionable. That being said, within a five-day window there might be some amount of dependency and non-stationarity across time, even within the null streams. With this in mind an option is to attempt to capture such global trends and dependencies, and monitor the residual errors of such a model instead, as suggested in Section~\ref{sec:intro}. Both approaches are discussed below.

Formally, our analysis and results pertain a window of $t=5$ consecutive days, starting on day $w\in\{1,\ldots,147\}$. We present results for all the possible windows. In short, the raw observations for the window indexed by $w$ are $\big(x^{(w)}_{ij} : i\in[n], j\in[t]\big)$ where $x^{(w)}_{i,j}=y_{i,j+w-1}$.

Taking into account the short time-windows and the first order epidemic dynamics a simple but sensible model to consider for this data is an AR(1) model, as suggested in \cite{Shtatland2007, Shtatland2008}. Note, however, that application of our methodology is crucially dependent on the validity of the model, but for presentation purposes we stick with this relatively simple model. Specifically, the model assumes that for null streams the observations $x^{(w)}_{i,j}$ are obtained as a sample from
\[
X^{(w)}_{i,j}-\mu^{(w)}=a^{(w)} \left(X^{(w)}_{i,j-1}-\mu^{(w)}\right)+\varepsilon_{i,j}^{(w)}\ ,
\]
where $i\in[n]$, $j\in[t]$, $a^{(w)},\mu^{(w)}\in\mathbb{R}$ are (unknown) parameters of the model (common to all null streams), and $\varepsilon_{i,j}^{(w)}$ are i.i.d. samples from an unknown zero-mean distribution. Despite its simplicity, this model can capture some of the epidemic dynamics when applied to very short time frames, in contrast with more sophisticated epidemiological models (like the ones described in \cite{held2019handbook}).

The first step in this approach is to estimate the unknown parameters of the model. Given that we do not have knowledge of the distribution of the errors a natural choice is to use the ordinary least squares estimator
\[
(\hat a^{(w)}, \hat \mu^{(w)}) = \arg\min_{a, \mu\in\mathbb{R}}\left\{\sum_{i=1}^n\sum_{j=1}^t \Big( (x^{(w)}_{i,j}-\mu) - a(x^{(w)}_{i,j-1}-\mu) \Big)^2\right\} \ .
\]
Finally, the methodology proposed in this paper is then applied to the residual errors, namely $(\tilde x^{(w)}_{i,j}: i\in[n],j\in[t])$ where
\[
\tilde x^{(w)}_{i,j}\equiv x^{(w)}_{i,j}-\hat\mu^{(w)}-\hat a^{(w)}\left(x^{(w)}_{i,j-1}-\hat\mu^{(w)}\right)\ ,
\]

We apply our testing methodology both to the raw data $(x^{(w)}_{i,j})_{i\in[n],j\in[t]}$ and to the residuals $(\tilde x^{(w)}_{i,j})_{i\in[n],j\in[t]}$, and contrast the obtained results. Note that often there are few very large observations. While these cases are important in the context of the application, one needs no powerful test to mark them as anomalous as their abnormality is so clear. Especially in the context of COVID-19, these ``clear'' outliers will be investigated regardless. A more interesting question is then if, apart from these ``clear'' outliers, we can still detect higher-than-usual values among the other municipalities.

To remove ``clear'' outliers in a rigorous way we use the max test as described in Theorem~\ref{th:max-perm-test}. First we obtain the $95\%$ quantile of the permutation maximum stream mean distribution. We then mark all streams with means exceeding this threshold as ``clear'' anomalies. Finally, we apply our permutation higher criticism test as in Theorem~\ref{th:perm-hc-discrete-test} on the remaining data to detect possible signals. We also use the higher criticism test using a normal approximation as in \secref{comp-approx} for a comparison. See also Remark~\ref{rem:alternatives_model} for a discussion on possible alternatives to consider when fitting the AR(1) model.

\begin{figure}
\centering
\begin{subfigure}[t]{0.49\textwidth}
\includegraphics[width=\textwidth]{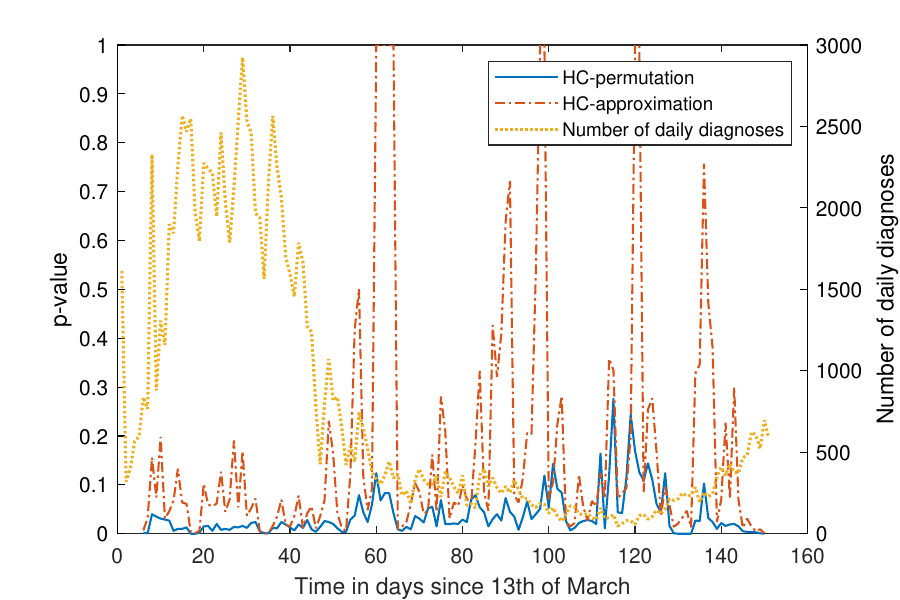}
\caption{Tests applied on the raw data.}
\end{subfigure}
\begin{subfigure}[t]{0.49\textwidth}
\includegraphics[width=\textwidth]{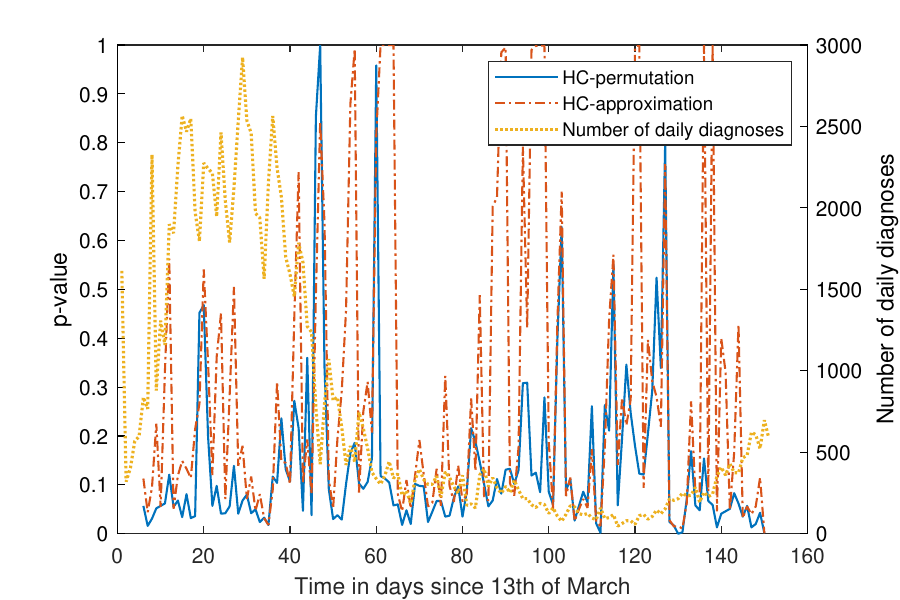}
\caption{Tests applied on the residuals.}
\end{subfigure}
\caption{The $p$-values of our permutation higher criticism test and the higher criticisms test using normal approximations for a window of the previous five days, along with the total number of daily diagnoses in the Netherlands. For each test $10^5$ permutations were used.}\label{fig:covid}
\end{figure}

We present the results obtained for each possible window of 5 consecutive days in Figure~\ref{fig:covid}. The $p$-value obtained by the testing procedures along with the virus' nationwide progression is depicted in that figure. Note that the time-windows are indexed by their starting day, so neighboring windows consider overlapping periods of time. Obviously, the obtained $p$-values are dependent and care must be taken if trying to interpret them jointly - the figure is given merely to aid the presentation.
As can be seen in Figure~\ref{fig:covid}, our test nearly always results in much smaller $p$-values than the approximation method at each window, both when using the raw data and the residuals of the AR(1) model. One can also see that, when the method is used on the residuals, larger $p$-values are typically observed. This indicates that in many of the windows considered, the use of the AR(1) model mitigates the effect of some dependencies and global trends that may unduly influence the conclusions. However, the autocorrelation parameter estimates were frequently small - across all windows considered, the estimates had a median value of 0.2243, and were smaller than 0.3 in 75\% of the windows. At 5\% significance, our test on the raw data rejects the null a total of 113 times out of 146, while the approximation test rejects a total of 49 times. On the residuals of our AR(1) model, our methodology rejects 43 times, while the approximation method rejects 20 times. Nevertheless, these figures should be interpreted with care, as the resulting $p$-values across different windows are dependent. 

There are cases where our test indicates anomalies in seemingly stable periods. In these periods, while hard to see in aggregated data, we thus have some evidence that some municipalities have larger-than-usual values. This does not necessarily lead to a nationwide outbreak, since local measures, such as a restricting access to specific nursery homes, might have been taken to prevent further spread. As our data is aggregated per municipality and local measures are very hard to identify, we cannot to take this into account in our analysis.

\begin{rem}
When the $p$-values of our test are small as above, one would ideally like to subsequently identify the anomalous municipalities. We refer to the ending of Section~\ref{sec:cdc} for a discussion on the possibilities when one would like to identify anomalous municipalities.
\end{rem}

\begin{rem}\label{rem:alternatives_model}
With respect to the exclusion of ``clear'' outliers, there are some natural alternatives to the approach above:

\begin{enumerate}[label=\Alph*.]
\item Remove obvious anomalies using the permutation distribution of the maximum stream mean on the raw data. Next, fit the AR(1) model on the remaining raw data, and apply the methodology on the residuals. In the results, we refer to this approach as ``approach A". \label{ap-raw}
\item First, fit an AR(1) model on the data. Then, identify obvious anomalous streams using the permutation distribution of the maximum stream mean on the residuals. If the stream mean of the residuals is larger than the 95\% quantile, we remove the corresponding stream from our original raw data. Fit a new AR(1) model on the remaining raw data streams. Use our methodology on the residuals following from the second model fit. In the results, we refer to this approach as ``approach B". \label{ap-refit}
\end{enumerate}

Compared to the previous approach, the first option avoids labeling streams as ``obvious'' anomalies based on the AR(1) model. The second option avoids using our methodology on residuals that arose from an AR(1) model fit which was unduly influenced by the presence of ``obvious'' anomalies.

The results do not qualitatively change compared to the results in the main text when these variations are considered. The results are presented in Figure~\ref{fig:covid-supp}.

\begin{figure}[!h]
\centering
\begin{subfigure}[t]{0.49\textwidth}
\includegraphics[width=\textwidth]{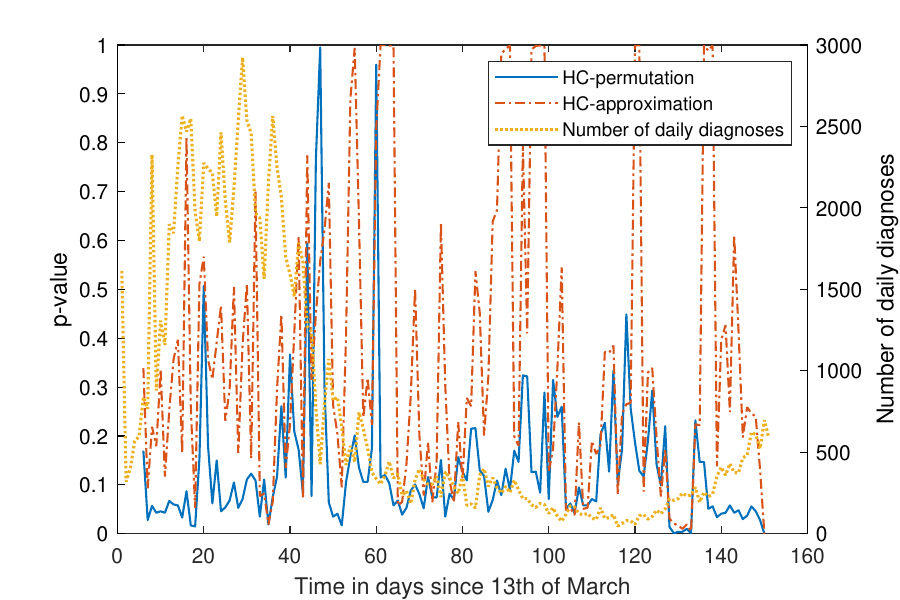}
\caption{Results following approach~\ref{ap-raw}}
\end{subfigure}
\begin{subfigure}[t]{0.49\textwidth}
\includegraphics[width=\textwidth]{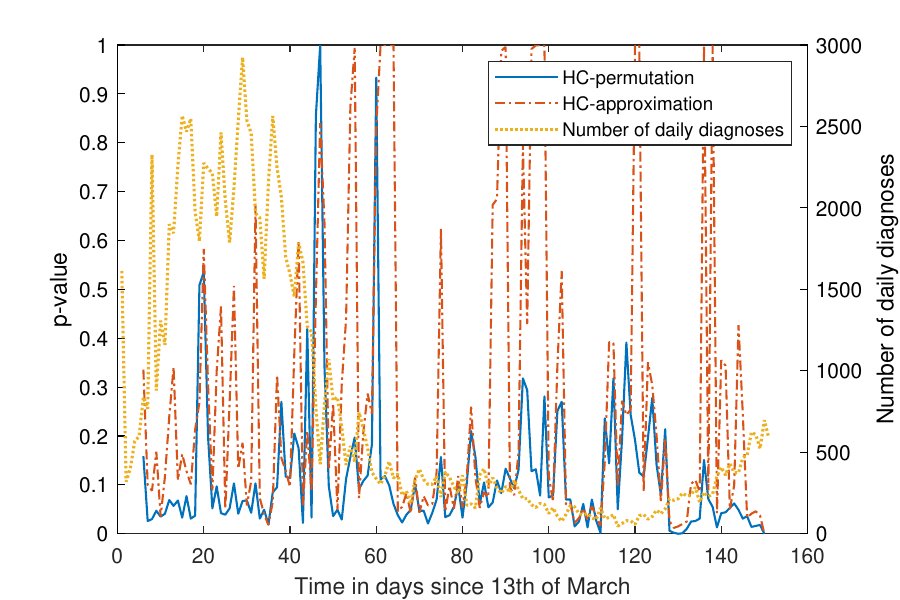}
\caption{Results following approach~\ref{ap-refit}}
\end{subfigure}
\caption{The $p$-values of our permutation higher criticism test and the higher criticisms test using normal approximations for a window of the previous five days, along with the total number of daily diagnoses in the Netherlands. For each test $10^5$ permutations were used.}\label{fig:covid-supp}
\end{figure}
\end{rem}

\newpage
\bibliographystyle{apalike}
\bibliography{rank-scan}

\end{document}